\documentclass[
    a4paper,
    notitlepage,
    prapplied,
    superscriptaddress,
    floatfix
]{revtex4-1}

\usepackage{mdframed}
\usepackage{amsmath,amssymb}
\usepackage{lipsum}
\usepackage{bm}
\usepackage{graphicx}
\usepackage{graphics}
\usepackage{amsmath}
\usepackage{array}
\usepackage{graphicx}
\usepackage[caption=false,farskip=0pt,labelfont={bf}]{subfig}
\usepackage{xcolor}
\usepackage{subfig}
\usepackage{hyperref}
\usepackage[capitalise]{cleveref}
\usepackage{textcomp}
\usepackage{epstopdf}
\usepackage{cases}
\usepackage{amssymb}
\usepackage{multirow}
\usepackage{siunitx}
\usepackage{cases}
\usepackage{tabularx}
\usepackage[top=1.1in, bottom=1.1in, left=1.0in, right=1.0in]{geometry}
\usepackage{tikz}
\usepackage{comment}
\usepackage{lineno,hyperref}
\modulolinenumbers[5]
\usepackage{booktabs}
\renewcommand{\arraystretch}{1.0} 

\begin{document}


\title{Synthetic iterative scheme for thermal applications in hotspot systems with large temperature variance}
\author{Chuang Zhang}
\email{zhangc520@hdu.edu.cn}
\affiliation{Department of Physics, Hangzhou Dianzi University, Hangzhou 310018, China}
\author{Qin Lou}
\email{louqin560916@163.com}
\affiliation{School of Energy and Power Engineering, University of Shanghai for Science and Technology, Shanghai 200093, China}
\author{Hong Liang}
\email{Corresponding author: lianghongstefanie@163.com}
\affiliation{Department of Physics, Hangzhou Dianzi University, Hangzhou 310018, China}
\date{\today}

\begin{abstract}

{\color{black}{A synthetic iterative scheme is developed for thermal applications in hotspot systems with large temperature variance.
Different from previous work with linearized equilibrium state and small temperature difference assumption, the phonon equilibrium distribution shows a nonlinear relationship with temperature and mean free path changes with the spatial temperature when the temperature difference of system is large, so that the same phonon mode may suffer different transport processes in different geometric regions.
In order to efficiently capture nonlinear and multiscale thermal behaviors, the Newton method is used and a macroscopic iteration is introduced for preprocessing based on the iterative solutions of the stationary phonon BTE.
Macroscopic and mesoscopic physical evolution processes are connected by the heat flux, which is no longer calculated by classical Fourier's law but obtained by taking the moment of phonon distribution function.
These two processes exchange information from different scales, such that the present scheme could efficiently deal with heat conduction problems from ballistic to diffusive regime.
Numerical tests show that the present scheme could efficiently capture the multiscale heat conduction in hotspot systems with large temperature variances.
In addition, a comparison is made between the solutions of the present scheme and effective Fourier's law by several heat dissipations problems under different sizes or selective phonon excitation.
Numerical results show that compared to the classical Fourier's law, the results of the effective Fourier's law could be closer to the BTE solutions by adjusting effective coefficients.
However, it is still difficult to capture some local nonlinear phenomena in complex geometries.}}

\end{abstract}


\maketitle

\section{INTRODUCTION}

With the continuous development of modern semiconductor process technology~\cite{IRDS2023}, the characteristic size of electronic devices decreases sharply from microns to nanometers and the hotspot issues become increasingly serious~\cite{warzoha_applications_2021,pop_energy_2010}, which have attracted great attention in academic~\cite{TANG2023123497,warzoha_applications_2021,pop_energy_2010} and industrial~\cite{TSMC_2023_self_heating,MBCFET2023_Samsung,intel_2023_GAAFET,BEOL2023_IMEC} communities.
Micro/nano scale thermal management in hotspot systems is particularly important because it will seriously reduce working efficiency and device life, but it faces huge challenge~\cite{warzoha_applications_2021,pop_energy_2010}.
Firstly, the heat generation in electronic devices mainly results from the interaction between electrons and phonons when the electric field drives the electrons to migrate from the source to the drain~\cite{pop2004analytic,ChenG05Oxford,PhysRevApplied.19.014007}.
The energy weights obtained by different phonon modes from electrons in this process are non-uniform, namely, this is a selective phonon excitation or non-thermal heating process~\cite{wan2024manipulating,PhysRevLett.126.077401,APLnonthermal2020,PhysRevApplied.19.014007}.
Secondly, when the system size is comparable to or smaller than phonon mean free path, ballistic phonon transport plays an important role on heat conduction~\cite{PhysRevApplied.10.054068,PhysRevApplied.11.024042,chuang2021graded,chen_non-fourier_2021,APLnonthermal2020}.
Many experimental studies in the past $30$ years have proven that the classical Fourier's law with bulk thermal conductivity cannot well predict the thermal behaviors in the solid materials at the micro/nano scales~\cite{ziabari2018a,beardo_general_2021,chang_breakdown_2008,lu2002size,xu_raman-based_2020,jiang_tutorial_2018,PhysRevApplied.10.054068,PhysRevApplied.11.024042}.

In order to efficiently solve the practical multiscale heat conduction problems, many models have been developed~\cite{xiaoyanliu_2022_review_thermal,PhysRevB.103.L140301,esee8c149,WANGMR15APPLICATION,de_tomas_kinetic_2014}.
{\color{black}{One of the most commonly used and simplest models is a combination of the macroscopic equations and effective empirical coefficients. 
For example, the effective Fourier's law,
\begin{align}
\bm{q}=-\kappa_{eff}(\bm{x}, T) \nabla T,
\label{eq:effFourierlaw}
\end{align}
is widely used in the practical multiscale thermal application~\cite{Three_Stacked_NanoplateFET2018,comparison_FIN_GAAFET,KUMAR2023100056} and the post-processing of most micro/nano scale thermal experiments~\cite{xu_raman-based_2020,jiang_tutorial_2018}, where a size- (or spatial) and temperature- dependent effective thermal conductivity is introduced instead of the bulk thermal conductivity.
Different effective thermal conductivity coefficients are used for different materials with various sizes, and external heat sources in the macroscopic equations usually default to uniformly heating all (quasi)particles.
The effective Fourier's law has been integrated into major commercial softwares, such as ANSYS, TCAD and COMSOL, and applied to multi-physics coupling simulations in actual 3D electronic devices (e.g., fin or gate-all-around field-effect transistors, FinFETs/GAAFETs)~\cite{Three_Stacked_NanoplateFET2018,comparison_FIN_GAAFET,KUMAR2023100056,TCAD_application_intel_2021_review,BEOL2023_IMEC}, where the choice of effective thermal conductivity is somewhat related to the experience of the engineers. 
Until today, there are few in-depth studies of how far the predictions of the effective Fourier's law deviate from the mesoscopic models or experiments at the micro/nano scale~\cite{ziabari2018a,beardo_general_2021}.}}

Another popular model is the mesoscopic Boltzmann transport equation (BTE), which ignores coherence but can still give reasonable predictions and capture some physical insights such as the self-heating~\cite{ni2012coupled,IEEE2023_5nm,comparison_FIN_GAAFET,KUMAR2023100056}, ballistic effects~\cite{MurthyJY05Review} and selective phonon excitations~\cite{PhysRevLett.126.077401,APLnonthermal2020,PhysRevApplied.19.014007}.
The self-heating and ballistic effects in transistors are widely studied under the framework of phonon BTE by many scientific research institutions~\cite{pop2004analytic,generation_thermalFET,JAP_2014_double_gate_BTE}, such as IMEC~\cite{BEOL2023_IMEC} and Intel~\cite{intel_2023_GAAFET}.
NanoHeat group leaded by Prof. Kenneth E. Goodson has made excellent progress on this topic in the past $30$ years~\cite{generation_thermalFET}.
They calculated detailed electron-phonon coupling processes, simulated self-heating effects and small scale heat dissipations in transistors~\cite{pop2004analytic,generation_thermalFET,SverdrupPG01subcontinuum}.
In addition, some researchers have paid attention to the effects of selective phonon excitation on micro/nano scale heat conduction in semi-conductor materials~\cite{PhysRevB.93.125432,wan2024manipulating,PhysRevLett.126.077401,APLnonthermal2020,PhysRevApplied.19.014007}.
Under the framework of phonon BTE, Chiloyan and Huberman $et~al$.~\cite{APLnonthermal2020} found that selective phonon excitation in silicon materials can lead to enhanced heat conduction beyond Fourier's predictions by targeted heating particular phonon modes, for example, only heating phonon modes with larger mean free path.
Huberman and Zhang $et~al$.~\cite{APS_second_sound} found that in quasi-1D frequency-domian thermoreflectance geometry the relationship between heating-frequency and phase lag in germanium materials is different under various selective phonon excitation.
Xu $et~al$.~\cite{PhysRevApplied.19.014007} found that the peak temperature rise in a silicon quasi-2D bulk FinFET is much larger than that predicted by classical Fourier's law if the non-equilibrium or selective phonon excitation effect is ignored.

Actually it is hard to obtain the analytical solutions of phonon BTE in practical thermal applications so that many numerical methods are developed for simulating micro/nano scale heat conduction problems.
One of the most popular methods is the ensemble Monte Carlo method, which uses statistical particles to represent the actual transport or scattering processes~\cite{MC_porous_gang2009JAP,CHEN2023108592,SHEN2023_IJHMT,PATHAK2021108003}.
It has been widely used for simulating the practical heat dissipations in three-dimensional transistors~\cite{3DFINFET_2014_mc,3DFINFETtransient,JAP_2022_PHONONTHERMAL}.
However, its time step and cell size have to be smaller than the relaxation time and mean free path due to the separate treatment of advection and scattering in single time step.
Besides the statistics errors and computational cost increase rapidly when system size increases significantly.
In order to reduce computational cost, the hybrid method is used~\cite{JAP_qinghao_2017,HAO2018496} whose basic idea is to discretize the whole computational domain into two or several parts.
The phonon BTE is solved in some domains where ballistic phonon transport dominates, and effective Fourier's law is used in other areas.
This method has also been applied successfully in the heat dissipations of electronic devices.
However, reasonable domain decomposition is very empirical and influences the final results significantly.
Another numerical method is to directly solve phonon BTE by discretizing the seven-dimensional phase space into a lot of small pieces~\cite{terris2009modeling,SyedAA14LargeScale,PhysRevApplied.19.014007}, which has no statistics errors but requires a huge number of computational memories, for example discrete ordinate method (DOM).
It adopts iterative strategies to solve the phonon BTE in the whole spatial space for a given point in wave vector space.
And then repeat the whole wave vector space to update the macroscopic distributions at the next step.
One of the main drawback of this method is the slow convergence speed in the (near) diffusive regime~\cite{ADAMS02fastiterative,Chuang17gray,ZHANG20191366,zhang2021e,ZHANG2023124715}.

In order to solve the drawback of DOM, we have proposed a synthetic iterative scheme to efficiently solve the stationary phonon BTE since $2017$~\cite{Chuang17gray,ZHANG20191366,zhang2021e,ZHANG2023124715}, which accelerates convergence significantly compared to DOM in the (near) diffusive regime.
It has also been made into an open source package GiftBTE by Prof Hua Bao's group~\cite{Hu_2024} {\color{black}{and applied for the heat dissipations in transistors~\cite{baohua_IEEE_2024}.}}
However, previous synthetic iterative scheme usually used a linearized equilibrium state with small temperature difference assumptions, which may not be suitable for the practical heat dissipations in real transistors due to large temperature rise.
{\color{black}{When the temperature difference in the whole domain is large, many thermal parameters including the phonon mean free path, scattering rates and thermal conductivity change with spatial position, which increases the nonlinear and multiscale characteristics of phonon transport.
Hence, it is very necessary to consider the temperature-dependent effects~\cite{zhang_discrete_2019} for practical thermal applications~\cite{TCAD_application_intel_2021_review,intel_2023_GAAFET}.  

In this work, a synthetic iterative scheme is developed for thermal applications with large temperature variance.
Numerical tests show that the present scheme could efficiently capture the multicale heat conduction in hotspot systems with large temperature variances.
In addition, the heat dissipations in hotspot systems with a variety of geometric structures, selective phonon excitation and sizes are simulated to estimate the deviations between the effective Fourier's law and phonon BTE. The rest of this article is organized as follows.
Phonon BTE and synthetic iterative scheme for large temperature variance
are introduced in Section~\ref{sec:bte} and Section \ref{sec:sis}, respectively. 
Numerical tests and discussions are conducted in Section~\ref{sec:results}. 
In Section \ref{sec:comparison}, a comparison is made between the solutions of phonon BTE and effective Fourier's law. Finally, a conclusion is made in Section \ref{sec:conclusions}.
}}

\section{Phonon Boltzmann transport equation}
\label{sec:bte}

To capture the non-Fourier heat conduction in hotspot systems, the stationary phonon Boltzmann transport equation (BTE) under the relaxation time approximation is used~\cite{ChenG05Oxford,ZHANG2023124715,PhysRevApplied.19.014007,APLnonthermal2020,BEOL2023_IMEC}
\begin{align}
\bm{v}_k \cdot \nabla_{\bm{x}}  g_k = \frac{ g_k^{eq} -g_k}{ \tau_k } +  p_k \dot{S} ,
\label{eq:pBTE}
\end{align}
where $g_k$ is the distribution function of energy density for phonon mode $k$, $\bm{x}$ is the spatial position, $\bm{v}_k$ is the phonon group velocity obtained by phonon dispersion and $\tau_k=\tau_k (T)$ is the temperature-dependent relaxation time which will be discussed later.
$g_k^{eq}$ is the equilibrium state, equal to the Bose-Einstein distribution multiplied by the energy $\hbar \omega$ possessed by each phonon mode~\cite{ChenG05Oxford},
\begin{align}
g_k^{eq} = \hbar \omega \frac{1}{ \exp{ \left(  \frac{\hbar \omega}{k_B T}  \right) }  -1  },
\label{eq:BEequilibrium}
\end{align}
where $\hbar$ is the Planck constant reduced by $2\pi$, $\omega$ is the angular frequency and $k_B$ is the Boltzmann constant.
$\dot{S}$ is the external heat source or heat generating power density at the macroscopic level, and $p_k$ is associated weight of energy absorbed by each phonon mode from the external heat source satisfying
\begin{align}
\int p_k d\bm{K} =1,
\end{align}
where $d\bm{K}$ represents an integral over the whole wave vector space.
Although it was usually assumed that the heat source energy is allocated to all phonon modes according to thermal equilibrium in macroscopic thermal simulations or most of previous non-Fourier thermal analysis based on phonon BTE, selective phonon excitation is widespread in thermal management of microelectronics~\cite{pop2004analytic,pop_energy_2010,APLnonthermal2020,PhysRevApplied.19.014007,ni2012coupled}.

The local energy density $U$ and heat flux $\bm{q}$ could be updated by taking the moment of phonon distribution functions,
\begin{align}
U &= \int g_k  d\bm{K} ,  \label{eq:energy} \\
\bm{q} &= \int \bm{v}_k g_k  d\bm{K}.  \label{eq:heatflux}
\end{align}
{\color{black}{The temperature cannot be well defined in non-equilibrium system. 
In order to deal with such non-equilibrium systems, the concept of local thermal equilibrium is introduced~\cite{Kubo1991statistical,MC_porous_gang2009JAP}, namely, there is a local equilibrium at which the energy density obtained by taking moment of the phonon equilibrium distribution with local equivalent equilibrium temperature $T$ is the same as that obtained by taking moment of the non-equilibrium phonon distribution,}}
\begin{align}
\int g^{eq}_k (T)  d\bm{K} = \int g_k  d\bm{K}.  \label{eq:temperature}
\end{align}
A pseudo-temperature $T_p$ is introduced to ensure the energy conservation of the phonon scattering term~\cite{MC_porous_gang2009JAP},
\begin{align}
\int \frac{ g_k^{eq} (T_p) -g_k }{ \tau_k (T)  } d\bm{K} =0.  \label{eq:pseudotemperature}
\end{align}
{\color{black}{When $\tau_k$ is a constant, $T_p =T$.}}

We mainly simulated the heat conduction in monocrystalline silicon~\cite{intel_2023_GAAFET,3DFINFET_2014_mc,Hu_2024} and phonon dispersion in the $[1~0~0]$ direction is chosen to represent the other directions.
Dispersion curves of one longitudinal acoustic/optical phonon branch (LA/LO) and two degenerate transverse acoustic/optical phonon branches (TA/TO) are approximated by empirical quadratic polynomial dispersions, as shown in Table.~\ref{dispersioncoe}~\cite{pop2004analytic}.
Relaxation time of optical phonon is $3.5$ ps~\cite{JAP_2014_double_gate_BTE} and Matthiessen's rule is used to coupled various scattering mechanisms of acoustic phonon together including the impurity scattering, umklapp (U) and normal (N) phonon-phonon scattering, 
\begin{equation}
\tau^{-1}=\tau_{{\text{impurity}}}^{-1}+\tau_{{\text{U}}}^{-1}+\tau_{{\text{N}}}^{-1}=\tau_{{\text{impurity}}}^{-1}+\tau_{{\text{NU}}}^{-1},
\label{eq:Matthiessen}
\end{equation}
as shown in Table.~\ref{relaxation}.
\renewcommand\arraystretch{1.2}
\begin{table}
\caption{Quadratic phonon dispersion coefficients of monocrystalline silicon~\cite{pop2004analytic}, where $\omega=c_{0}+ c_{1} k+c_{2} k^2$ and $k$ is the wave vector in $[1~0~0]$ direction. }
\centering
\begin{tabular}{*{4}{c}}
\hline
\hline
Branch & $c_{0}$ ($10^{13}$ rad/s) &  $c_{1}$ ($10^5$ cm/s)  & $c_{2}$ ($10^{-3}$ cm$^2$/s)  \\
 \hline
LA  & 0 & 9.01 & -2.0      \\
 \hline
TA  & 0 & 5.23 & -2.26      \\
 \hline
LO  &  9.88 & 0   & -1.60     \\
 \hline
TO  &  10.20 &  -2.57  & 1.11   \\
\hline
\hline
\end{tabular}
\label{dispersioncoe}
\end{table}
\begin{table}
\caption{Phonon scattering rates of monocrystalline silicon~\cite{terris2009modeling}, where $a=0.543$ nm is the lattice constant of silicon.  }
\centering
\begin{tabular}{*{2}{c}}
\hline
\hline
$\tau_{{\text{impurity}}}^{-1}$   &  $A_{i}\omega^{4}$, ~~$A_{i}=1.498\times10^{-45}~{\text{s}^{\text{3}}}$;       \\
 \hline
LA  & $\tau_{{\text{NU}}}^{-1}=B_{L}\omega^{2}T^{3}$,~~$B_{L}=1.180\times 10^{-24}~{\text{K}^{\text{-3}}}$;      \\
 \hline
\multirow{4}{*}{{\shortstack{TA }}}  & $\tau_{{\text{NU}}}^{-1}=B_T\omega T^4$,~~$0 \leq k <  \pi /a$;      \\
   & $\tau_{{\text{NU}}}^{-1}=B_U\omega^{2}/{\sinh(\hbar\omega/k_{B}T)}$,~~$ \pi /2 \leq k \leq 2\pi /a$;     \\
   & $B_T=8.708\times 10^{-13}~{\text{K}^{\text{-3}}}$,~~ $B_{U}=2.890\times10^{-18}~{\text{s}}$.    \\
\hline
\hline
\end{tabular}
\label{relaxation}
\end{table}

\section{Synthetic iterative scheme with large temperature variance}
\label{sec:sis}

To solve the phonon BTE iteratively, a semi-implicit scheme is firstly introduced,
\begin{align}
\bm{v}_k \cdot \nabla  g_k^{n+1/2} = \frac{ g_k^{eq}(T^{n}_p ) -g_k^{n+1/2} }{ \tau_k (T^n) } ,
\label{eq:DBTE}
\end{align}
where the equilibrium state $g_k^{eq}(T^{n}_p )$ is at the $n$-th iteration step and the distribution function is at the $n+1/2$-th iteration step $g_k^{n+1/2}$.
For a given phonon mode $k$, the finite volume method is used to discretize the whole spatial space,
\begin{align}
\frac{1}{V_i} \sum_{j\in N(i)}  S_{ij} \mathbf{n}_{ij}  \cdot  \bm{v}_{k}  g^{n+1/2}_{ij,k}
=\frac{ g^{eq}_{i,k} (T_p^n) - g^{n+1/2}_{i,k}  }{ \tau_k (T_i^n)   } ,
\label{eq:DiSIBTE}
\end{align}
where $V_i$ is the volume of the spatial cell $i$, $N(i)$ is the sets of face neighbor cells of cell $i$, $ij$ is the interface between the cell $i$ and cell $j$, $S_{ij}$ is the area of the interface $ij$, and $\mathbf{n}_{ij}$ is the normal of the interface $ij$ directing from the cell $i$ to the cell $j$. 
{\color{black}{$g^{n+1/2}_{i,k}$ (or $g^{n+1/2}_{ij,k}$) is the phonon distribution function at discretized spatial cell $i$ (or cell interface $ij$).
Usually a first-order upwind or step scheme is used,
\begin{align}\label{eq:upwind}
\begin{split}
g_{ij,k}= \left \{
\begin{array}{lr}
    g_{i,k},   & \mathbf{n}_{ij}  \cdot  \bm{v}_{k}  \geq  0   \vspace{2ex}   \\
    g_{j,k},    & \mathbf{n}_{ij}  \cdot  \bm{v}_{k} < 0
\end{array}
\right.
\end{split}
\end{align}
It is numerically stable and efficient in the ballistic regime, but has poor spatial accuracy in the (near) diffusive regime.

In order to realize a higher-order spatial accuracy in the (near) diffusive regime, make a transformation of Eq.~\eqref{eq:DiSIBTE},
\begin{align}
\frac{ \Delta  g^{n}_{i,k}  }{\tau_k (T_i^n)  } + \frac{1}{V_i} \sum_{j\in N(i)}  S_{ij} \mathbf{n}_{ij}  \cdot  \bm{v}_{k}  \Delta g^{n}_{ij,k}
=\frac{ g^{eq}_{i,k} (T_p^n) - g^{n}_{i,k}  }{  \tau_k (T_i^n)   } - \frac{1}{V_i} \sum_{j\in N(i)}  S_{ij} \mathbf{n}_{ij}  \cdot  \bm{v}_{k}  g^{n}_{ij,k} ,
\label{eq:DiSIBTEdelta}
\end{align}
where $\Delta g^n= g^{n+1/2} -g^{n}$ and the first-order upwind scheme is used to deal with $\Delta g^{n}_{ij,k}$.
It is found that the final numerical accuracy is totally controlled by the right-hand side of Eq.~\eqref{eq:DiSIBTEdelta} since at steady state $\Delta g \rightarrow 0$.
Hence, the reconstruction of the distribution function $g^{n}_{ij,k} $ at the cell interface determines the spatial accuracy of a numerical scheme.
The phonon BTE is solved again at the cell interface along the group velocity direction with a certain length $| \bm{v}_k \Delta t|$,
\begin{align}
 \bm{v}_k  \frac{ g^{n} (\bm{x}_{ij})  - g^{n} (\bm{x}_{ij} -\bm{v}_k \Delta t) }{ \bm{v}_k \Delta t } &= \frac{g^{eq} (T_{p, ij}^{n} ) -g^{n} (\bm{x}_{ij})  }{ \tau_k (T_{ij}^n) },  \label{eq:BTEinterface}  \\
 \Longrightarrow  g^{n} (\bm{x}_{ij}) &= \frac{ \tau_k g^{n} (\bm{x}_{ij} -\bm{v}_k \Delta t) + \Delta t g^{eq} (T_{p, ij}^{n} ) }{ \Delta t + \tau_k (T_{ij}^n) },  \label{eq:ReconstructFace}
\end{align}
where $g^{n} (\bm{x}_{ij} -\bm{v}_k \Delta t)$ or $T_{p, ij}^{n}$, $T_{ij}^{n}$ is obtained by the numerical interpolation of $g^{n} (\bm{x}_i)$  or $T_{p, i}^{n}$, $T_{i}^{n}$, and van Leer limiter is used for the spatial gradients.
$\Delta t$ is a time step and satisfies
\begin{align}
\Delta t = c_1  \times  \frac{ \Delta x_{min} }{2  v_{max} },
\end{align}
where $\Delta x_{min}$ is the minimum discretized cell size, $v_{max}$ is the maximum group velocity, $c_1$ is a dimensionless parameter. In the present simulations, $c_1=0.95$.
}}
Combining Eqs.~(\ref{eq:upwind},\ref{eq:DiSIBTEdelta}) or Eqs.~(\ref{eq:ReconstructFace},\ref{eq:DiSIBTEdelta}), the phonon distribution function $g^{n+1/2}_{i,k}$ or $g^{n+1/2}_{ij,k}$ can both be calculated.
More details can be found in our previous paper~\cite{ZHANG2023124715}.

Repeat the above process and traverse all phonon modes, then the phonon distribution functions $g_k^{n+1/2}$ in the whole phase space can be obtained.
The macroscopic pseudo-temperature at the $n+1/2$-th iteration step and temperature at the next iteration step are updated {\color{black}{based on Eqs.~(\ref{eq:pseudotemperature},\ref{eq:temperature})}},
\begin{align}
\sum w_k \left( \frac{ g_k^{eq} (T_p^{n+1/2}) - g_k^{n+1/2} }{ \tau_k (T^n)  }   \right)  & =0, \label{eq:DBTETp}  \\
\sum w_k  \left(  g_k^{n+1/2} -   g_k^{eq} (T^{n+1} )   \right)  &=0,  \label{eq:DBTET}
\end{align}
where $w_k$ is the associated weight of numerical quadrature $\sum$ in the first Brillouin zone.
Above two nonlinear equations are solved iteratively by Newton method~\cite{zhang_discrete_2019} to update the temperature.
{\color{black}{If setting $T_p^{n+1}=T_p^{n+1/2}$, then it is the typical implicit DOM which converges fast in the ballistic regime but converges much slowly in the (near) diffusive regime~\cite{ADAMS02fastiterative,Chuang17gray}.}}

Secondly, a macroscopic iteration is introduced to accelerate convergence in the (near) diffusive regime.
{\color{black}{Considering first law of the thermodynamics at steady state,
\begin{equation}
\nabla \cdot \bm{q}  = \dot{S},
\label{eq:first}
\end{equation}
then a macroscopic residual is defined as
\begin{equation}
\text{RES} = -\nabla \cdot \bm{q} + \dot{S} = -Q(T, T_p),
\label{eq:RESMacro}
\end{equation}
where $Q(T,T_p)$ is a functional to represent the real relationship between the heat flux and temperature.
Eq.~\eqref{eq:RESMacro} is valid from the ballistic to diffusive regime.
Note that the real relationship $Q$ between the heat flux and temperature are unknown so that the inexact Newton method~\cite{Moore68Newton} is used to solve it iteratively.
More details can be found in our previous paper~\cite{Chuang17gray}.}}
An approximated linear operator $\tilde{Q}$ is introduced to find an increment of pseudo-temperature $\Delta T_p$ for diminishing the macroscopic residual so that a macroscopic iteration can be constructed,
\begin{align}
\tilde{Q} ( T_p^{n+1 } ) - \tilde{Q} ( T_p^{n+1/2 } )   &= \text{RES}^{n+1/2} \\
\tilde{Q} (\Delta T_p^{n+1/2} ) &= \text{RES}^{n+1/2}  ,
\label{eq:RESMacronewton}
\end{align}
where
\begin{align}
\Delta T_p^{n+1/2} & = T_p^{n+1}- T_p^{n+1/2} ,\\
\tilde{Q}(\Delta T_p) &= \nabla \cdot (- \beta  \cdot \nabla (\Delta T_p) ),
\label{eq:fourier22}
\end{align}
where $\beta$ is a constant for simplicity with the same dimension as thermal conductivity.
Usually $\beta$ is equal to the bulk thermal conductivity at the initial ambient temperature.
Equation~\eqref{eq:first} is satisfied when the macroscopic residual vanishes, {\color{black}{and the approximated linear operator does not affect the final convergent solutions based on the theorem of inexact Newton method~\cite{Moore68Newton,Chuang17gray}.
Theoretically the convergence speed is faster if the approximated linear operator is closer to the real operator~\cite{Moore68Newton,Chuang17gray}, so that an effective Fourier's law is used for the linear operator.}}
Similarly finite volume method is used for the discretization of Eq.~\eqref{eq:RESMacronewton},
\begin{equation}
-\sum_{j\in N(i)} S_{ij} \mathbf{n}_{ij}   \cdot  \left(  \bm{\kappa}_{\text{bulk}}  \cdot  \nabla ( \Delta T_{p,ij}^{n+1/2})  \right) =\text{RES}_i^{n+1/2}  =
- \sum_{j\in N(i)}    S_{ij}  \mathbf{n}_{ij}  \cdot   \bm{q}_{ij}^{n+1/2}  + \dot{S}_i  ,
\label{eq:dvgoverningE}
\end{equation}
where
\begin{align}
\bm{q}_{ij}^{n+1/2} & =  \sum w_k \bm{v}_{k}  g^{n+1/2}_{ij,k} .
\label{eq:dis_flux}
\end{align}
Conjugate gradient method is used to solve the above diffusion equation~\eqref{eq:dvgoverningE}.

{\color{black}{The heat flux in the macroscopic equation~\eqref{eq:dvgoverningE} is obtained by taking the moment of phonon distribution function, rather than the classical or effective Fourier’s law.
The heat flux obtained by this strategy is valid from ballistic to diffusive regime. 
Macroscopic and mesoscopic iterative processes exchange information from different scales, such that the present scheme could efficiently deal with heat conduction problems from ballistic to diffusive regime.
The final convergent solutions are totally controlled by the phonon BTE, and the macroscopic operator only plays an accelerate role.
Compared to the iterative solutions of the phonon BTE, the computational cost of the macroscopic iteration is very small due to the less degree of freedom.
For details of isothermal or diffusely/specular reflecting adiabatic boundary conditions, please see \textit{3.3} $Boundary~conditions$ in our previous paper~\cite{ZHANG2023124715}.

At the end of this section, let's briefly analyze, why the above discrete approach is used and why this treatment can accelerate convergence from a physical point of view.
Steady-state phonon BTE mainly describes the evolution of phonon distribution function in the spatial space and wave vector space. 
Different phonon modes migrate freely or scatter with other phonons in six-dimensional phase space, and the average distance between two adjacent phonons scattering is the phonon mean free path.
For the mesoscopic iteration~\eqref{eq:DBTE}, it assumes that the macroscopic distributions or equilibrium states are fixed for a given phonon mode, and then updates the phonon distribution function in the whole spatial space. 
In other words, it is assumed that phonons do not interact with other phonon modes and the corresponding equilibrium states remain unchanged when the phonon distribution function is updated and evolved in the spatial space. 
When all phonon distribution functions are updated, then the macroscopic distributions or equilibrium states are updated based on the conservation principle of phonon scattering~(\ref{eq:DBTET},\ref{eq:DBTETp}). 
This method decouples phonon migration from scattering in one iteration step, which indicates that it efficiently updates the distribution function in the spatial space within the range of a mean free path in one iteration step.
In the ballistic regime, phonon scattering is rare and the phonon mean free path is comparable to or larger than system size.
This method is in good agreement with the actual physical evolution process, so the convergence speed is fast. 
However, in the (near) diffusive regime, phonon scattering is much frequent and the phonon mean free path is much smaller than system size, so that it converges very slowly.

In order to accelerate convergence in the (near) diffusive regime, we need to make the numerical treatments follow the actual laws of physical evolution~\cite{ADAMS02fastiterative,Chuang17gray,ZHANG20191366,zhang2021e,ZHANG2023124715}.
Note that first law of thermodynamics is valid at any scales, which contains an unknown relationship between the heat flux and temperature.
Fortunately, this relationship is close to effective/classical Fourier's law and temperature propagation follows the law of diffusion in the (near) diffusive regime. 
Furthermore, both the heat flux~\eqref{eq:heatflux} and temperature~(\ref{eq:temperature},\ref{eq:pseudotemperature}) could be obtained by taking the moment of distribution function under the framework of BTE from the ballistic to diffusive regime.
In other words, this complex relationship between (pseudo-)temperature and heat flux could be explicitly constructed by the mesoscopic distribution function.
In the (near) diffusive regime, although the convergence speed of mesoscopic iteration is very slow, it correctly gives a phonon distribution function in the spatial space within the range of a mean free path, from which the corresponding heat flux distribution~\eqref{eq:dis_flux} or macroscopic residual~\eqref{eq:RESMacronewton} can be obtained.
Then a macroscopic iteration can be invoked on the basis of the mesoscopic iteration and connected by the heat flux, which is calculated by taking the moment of phonon distribution function.
The drawback of mesoscopic iteration in the (near) diffusive regime is compensated by the macroscopic iteration, by introducing an approximate Fourier operator to achieve fast convergence based on the theorem of inexact Newton method~\cite{Chuang17gray}. 
The unknown relationship between the heat flux and temperature in the macroscopic iteration is compensated by the mesoscopic iteration. }}

\section{Results and discussions}
\label{sec:results}

Without special statements, the isotropic wave vector space is divided into $N_B$ equal parts using the rectangular integration rule along the radial direction from $0$ to $2 \pi/a$ for each phonon branch, where $a=0.543$ nm is the lattice constant of silicon, and the solid angle space is discretized into $N_{dir}$ pieces using the Gauss-Legendre integration rule.
$N_B=20$ (or $10$) for each acoustic (or optical) phonon branch, respectively.
$N_{dir}=40\times 8$ for quasi-1D simulations and $N_{dir}=32 \times 32$ for quasi-2D or 3D simulations.
All numerical results are obtained by a three-dimensional C/C++ program.
MPI parallelization computation with $40- 320$ CPU cores based on the decomposition of wave vector space is implemented.
In order to reduce the communication times and improve the parallel efficiency, the discrete points corresponding to the specular reflection should be ensured in the same CPU core when the discrete points of the wave vector space are partitioned.
The iteration reaches convergence when the difference of the macroscopic variables $W$ between two successive iteration step is smaller than a threshold, for example,
\begin{equation}
\epsilon=\frac{ \sqrt {\sum_{i}^{N_{cell}}{(W_i^{n}-W_i^{n+1})^2} } } {  \sqrt {\sum_{i}^{N_{cell}}{(W_i^{n})^2} } } < 10^{-6},
\label{eq:epsilon}
\end{equation}
where $N_{cell}$ is the number of discretized cells.

\subsection{Quasi-1D heat conduction}
\label{sec:quasi1dsec}

\subsubsection{Accuracy and efficiency test}

\begin{figure}[htb]
\centering  
\subfloat[]{\includegraphics[width=0.4\textwidth]{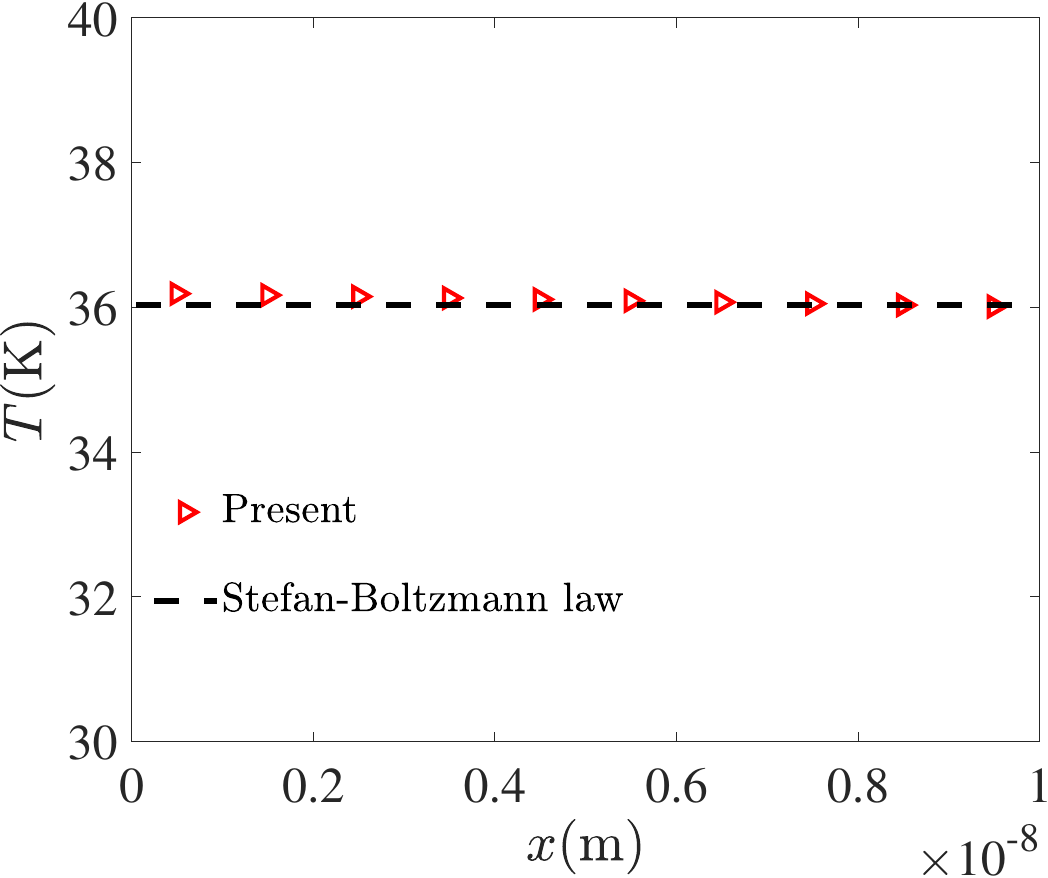}}~~
\subfloat[]{\includegraphics[width=0.4\textwidth]{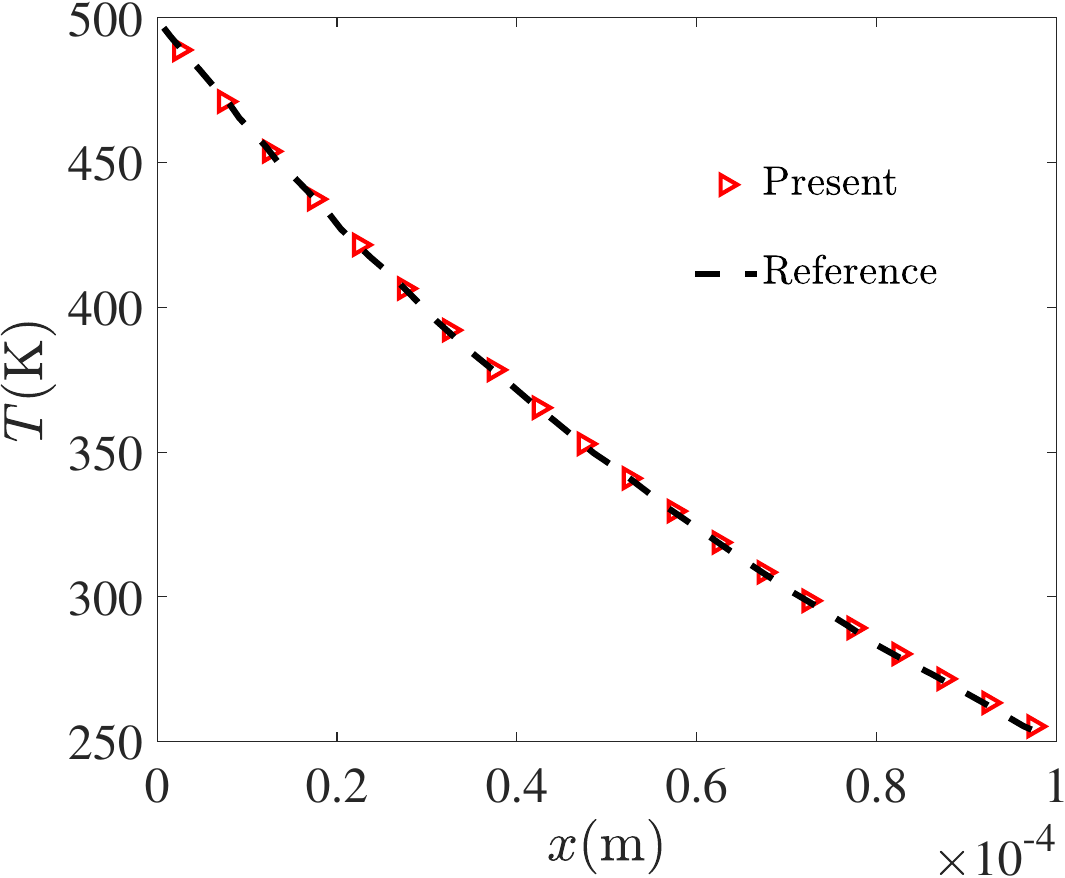}} \\
\subfloat[]{\includegraphics[width=0.4\textwidth]{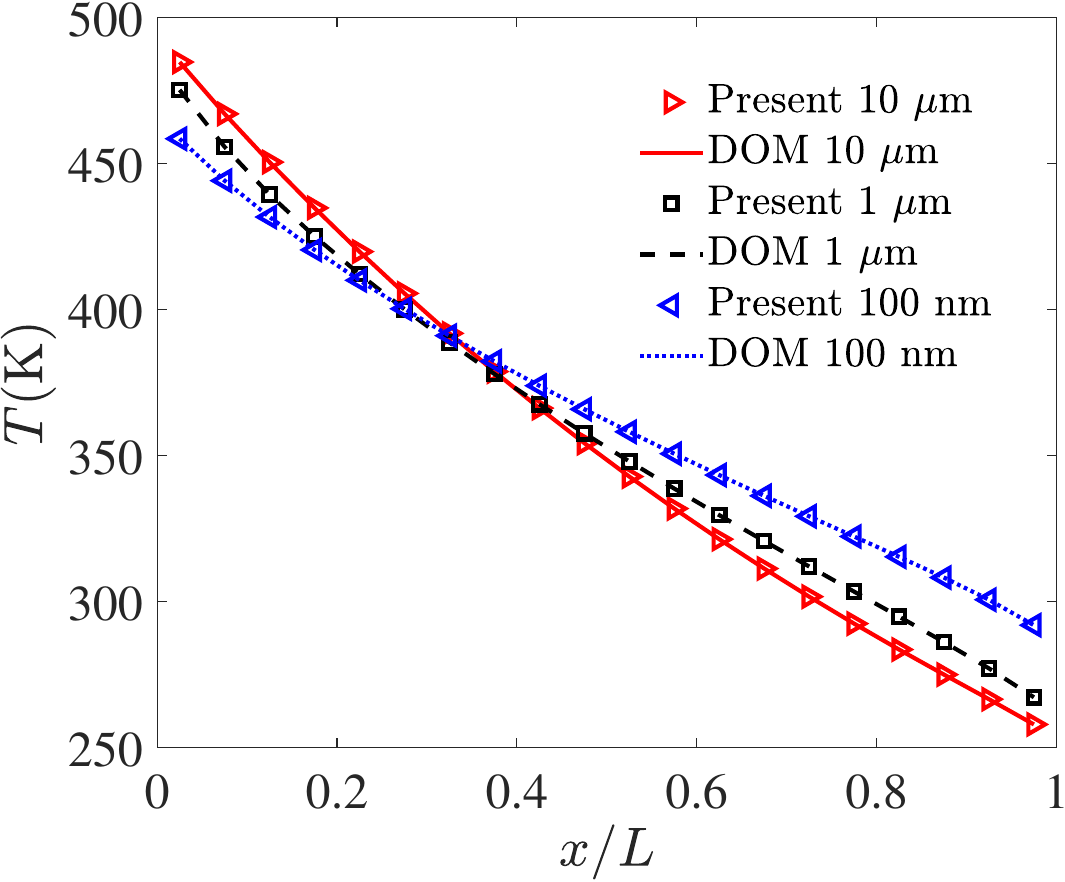}}~~ 
\subfloat[]{\includegraphics[width=0.4\textwidth]{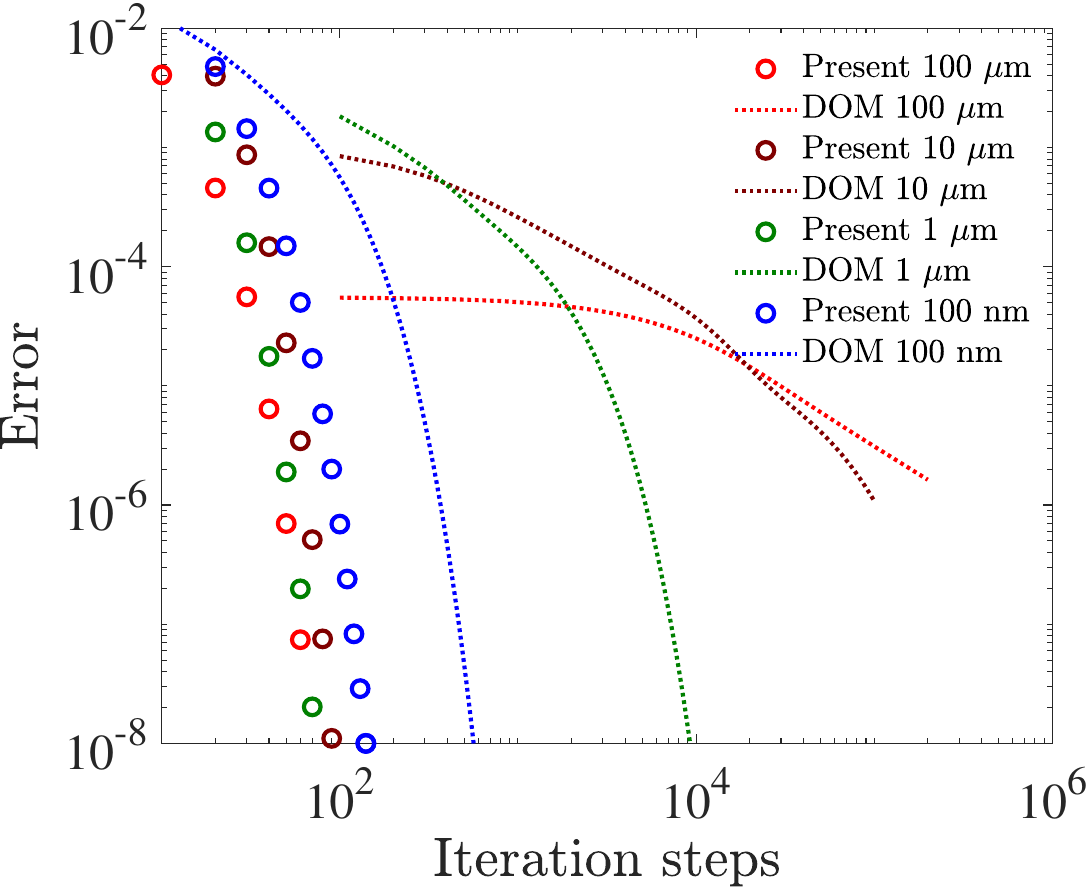}} \\
\caption{
(a) Ballistic regime, $L=10$ nm, $T_h=40$ K, $T_c=30$ K, $10$ discretized uniform cells are used. (b,c,d) $T_h=500$ K, $T_c=250$ K, $20$ discretized uniform cells are used. (b) Diffusive regime, $L=100$ $\mu$m, `Reference' comes from Ref.~\cite{Lacroix05}. A comparison is made between the present scheme and DOM in (c) numerical accuracy and (d) computational efficiency, {\color{black}{where `Error' is the difference of the macroscopic heat flux between two successive iteration step. It takes less than $1$ second and about $100$ iteration steps to reach convergence for the present scheme regardless of the system sizes.}}   }
\label{validation}
\end{figure}
Quasi-1D steady heat conduction without external heat source is simulated firstly.
System size is $L$ and the temperature at each end side of the system is fixed as $T_h$ and $T_c$, respectively, where isothermal boundary condition is used for the two side of the system.
Initial temperature inside the domain is $T_c$ and $40$ CPU cores are used.

When $L=10$ nm, $T_h=40$ K and $T_c=30$ K, phonon mean free path is much larger than system size and ballistic phonon transport dominates heat conduction.
{\color{black}{Phonons emitted from one boundary are absorbed directly by the other boundary without scattering.
The specific heat is proportional to $T^3$ in the low temperature limit so that the temperature in the ballistic limit satisfies Stefan-Boltzmann law~\cite{MajumdarA93Film}, i.e., $T^4= (T_h^4 +T_c^4)/2$.}}
Numerical results in~\cref{validation}(a) show that the present results well follow the
Stefan-Boltzmann law.
When $T_h=500$ K, $T_c=250$ K and the system size is larger than phonon mean free path, e.g., $L=100~\mu$m, as shown in~\cref{validation}(b), our numerical solutions are in excellent agreement with the reference data which is obtained by solving classical Fourier's law with the temperature dependent thermal conductivity~\cite{Lacroix05}.
{\color{black}{Note that when the temperature difference in the domain is large, the thermal conductivity varies with the spatial position so that the temperature distributions are nonlinear.
In addition, $20$ discretized uniform cells are enough for the present scheme to accurately recover the classical Fourier's law.
When system size decreases from $100~\mu$m to $100$ nm, numerical results in ~\cref{validation}(c) show that the present results agree well with the typical DOM.
Furthermore, the present scheme significantly reduce the iteration steps compared to the DOM when the system size is large, as shown in~\cref{validation}(d), where `Error' is the difference of the macroscopic heat flux between two successive iteration step.
It takes less than $1$ second and tens of iteration steps to reach convergence for the present scheme regardless of the system sizes.
Above results show that the present scheme could efficiently describe the heat conduction from the ballistic to diffusive regime.}}

\subsubsection{Selective phonon excitation}
\label{sec:quasi1dSPE}

Effects of selective phonon excitation on micro/nano scale heat conduction is investigated.
Several ways of selective phonon excitation are considered, including `only heating LO', `only heating TO', `only heating LA', `only heating TA', `equal heating' and `Joule heating'.
`equal heating' represents that $p_k$ in Eq.~\eqref{eq:pBTE} is equal for all phonon modes.
`only heating X' represents that $p_k =0$ for other phonon branches and $p_k$ is equal for all phonon modes in X branch.
In `Joule heating', the energy weights $p_k$ of various phonon modes absorbing from Joule heating in microelectronics are considered, which theoretically results from complicated electron-phonon coupling process.
{\color{black}{The scattering strengths between electrons with various energy level and phonons with different modes are different and scattering rates depend on the spatial temperature, too. 
It is better to simultaneously solve the electron and phonon BTE with full scattering kernel. 
However, it is very expensive and inefficient. Hence some empirical models/parameters or artificial heat source has to be used.
In this work, the energy weights are obtained from previous references~\cite{PhysRevApplied.19.014007,ni2012coupled}, where $16\%$ of total Joule heating energy is absorbed by `LA', $59\%$ is absorbed by `LO', $4.4\%$ is absorbed by `TA', $20.6\%$ is absorbed by `TO'. For phonon modes in the same phonon branch, we assume that $p_k$ is equal.}}

\begin{figure}[htb]
\centering
\subfloat[$P_{power}=5 \times 10^{18}$ W/m$^3$, $d_{pump}=5$ nm, $L=100$ nm ]{\includegraphics[width=0.4\textwidth]{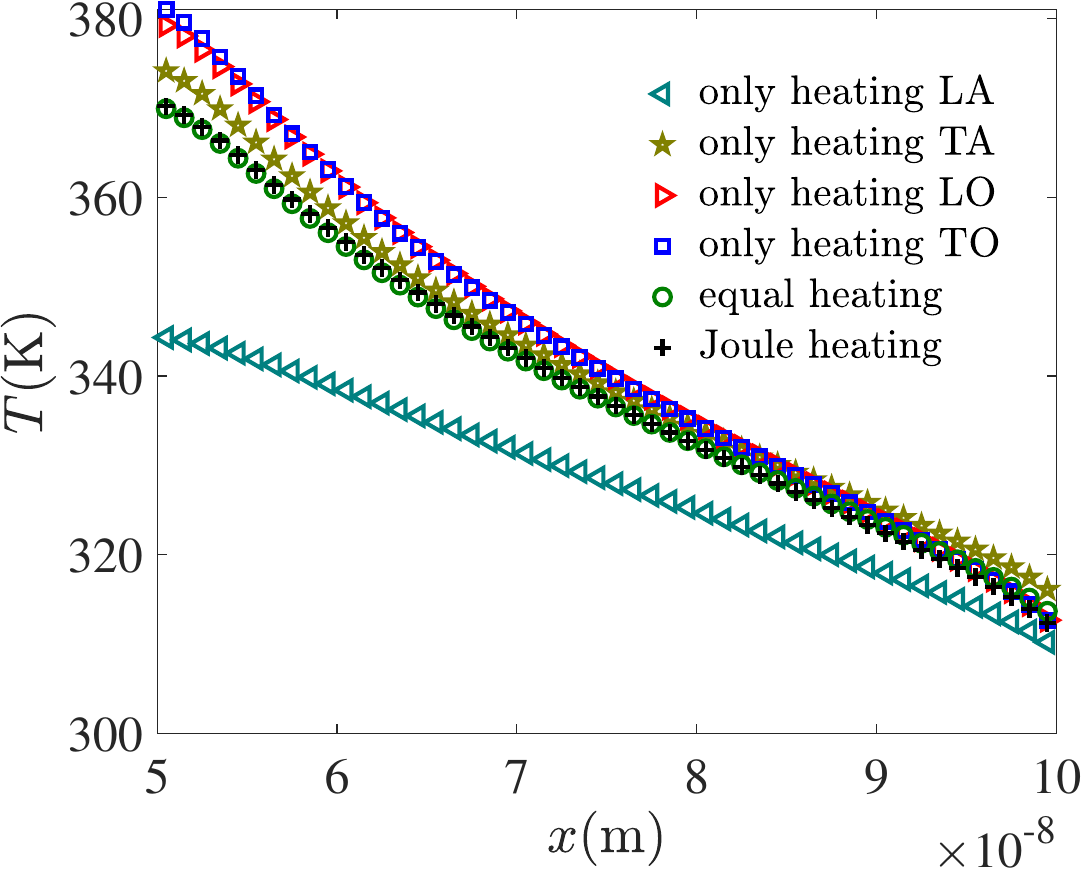}}~~
\subfloat[$P_{power}=5 \times 10^{18}$ W/m$^3$, $d_{pump}=5$ nm, $L=1~\mu$m]{\includegraphics[width=0.4\textwidth]{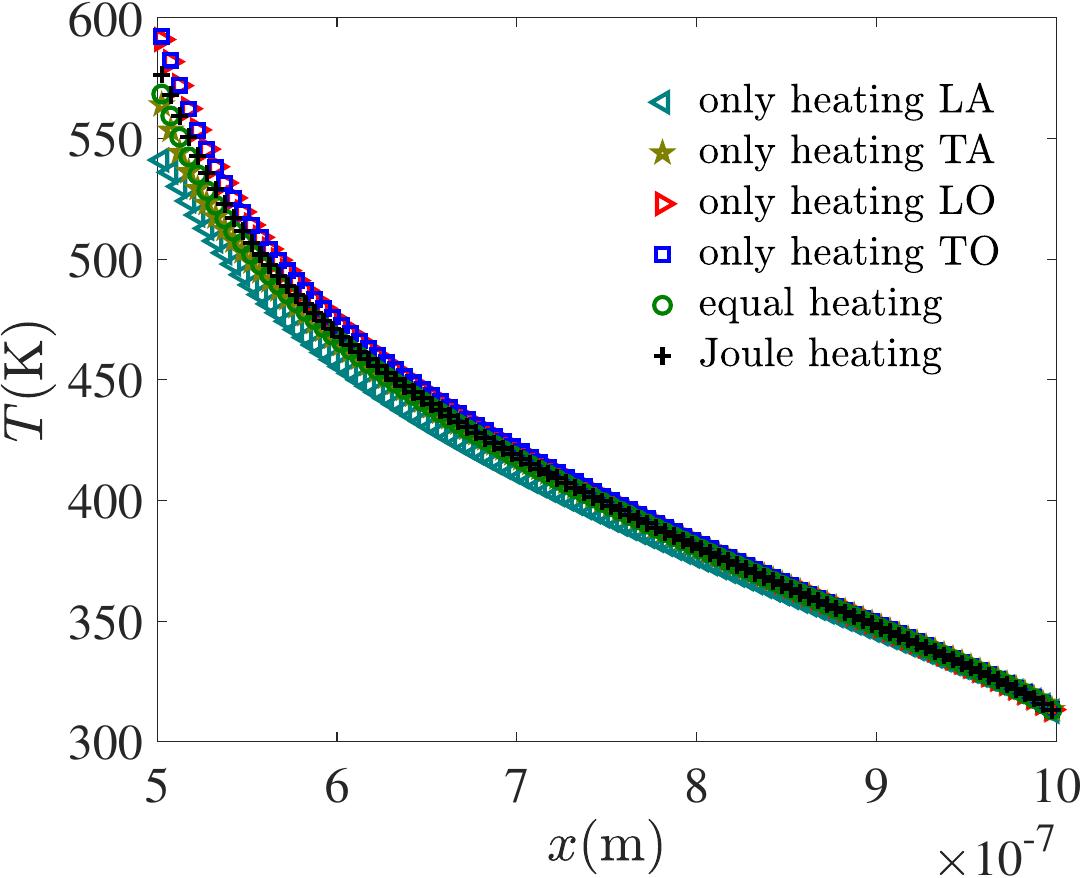}}~~ \\
\subfloat[$P_{power}=5 \times 10^{17}$ W/m$^3$, $d_{pump}=20$ nm, $L=100$ nm]{\includegraphics[width=0.4\textwidth]{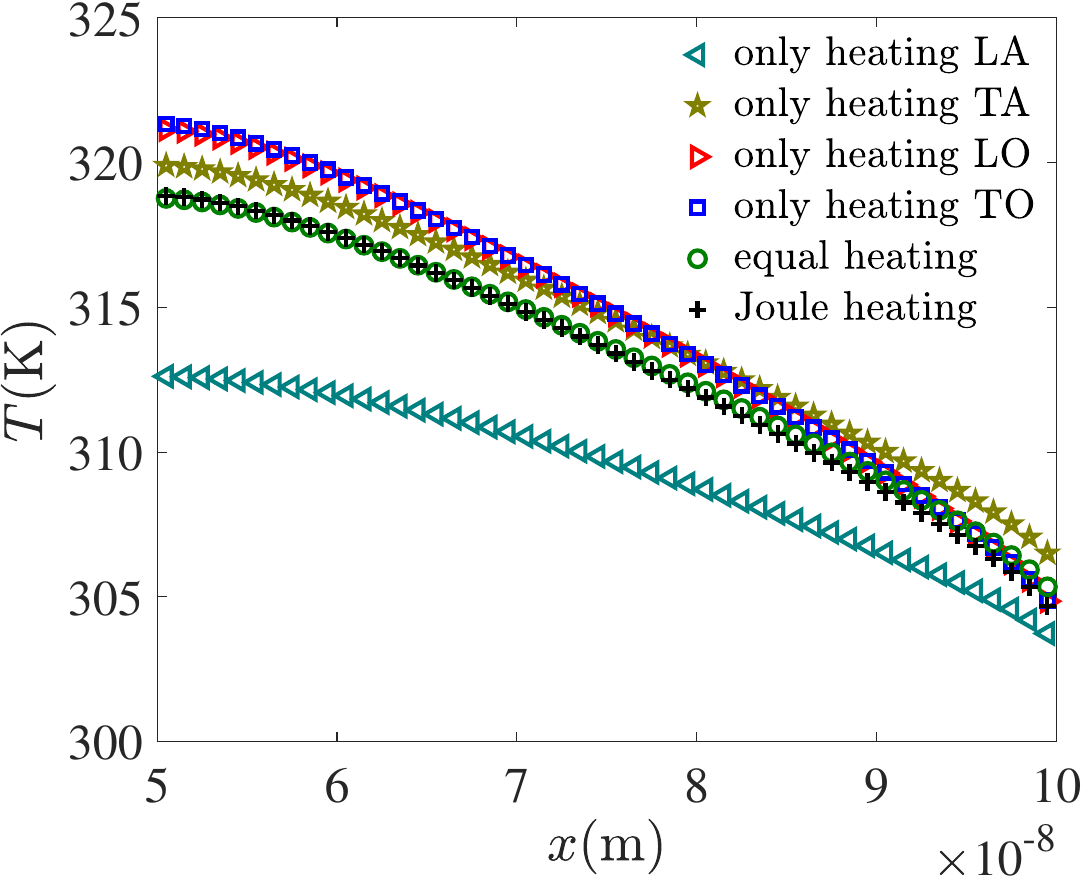}}~~
\subfloat[$P_{power}=5 \times 10^{17}$ W/m$^3$, $d_{pump}=20$ nm, $L=1~\mu$m]{\includegraphics[width=0.4\textwidth]{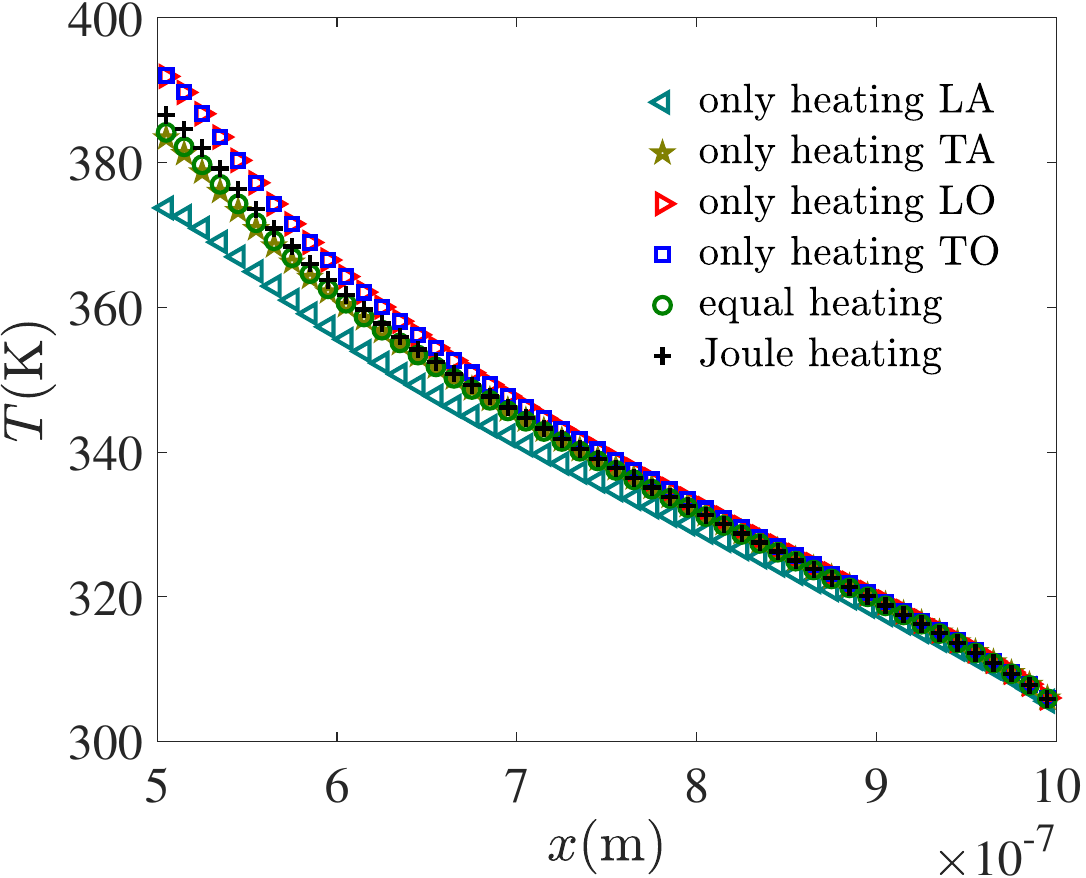}}~~ \\
\caption{Temperature distributions under different selective phonon excitations, maximum heat generating power density $P_{power}$, radius of heat source $d_{pump}$. Half the spatial distribution is drawn due to symmetry.}
\label{target_heating1D}
\end{figure}
A Gaussian heat source is implemented inside a quasi-1D system with size $L$,
\begin{align}
\dot{S} = P_{power} \exp{ \left( - \frac{ d }{d_{pump}} \right) },
\end{align}
where $P_{power}$ is the maximum heat generating power density, $d$ is the distance from the center and $d_{pump}$ is the radius of heat source.
Temperature at each end side of the system is fixed as $T_c= 300$ K with isothermal boundary conditions and initial temperature inside the domain is $T_c$.

Steady thermal transport under different selective phonon excitations, maximum heat generating power density $P_{power}$, radius of heat source $d_{pump}$ is simulated and half of spatial temperature distributions are plotted in~\cref{target_heating1D}.
In order to accurately capture the spatial distribution of heat source, $100-200$ discretized cells are used.
{\color{black}{Note that the external heat source continuously heats the system, which results in that the system temperature keeps rising and the phonon scattering rates keep changing during each iteration step, hence theoretically it needs more iterative steps to reach convergence compared with that without external heat source.
For all cases, it takes $64$ CPU cores, $60-140$ iteration steps and $5-10$ seconds to reach convergence.}}

When $L=100$ nm, numerical results in~\cref{target_heating1D}(a,c) show that the temperature rises are different under various selective phonon excitations and large temperature slip appears near the right boundary.
{\color{black}{Actually when the system size is comparable to or smaller than phonon mean free path, there is insufficient phonon-phonon scattering so that the phonon BTE could be approximated as
\begin{align}
\bm{v}_k \cdot \nabla_{\bm{x}}  g_k  &=   p_k \dot{S} ,  \\
\Longrightarrow  \bm{v}_k \left( g_k (\bm{x}_1 + \bm{d} ) -g_k(\bm{x}_1  ) \right) &=  \int_{\bm{x}_1}^ { \bm{x}_1 + \bm{d} }  p_k \dot{S} (x')dx',
\end{align}
where $\bm{d}$ is a distance and $\bm{x}_1$ is the spatial position.
Above two equations indicate that when a phonon mode absorbed energy from the external heat source and transferred thermal energy to other spatial domain, the heat dissipation efficiency of this process depends on the phonon group velocity and its absorbed energy in the ballistic regime. 

Note that the local temperature is a sum of the phonon distribution function over the wave vector space so that the temperature rise in the heat source areas depends on the phonon group velocity in the ballistic regime, rather than scattering. }}
The group velocity of LA phonon is large so that its temperature rise is relatively small when the heat source only heats LA phonon.
The optical phonon has small group velocity so that the temperature rise of `only heating LO' and `only heating TO' is high.
Temperature differences among these three selective phonon excitations (`Only heating TA', `Only heating TO' and `Only heating LO') are relatively small.
{\color{black}{Above results also clarify that the heat dissipations at the micro/nano scale will be more efficient if more external heat source is absorbed and transferred by LA phonons.
In addition, the temperature rises of `equal heating' and `Joule heating' are almost the same.
That's because the differences of selective phonon excitation on micro/nano scale heat conduction are mainly reflected in the energy weight absorbed by LA phonons from external heat source based on the results shown in~\cref{target_heating1D}(a).
And about $16\%$ heating source energy is absorbed by LA phonons for both `equal heating' and `Joule heating'.}}

When system size increases, it can be found that the temperature distributions under different selective phonon excitation tend to the same in~\cref{target_heating1D}(b,d) and temperature slip near the right boundary decreases.
Because when the system size increases, frequent phonon-phonon scattering drives all phonon modes to local equilibrium with the same temperature.

{\color{black}{
\begin{figure}[htb]
\centering
\subfloat[]{\includegraphics[width=0.45\textwidth]{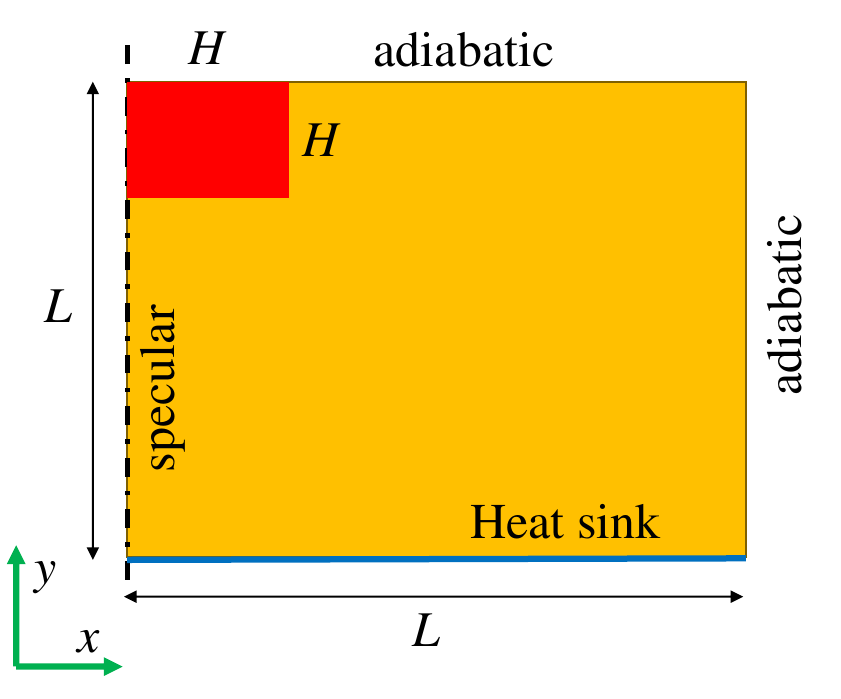}}~~
\subfloat[]{\includegraphics[width=0.45\textwidth]{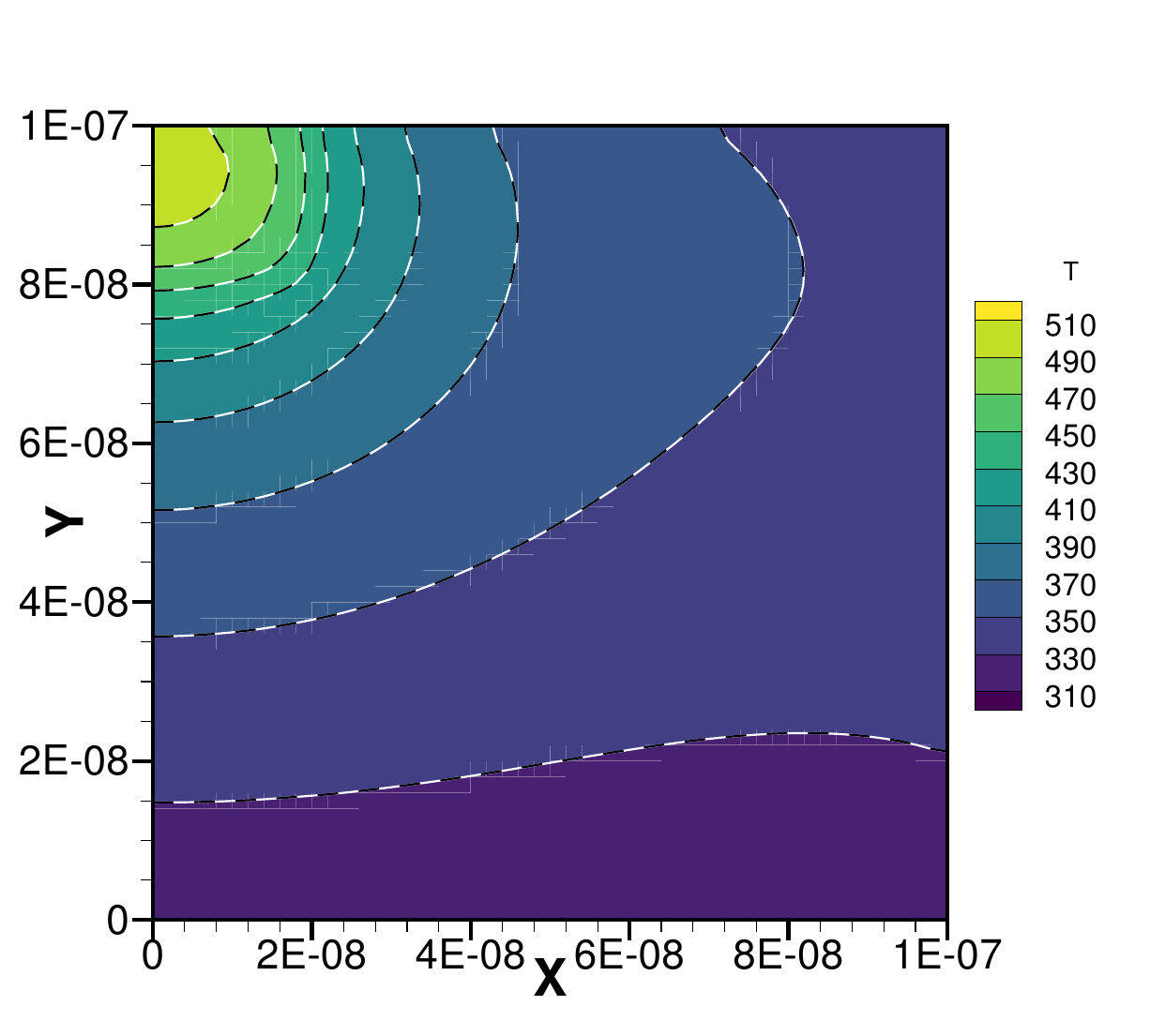}}~~ \\
\subfloat[]{\includegraphics[width=0.45\textwidth]{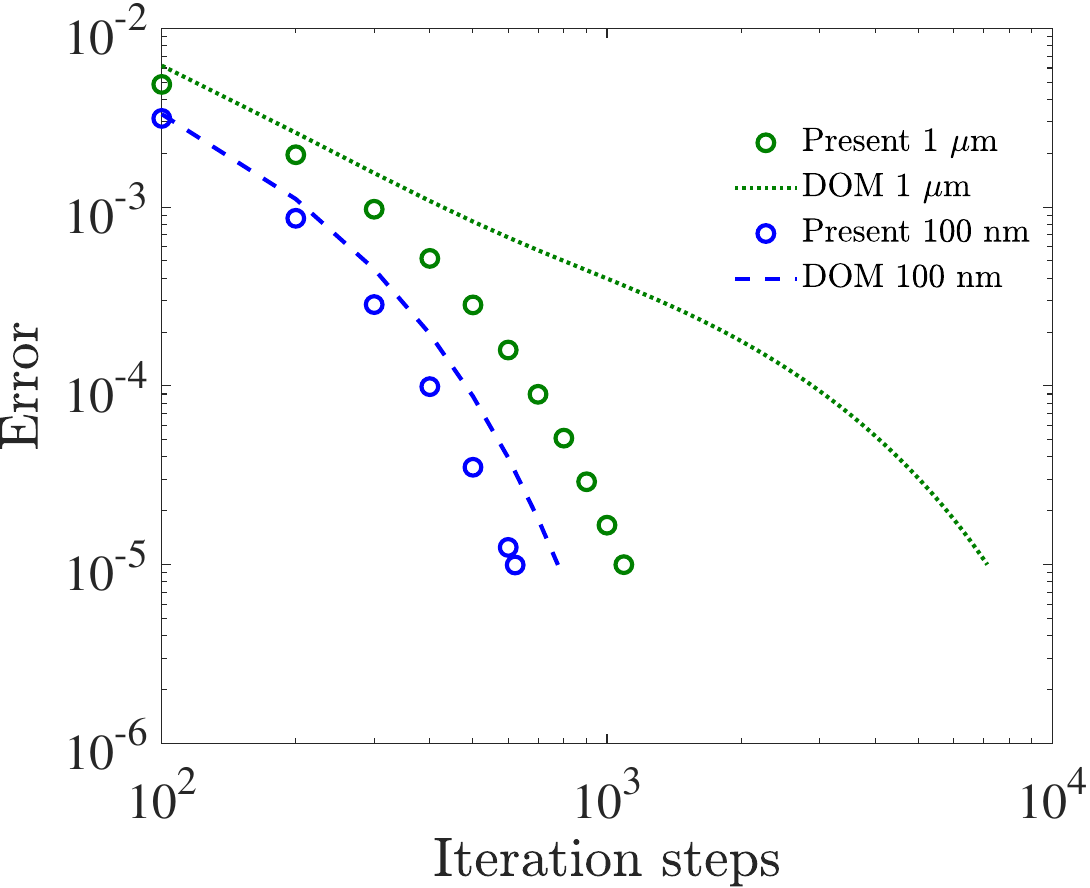}}~~
\subfloat[]{\includegraphics[width=0.45\textwidth]{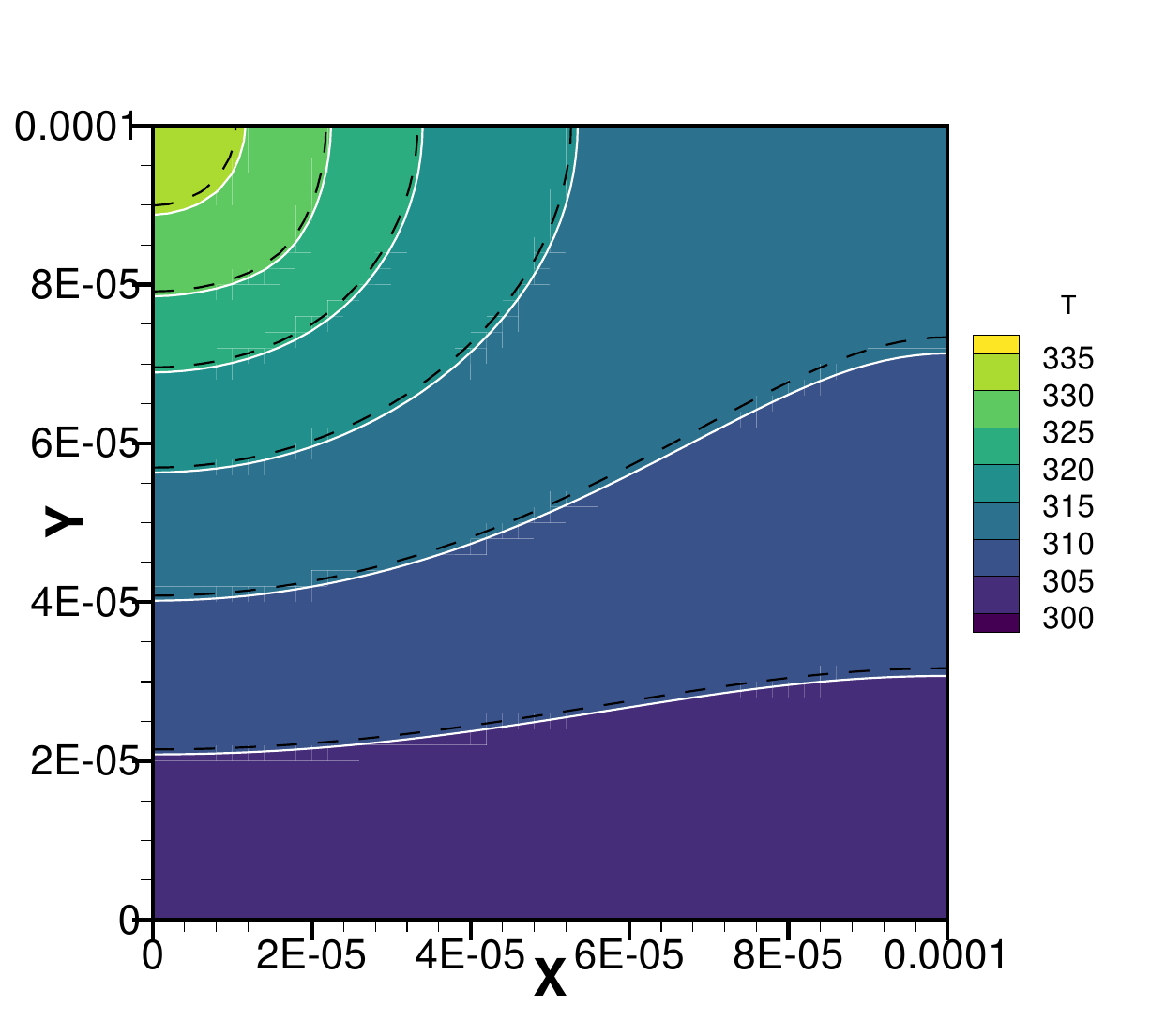}}~~ \\
\caption{(a) Schematic of a quasi-2D hotspot system. (b) Temperature contour with system size $L=100$ nm, where the background and white solid line is the present results and the black dashed line is the DOM solution. (c) {\color{black}{Convergence efficiency. Iteration is finished when (`Error') the difference of the macroscopic heat flux between two successive iteration step is smaller than $10^{-5}$.}} (d)  Temperature contour with system size $L=100~\mu$m, where the background and white solid line is the present results and the black dashed line is the solution of classical Fourier's law.}
\label{2Dhotspot}
\end{figure}
\begin{figure}[htb]
\centering
\subfloat[$L=100$ $\mu$m]{\includegraphics[width=0.4\textwidth]{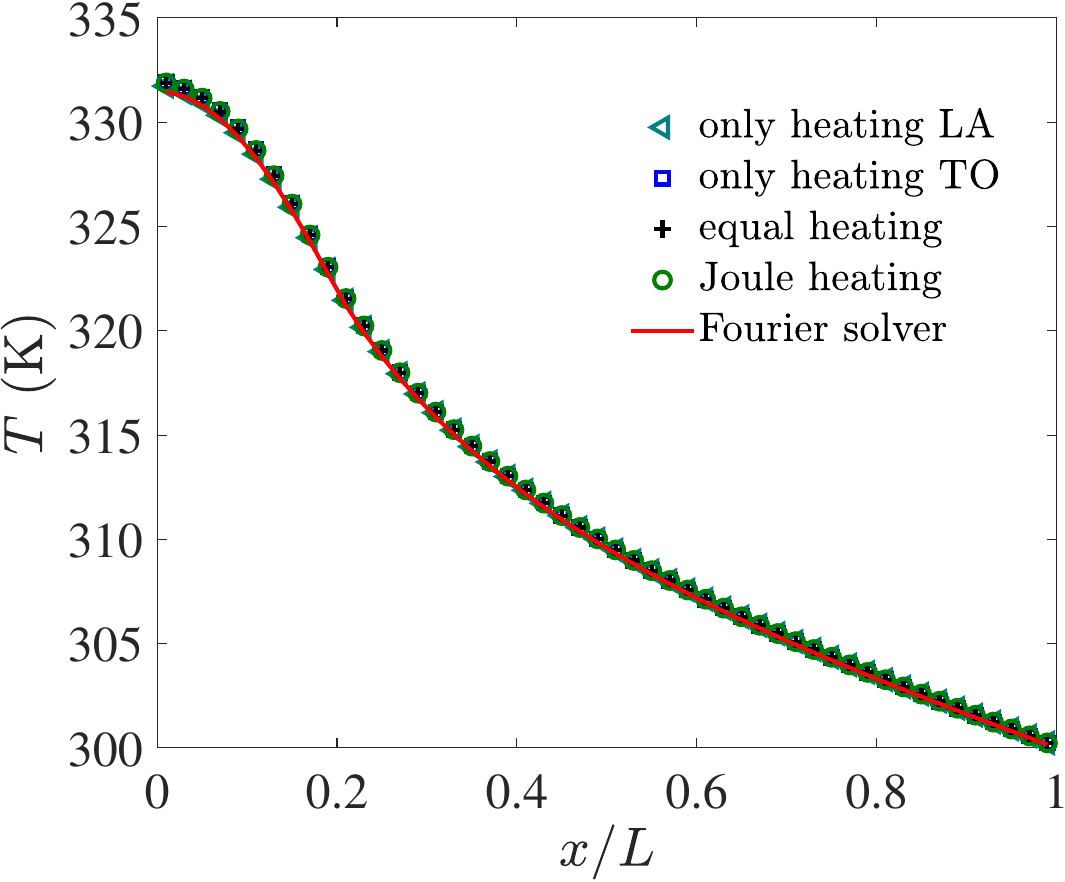}}~~
\subfloat[$L=1$ $\mu$m]{\includegraphics[width=0.4\textwidth]{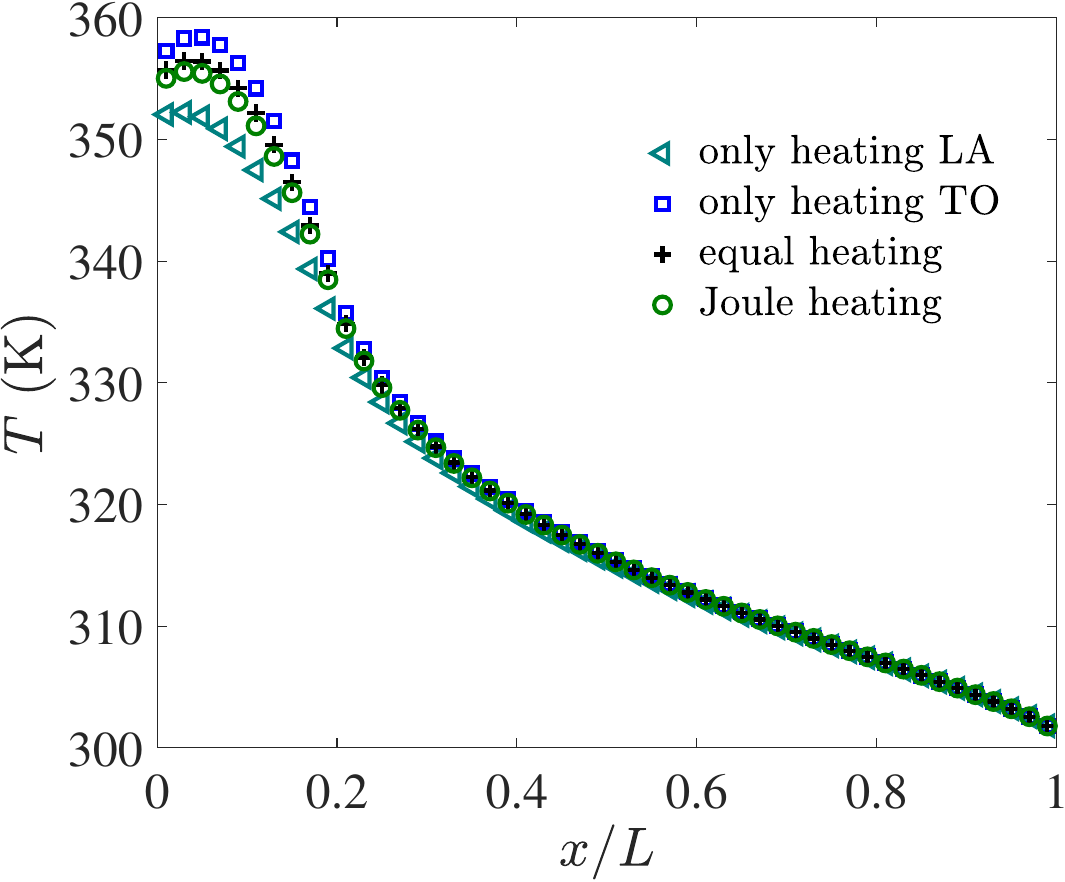}}~~ \\
\subfloat[$L=100$ nm]{\includegraphics[width=0.4\textwidth]{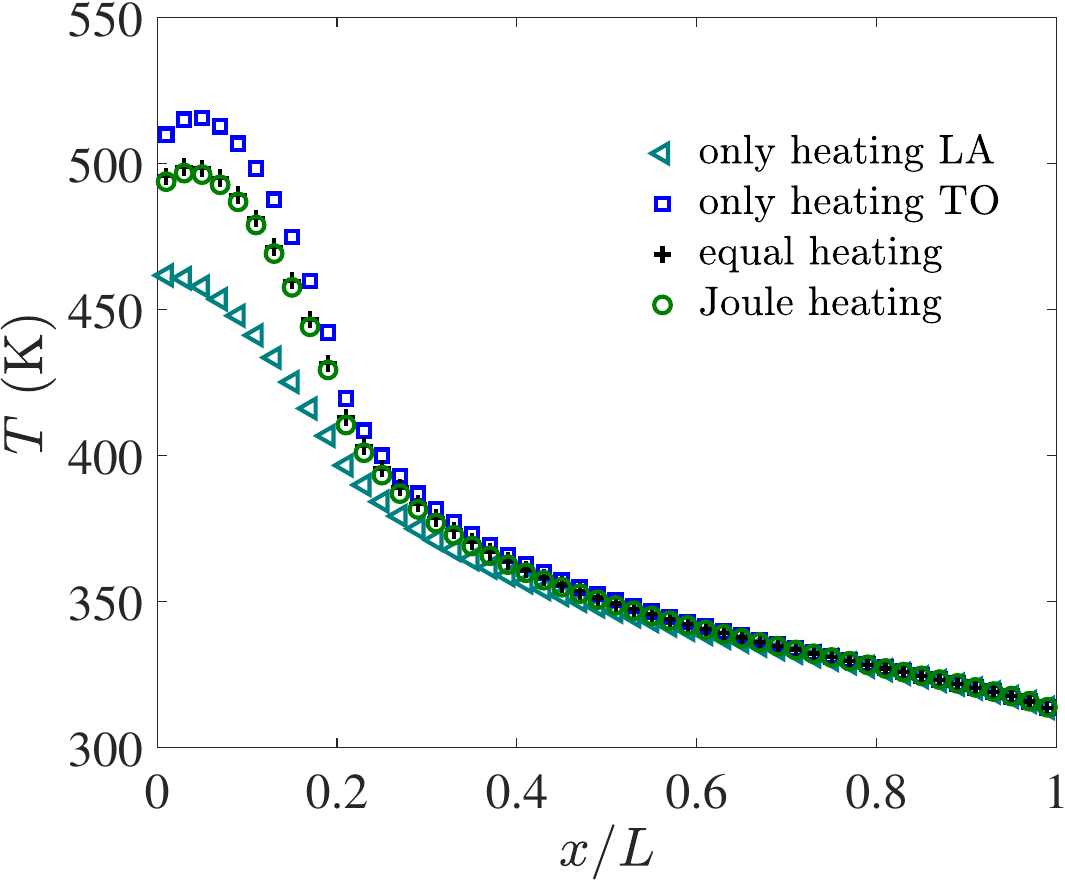}}~~
\subfloat[$L=50$ nm]{\includegraphics[width=0.4\textwidth]{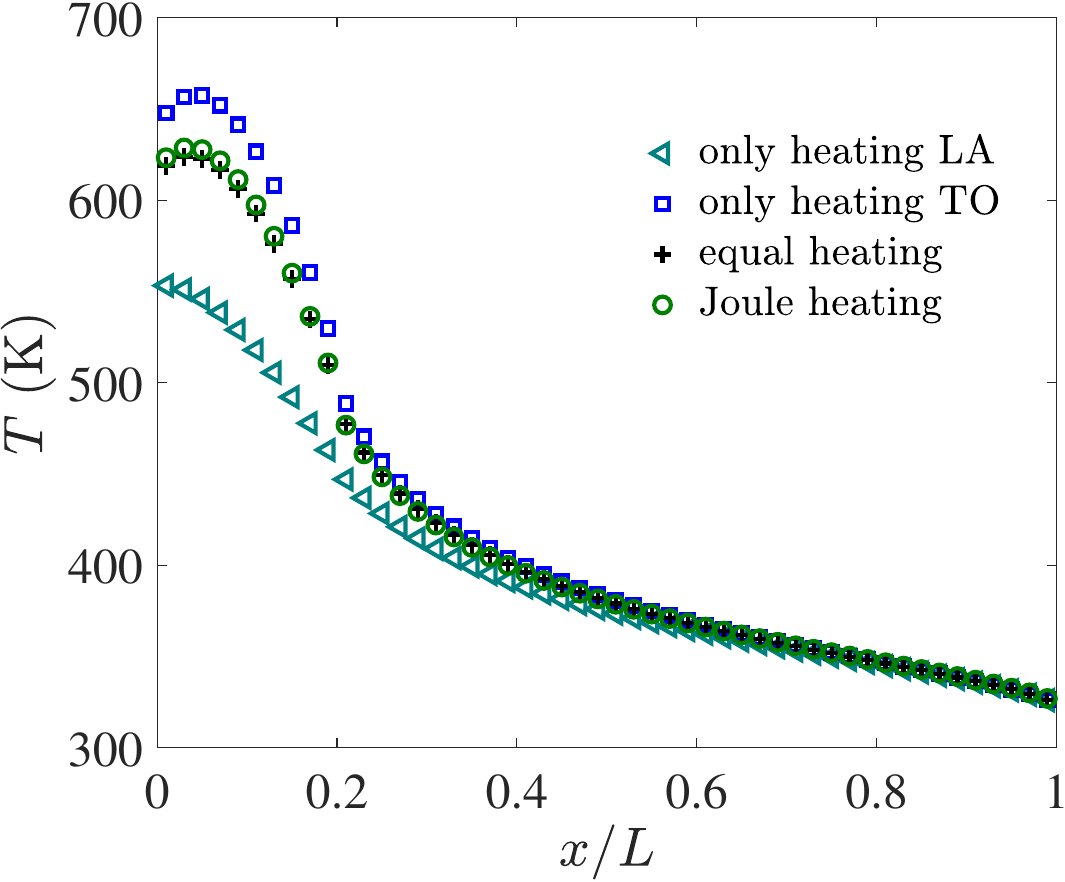}}~~ \\
\caption{Spatial distributions of the temperature along a diagonal line from the upper left to the lower right with various system sizes.}
\label{2Dhotspotthermal}
\end{figure}

\subsection{Heat conduction in a quasi-2D hotspot system}

As shown in~\cref{2Dhotspot}(a), a heat source with side length $H$ is embedded in the silicon substrate, which is similar to the heat generation and dissipations in a planar MOSFET device~\cite{YangRg05BDE}.
System size of substrate is $L= 5 H $ and the bottom is the heat sink with fixed temperature $T_c= 300$ K.
Left boundary is symmetric, and the top/right boundary is diffusely reflecting adiabatic boundary.
We control the total thermal input power to be equal, and heat generation term is $\dot{S}= 7.1 \times 10^{12}$ W/m$^3$, $7.1 \times 10^{16}$ W/m$^3$, $7.1 \times 10^{18}$ W/m$^3$ and $4 \times 7.1 \times 10^{16}$ W/m$^3$ for system size $L=100~\mu$m, $1~\mu$m, $100$ nm and $50$ nm, respectively.

Firstly, heat conduction with `Joule heating' is simulated by the synthetic iterative scheme and the predicted results are compared with the typical DOM or classical Fourier's law.
$50 \times 50$ discretized cells are used regardless of system sizes and $320$ CPU cores are used for paralleling.
From~\cref{2Dhotspot}(b), it can be found that when the system size is $L=100$ nm, the present results are in excellent agreement with the typical DOM and there are temperature slip near the boundaries.
When the system size increases, the present scheme has a faster convergence speed compared to the typical DOM, and it is very hard for DOM to quickly converge when the system size is larger than one micron, as shown in~\cref{2Dhotspot}(c). 
It needs more iteration steps compared to the cases in Sec.~\ref{sec:quasi1dsec}.
That's because 1) the relaxation time changes significantly in each iteration step due to the large heat generation. 2) the temperature at the diffusely/specular reflecting boundaries are unknown and changes with the interior macroscopic distributions in each iteration steps.
These input parameters and boundary fields change with the iterative process, which indicates that the targeted discretized equations we numerically solve in each iteration step is changing.
When system size is $100~\mu$m (\cref{2Dhotspot}(d)), our results could recover the classical Fourier's law with bulk thermal conductivity.
These results show that the present scheme could capture the heat conduction from the ballistic to diffusive regime.

Secondly, the effects of selective phonon excitation (e.g., `only heating LA', `only heating TO', `equal heating', `Joule heating') on heat conduction in quasi-2D hotspot system is studied.
Basic heat conduction process is shown as follows: when a phonon mode absorbs thermal energy from the external heat source, on one hand, it can transfer energy to other phonon modes at the same geometric location through phonon-phonon scattering, or transfer energy from the heat source region to other geometric regions through ballistic transport. 
These two energy transfer processes almost occur simultaneously until the phonon energy is completely absorbed by the heat sink at the bottom.

When system size is large enough, sufficient phonon-phonon scattering dominates heat conduction.
When phonons absorb the thermal energy from the heat source, they will suffer frequent energy exchange with other phonons before they are absorbed by the bottom heat sink.
Sufficient energy exchange leads to a similar temperature distribution regardless of selective phonon excitations.
Numerical results are in excellent agreement with classical Fourier's solutions, as shown in~\cref{2Dhotspotthermal}(a).
When the system size decreases, phonon-phonon scattering becomes insufficient and ballistic phonon transport dominates heat conduction.
On one hand, the decreasing system size reduces the effective thermal conductivity, which results in a higher temperature rise.
On the other hand, the insufficient phonon-phonon scattering process will allow different phonon modes to transfer energy from the heat source region to the heat sink independently of each other.
Note that different phonon modes have different specific heat and group velocity, namely, they have different heat conduction abilities. 
As can be seen from the numerical results shown in~\cref{2Dhotspotthermal}, the temperature rise is smallest when the heat source only heats LA phonons, while it is highest when only heats TO phonons. 
It indicates that LA phonons have strongest heat conduction ability.
In addition, it can be found that the spatial distributions of temperature for `equal heating' and `Joule heating' are almost the same, which is similar to that predicted in the above subsection~\ref{sec:quasi1dSPE}.
The present results can provide a theoretical basis for regulating the heat dissipation efficiency of semiconductor devices or hotspot systems by selective phonon excitation.
}}

\section{Comparison between the effective Fourier's law and phonon BTE}
\label{sec:comparison}

{\color{black}{In this section, thermal dissipations in FinFET/GAAFET structures are simulated, where most of geometrical parameters are obtained from previous references~\cite{3DFINFET_2014_mc,bulkfinfet2019,3DFINFETtransient,Nanosheet_FINFET2018,IBM_2017_nanosheet}.
Discretization of the wave vector space and the parallel computing strategy of the whole solution process are set up in the same way as mentioned in the above section \ref{sec:results}.
We mainly want to investigate how far the predictions of effective Fourier's law \eqref{eq:effFourierlaw} deviate from the phonon BTE.
This research will be of great significance for thermal applications.

Note that the substrate size is larger than $5~\mu$m or tens of microns in some references~\cite{gaafet2020,Three_Stacked_NanoplateFET2018}, while there are only hundreds of nanometers in some references~\cite{finfet2018sub14,comparison_FIN_GAAFET,3DFINFET_2014_mc}.
Obviously the peak temperature will be lower and the computational cost will be smaller if the heat sink is closer to the Joule heating areas.
In the following simulations, we fixed the substrate size at $200$ nm according to the previous paper~\cite{3DFINFET_2014_mc}.}}

\subsection{Heat conduction in a quasi-2D vertical cross-section FinFET}

\begin{figure}[htb]
\begin{minipage}[b]{0.3\linewidth}
\centering
\subfloat[Quasi-2D FinFET]{\includegraphics[scale=0.25,clip=true]{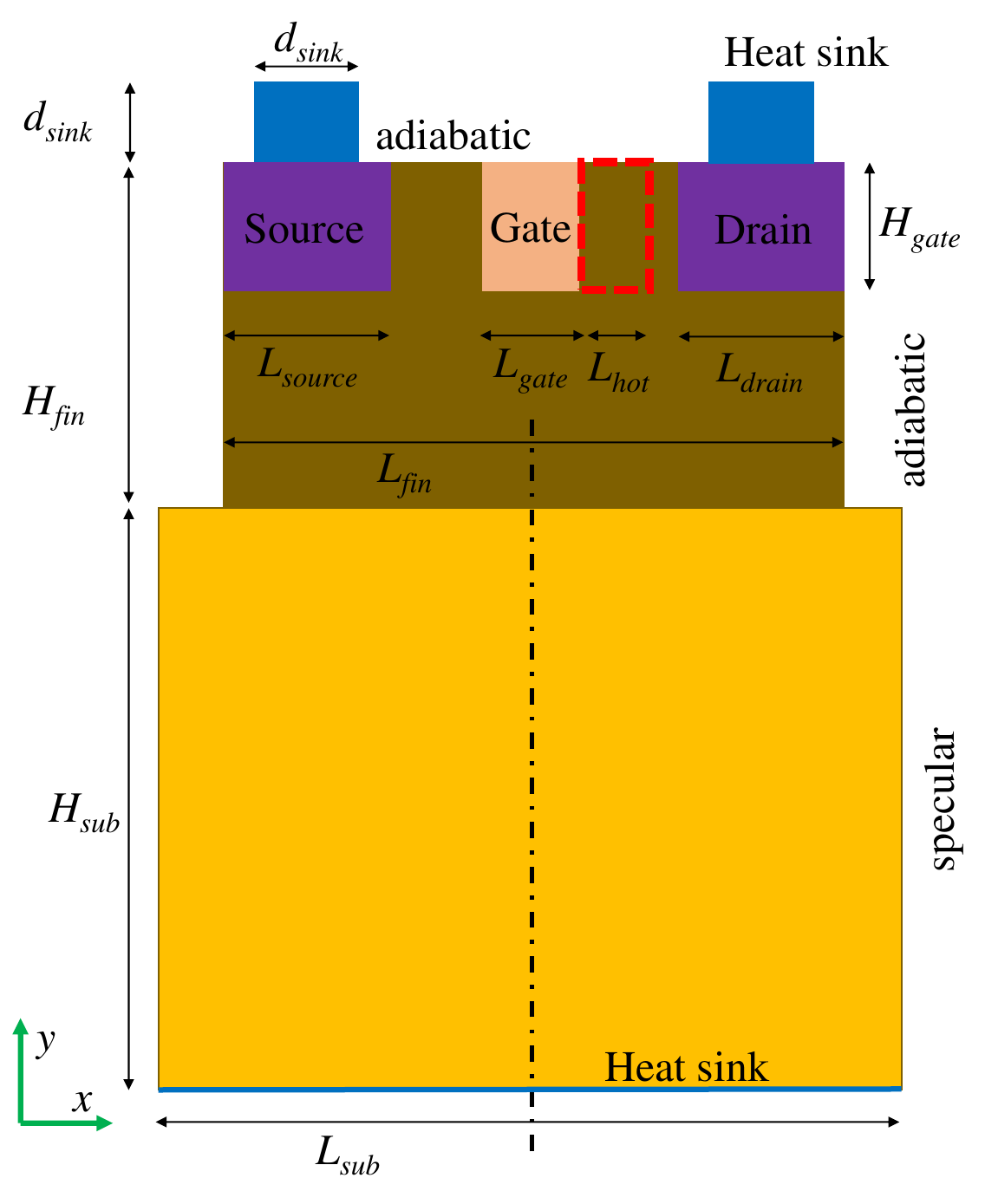}} \\
\subfloat[Discretized cell]{\includegraphics[scale=0.30,viewport=130 10 640 550,clip=true]{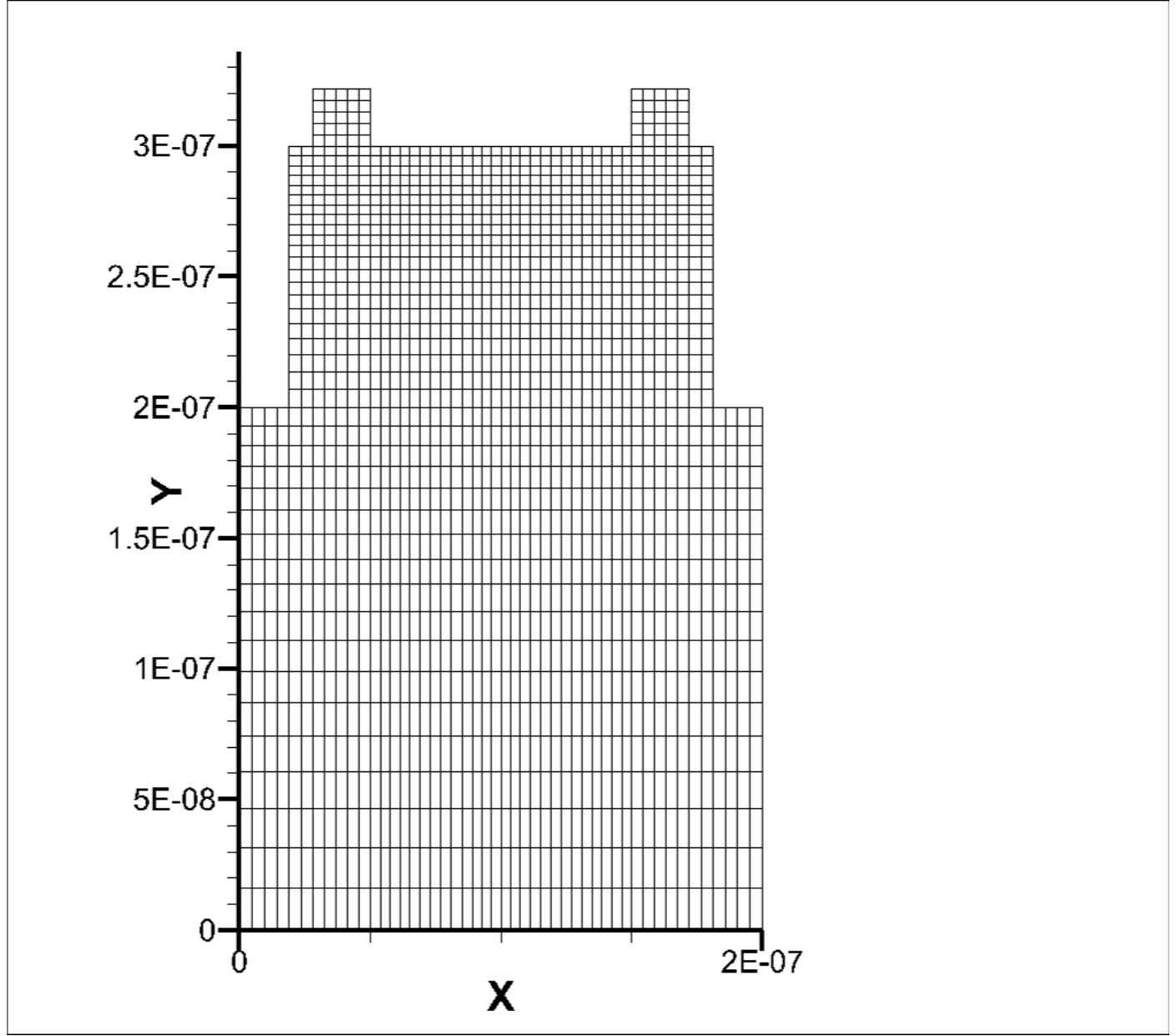}} 
\end{minipage}
~
\begin{minipage}[b]{0.3\linewidth}
\centering
\subfloat[Only heating TO, BTE]{\includegraphics[scale=0.30,viewport=130 10 640 550,clip=true]{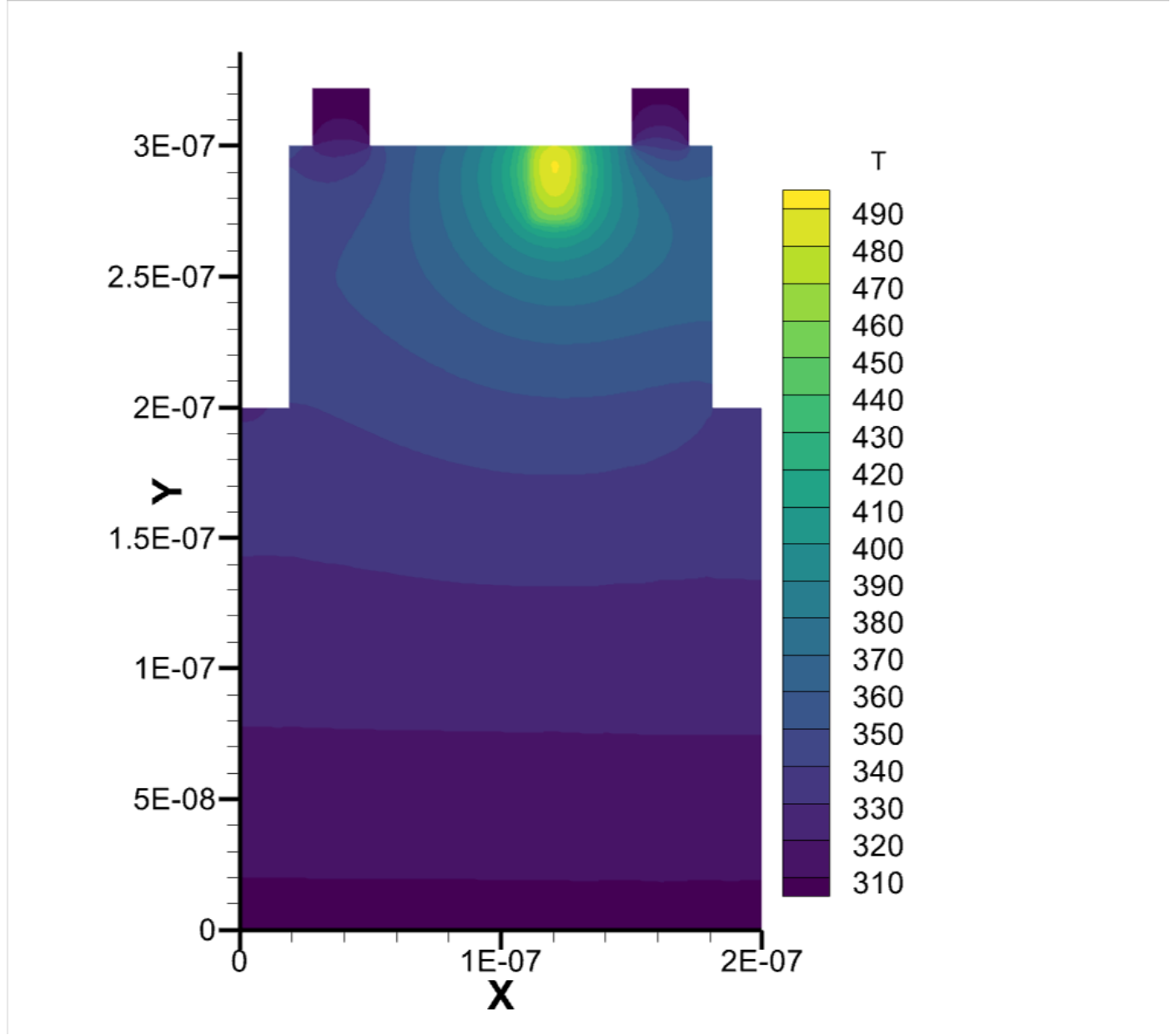}}   \\
\subfloat[Joule heating, BTE]{\includegraphics[scale=0.30,viewport=130 10 640 550,clip=true]{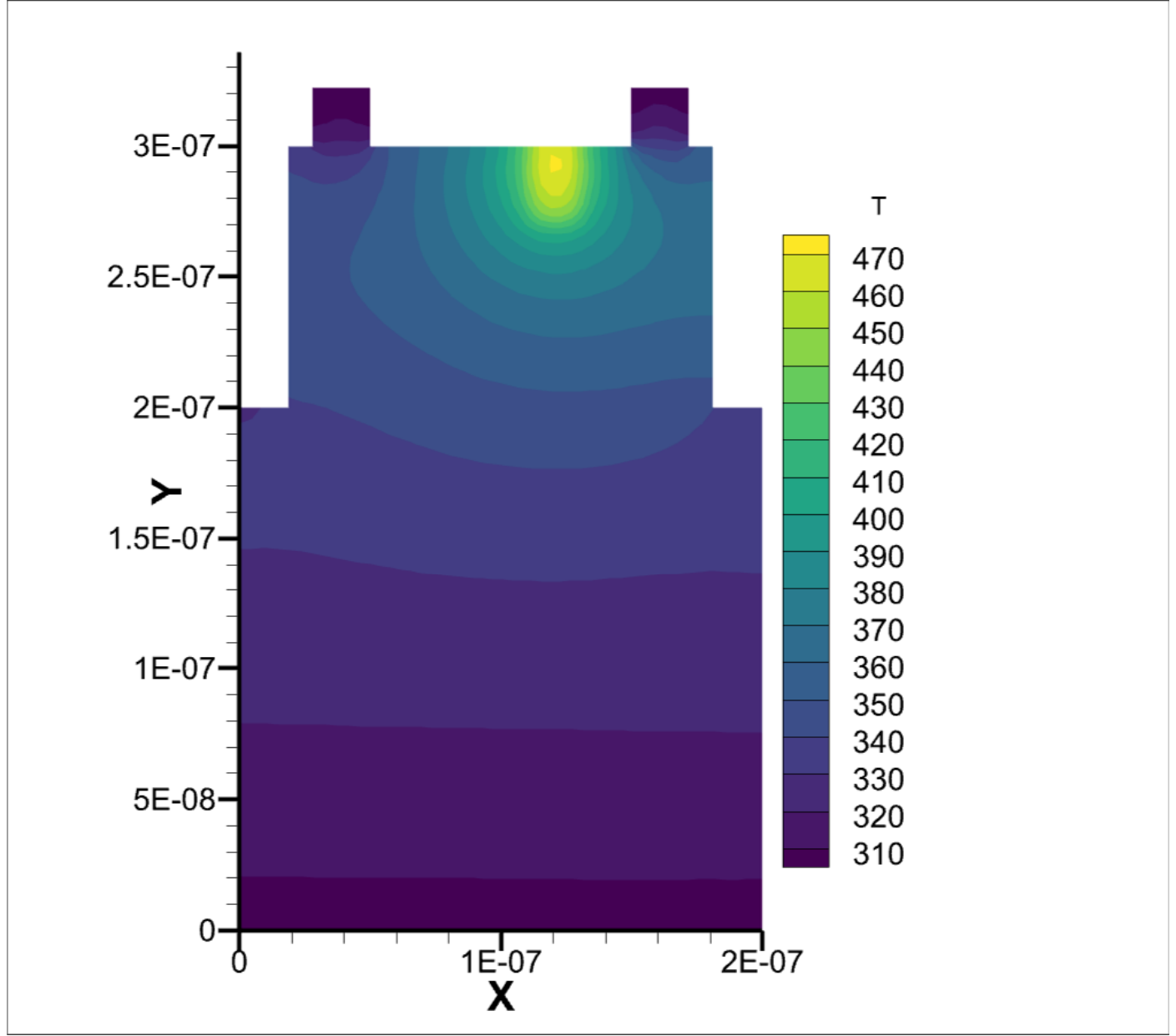}}
\end{minipage}
~
\begin{minipage}[b]{0.3\linewidth}
\centering
\subfloat[Equal heating, BTE]{\includegraphics[scale=0.30,viewport=130 10 640 550,clip=true]{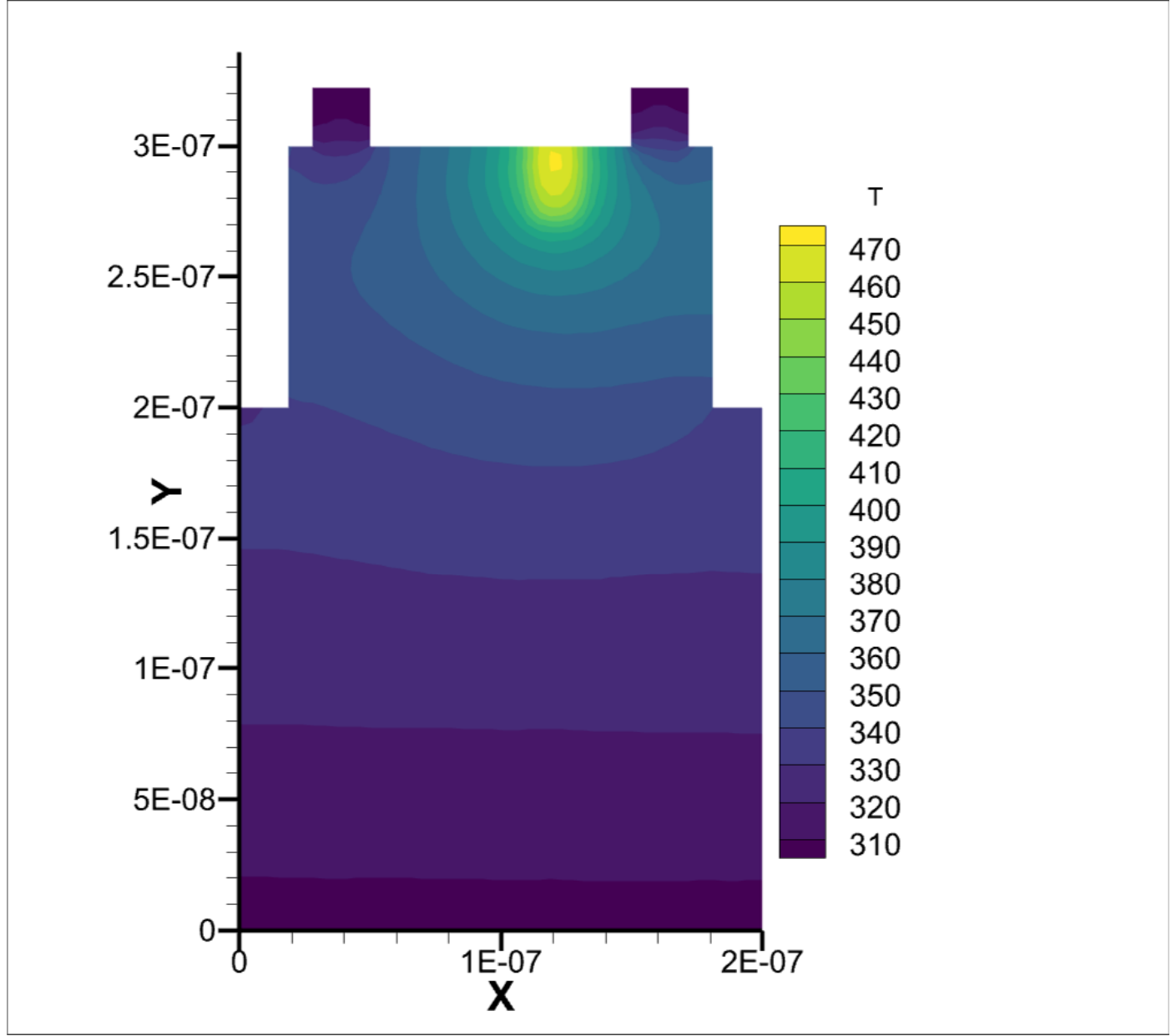}}  \\
\subfloat[Effective Fourier's law]{\includegraphics[scale=0.30,viewport=130 10 640 550,clip=true]{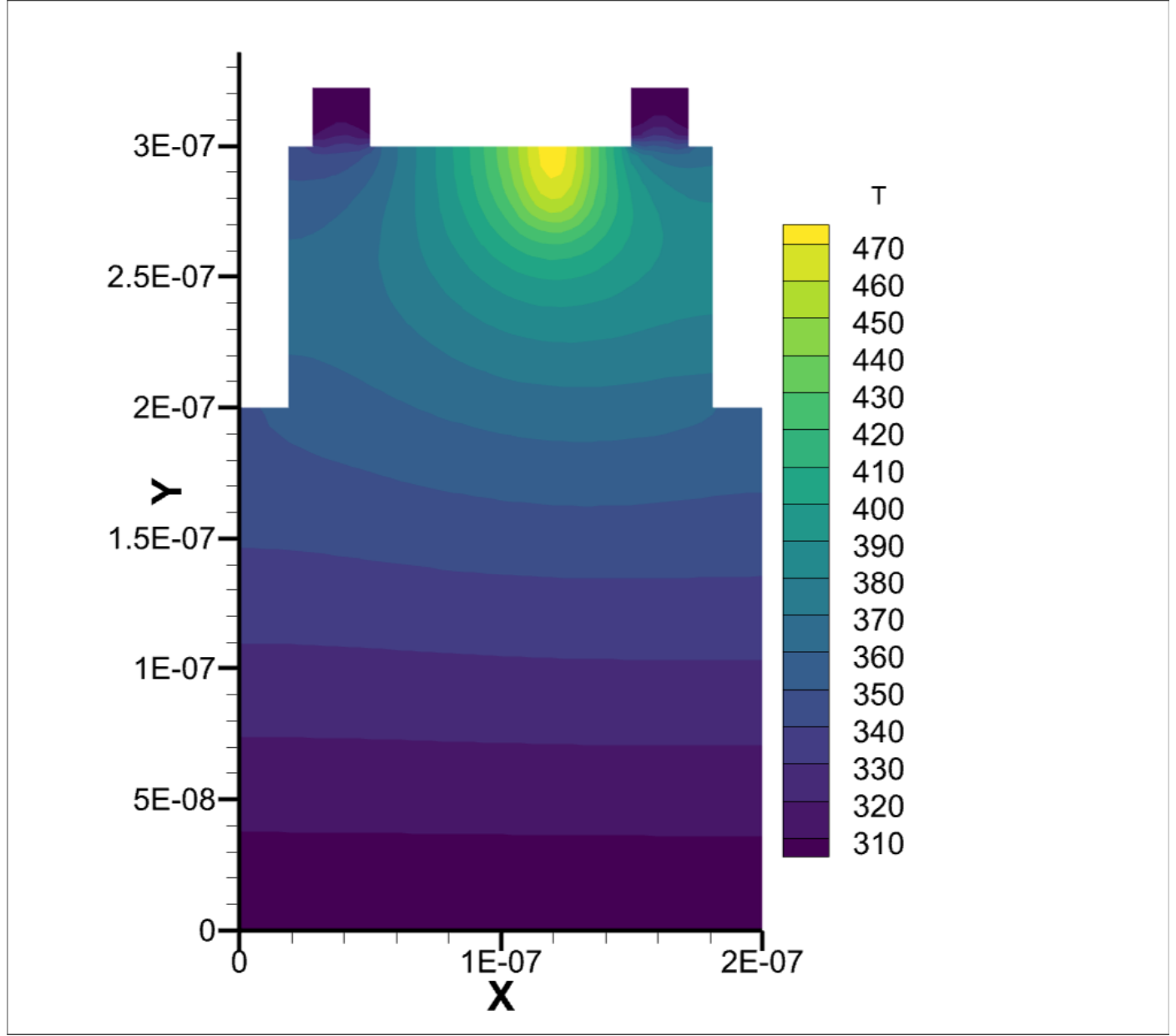}} 
\end{minipage}
\caption{(a) Schematic view of a quasi-2D vertical cross-section FinFET geometry. The Joule heating area is generated at the right edge of the gate with power density $7.1\times 10^{18}$ W/m$^3$. (b) $48 \times 44$ non-uniform discretized cells are used for the computational domain. We also simulated the numerical results under a denser grid (grid density in the $x$ direction is 4 times denser and grid density in the $y$ direction is 2 times denser), which is basically no different from the current results. (c-f) Steady temperature distributions predicted by synthetic iterative scheme with (c,d,e) various selective phonon excitation and predicted by (f) the effective Fourier's law with temperature- and size-dependent effective thermal conductivity. }
\label{FINFET2022_thermal}
\end{figure}
Non-Fourier heat conduction in a quasi-2D vertical cross-section FinFET~\cite{3DFINFET_2014_mc} is studied and schematic of geometry is shown in~\cref{FINFET2022_thermal}(a).
A fin is placed on the top of a substrate, where both height and length of the substrate are $H_{sub}=L_{sub}=200$ nm.
The bottom of substrate is a heat sink with isothermal temperature $300$ K and its left and right boundaries are symmetric because many FinFETs are arranged on the substrate by periodic arrays.
The height and length of fin is $H_{fin}= 100$ nm and $L_{fin}=162$ nm, respectively.
Source and drain are placed at the two end side of the fin with length $L_{source}=L_{drain}=40$ nm.
Gate length is $L_{gate}=22$ nm, and its position is at the center between source and drain.
The heights of source, drain and geat are all $H=30$ nm, and the spacer lengths $L_{sp}$ between source and gate, drain and gate are both $30$ nm.
Two square heat sinks with edge length $d_{sink} = 22$ nm, whose top, left and right boundaries are isothermal boundary conditions with temperature $300$ K, are placed directly above the center of the source and drain respectively.
Other boundaries of this structure are all diffusely reflecting adiabatic boundary conditions.
The external heat source is at the right edge of the gate with power density $\dot{S}=7.1\times 10^{18}$ W/m$^3$~\cite{3DFINFET_2014_mc} and length $L_{hot}=20$ nm.

Three selective phonon excitations are considered, i.e., `equal heating', `only heating TO' and `Joule heating'.
Non-uniform discretized cells are used and shown in~\cref{FINFET2022_thermal}(b).
Numerical results with various selective phonon excitation obtained by the present scheme are shown in~\cref{FINFET2022_thermal}(c,d,e).
It can be found that the temperature rises near Joule heating spot are highest, larger than $150$ K.
The temperature rises of `Only heating TO' are higher than `equal heating' and `Joule heating', which is in consistent with the results in~\cref{target_heating1D}.

{\color{black}{Effective Fourier's law with temperature- and size-dependent effective thermal conductivity
\begin{align}
\kappa_{eff} = \alpha(\bm{x},L) \kappa_{bulk}(T)=  \alpha(\bm{x},L) \times  \exp(12.570) / T^{1.326}
\label{eq:effectivekappa}
\end{align}
is also simulated, where $\kappa_{bulk}(T)= \exp(12.570) / T^{1.326} $ is the temperature-dependent bulk thermal conductivity of silicon given in Ref.~\cite{Lacroix05} and $\alpha$ is a dimensionless parameter depends on the spatial position $\bm{x}$ or system characteristic length $L$.}}
Based on previous studies of size effects in room temperature silicon~\cite{ZHANG20191366,zhang2021e}, for example, the thermal conductivity of in-plane heat conduction (the length in the other two directions is infinite) at film thickness $100~\mu$m, $100$ nm and $10$ nm is $146.0$ W/(m$\cdot$K), $57.6$ W/(m$\cdot$K), $16.3$ W/(m$\cdot$K), respectively.  
We set $\alpha \approx 10.0/146.0$ in the top square heat sink areas.
Considering the similar size of the fins and the substrate areas, we approximately assume that the thermal conductivity of the two regions is equal in the present simulation and $\alpha \approx 30.0/146.0$.
Numerical results in~\cref{FINFET2022_thermal}(f) show that the temperature rise predicted by effective Fourier's law deviates a little from the numerical results of phonon BTE with `equal heating'.
One of reasons for deviations is that there is no temperature or heat flux slip near the boundaries in the solutions of effective Fourier's law.
Similar thermal distributions predicted by the effective Fourier's law and phonon BTE show that the macroscopic equations with several empirical coefficients could coarsely capture the micro/nano scale heat conduction in this case.

\begin{figure}[htb]
\begin{minipage}[b]{0.3\linewidth}
\centering
\subfloat[Single-fin bulk FET]{\includegraphics[scale=0.19, clip=true]{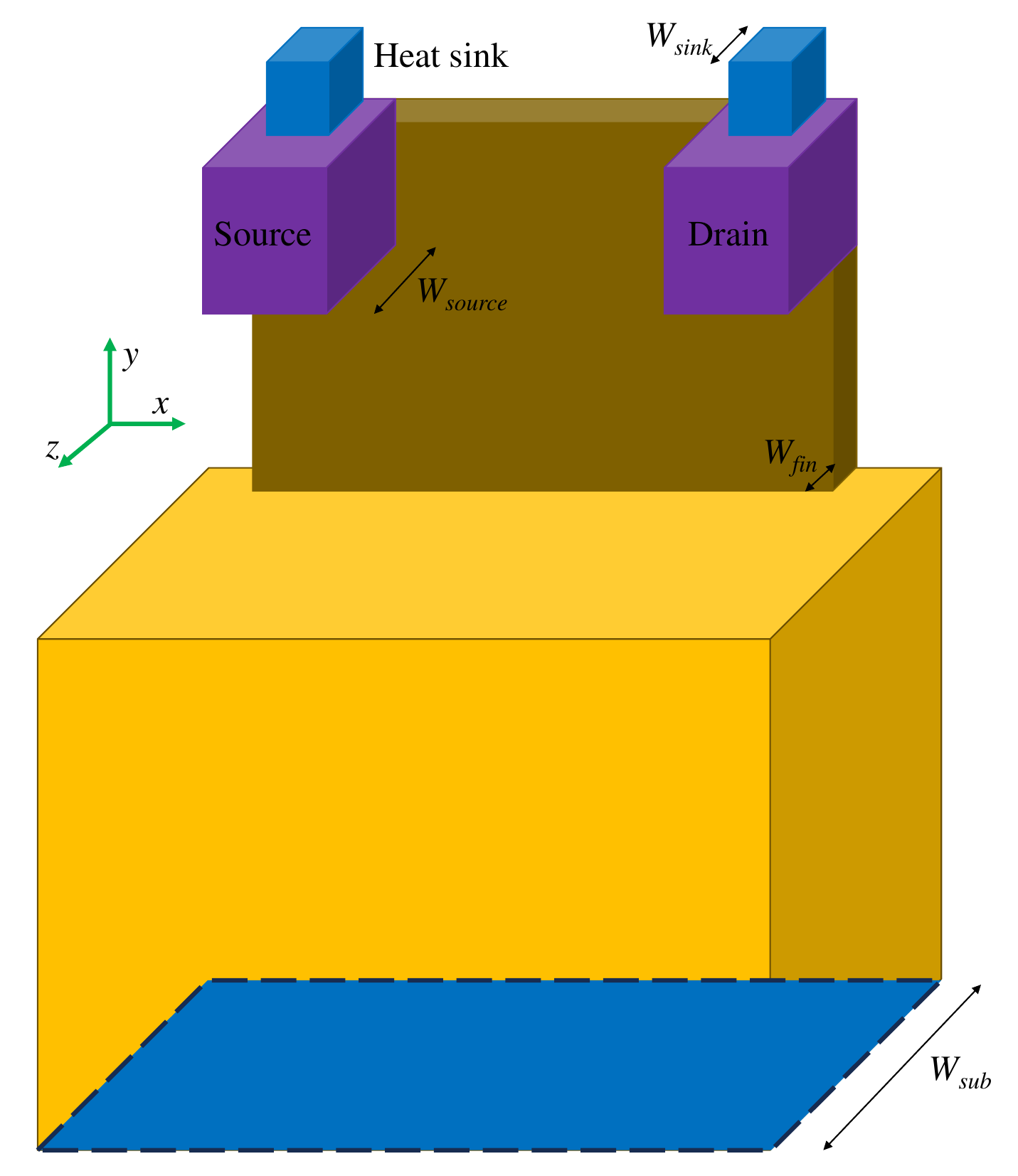 }} \\
\subfloat[Discretized cells]{\includegraphics[scale=0.19, clip=true]{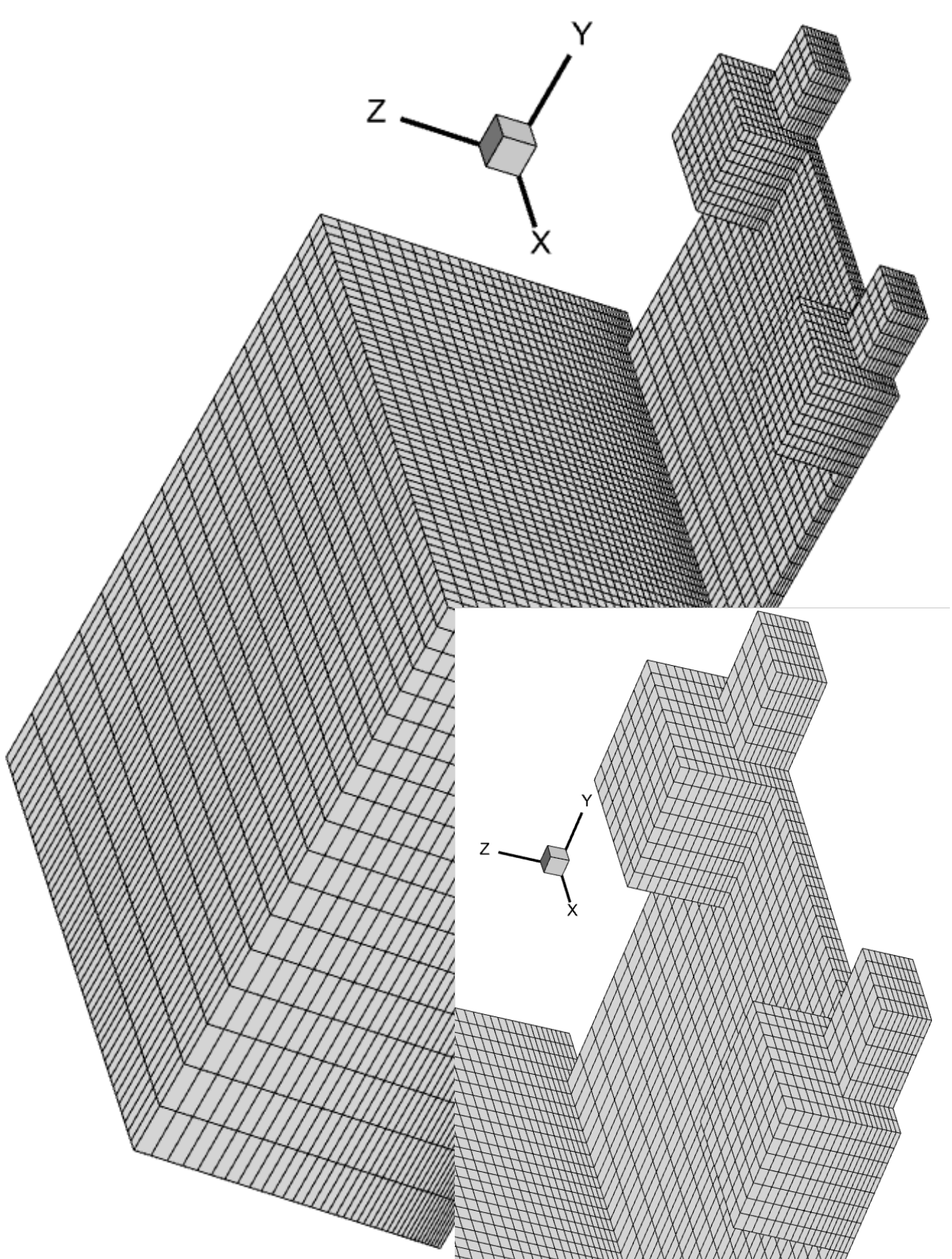 }} 
\end{minipage}
~
\begin{minipage}[b]{0.3\linewidth}
\centering
\subfloat[BTE solution]{\includegraphics[scale=0.28,viewport=130 10 640 580,clip=true]{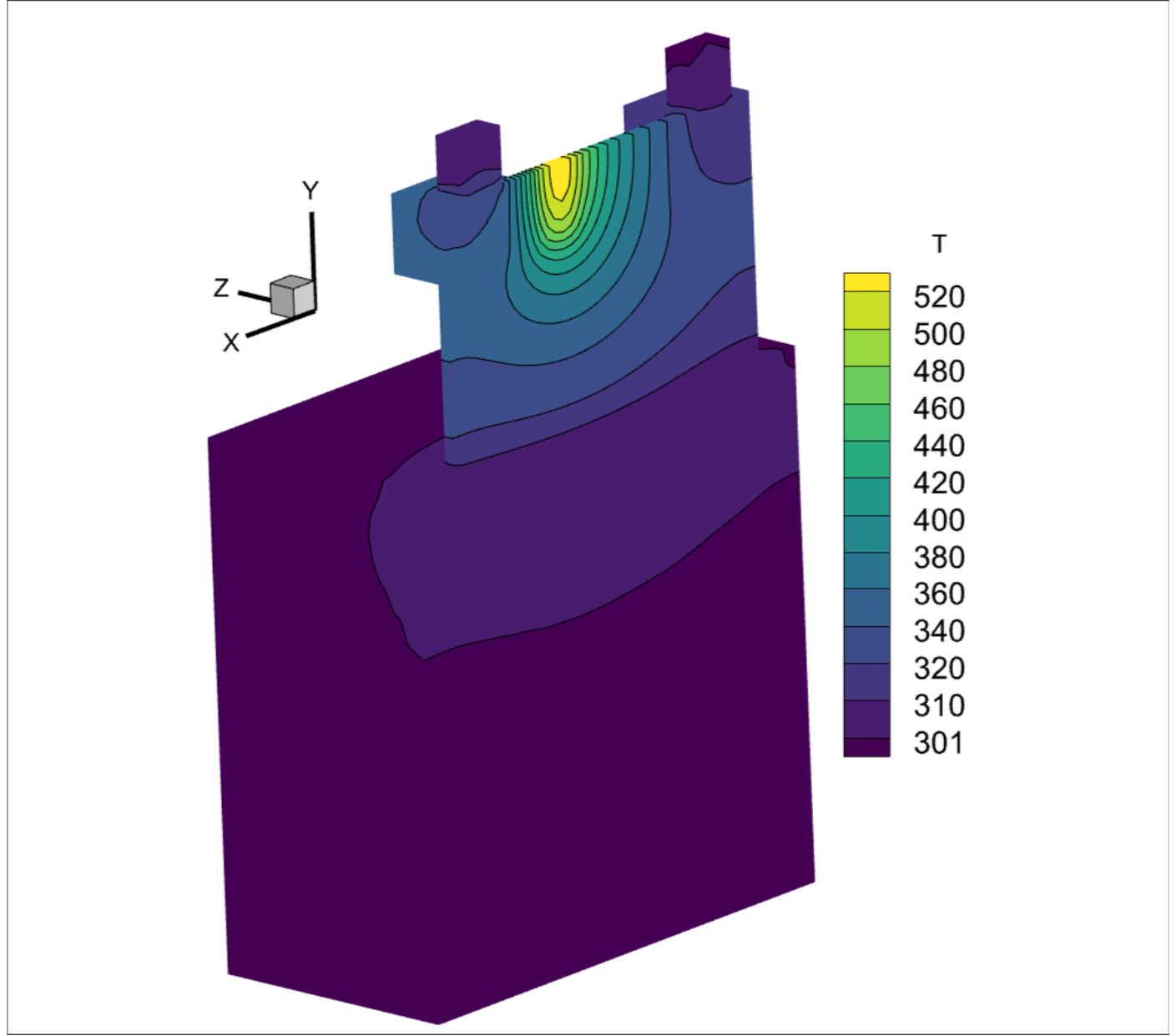}}  \\
\subfloat[Effective Fourier's law]{\includegraphics[scale=0.28,viewport=170 10 680 580,clip=true]{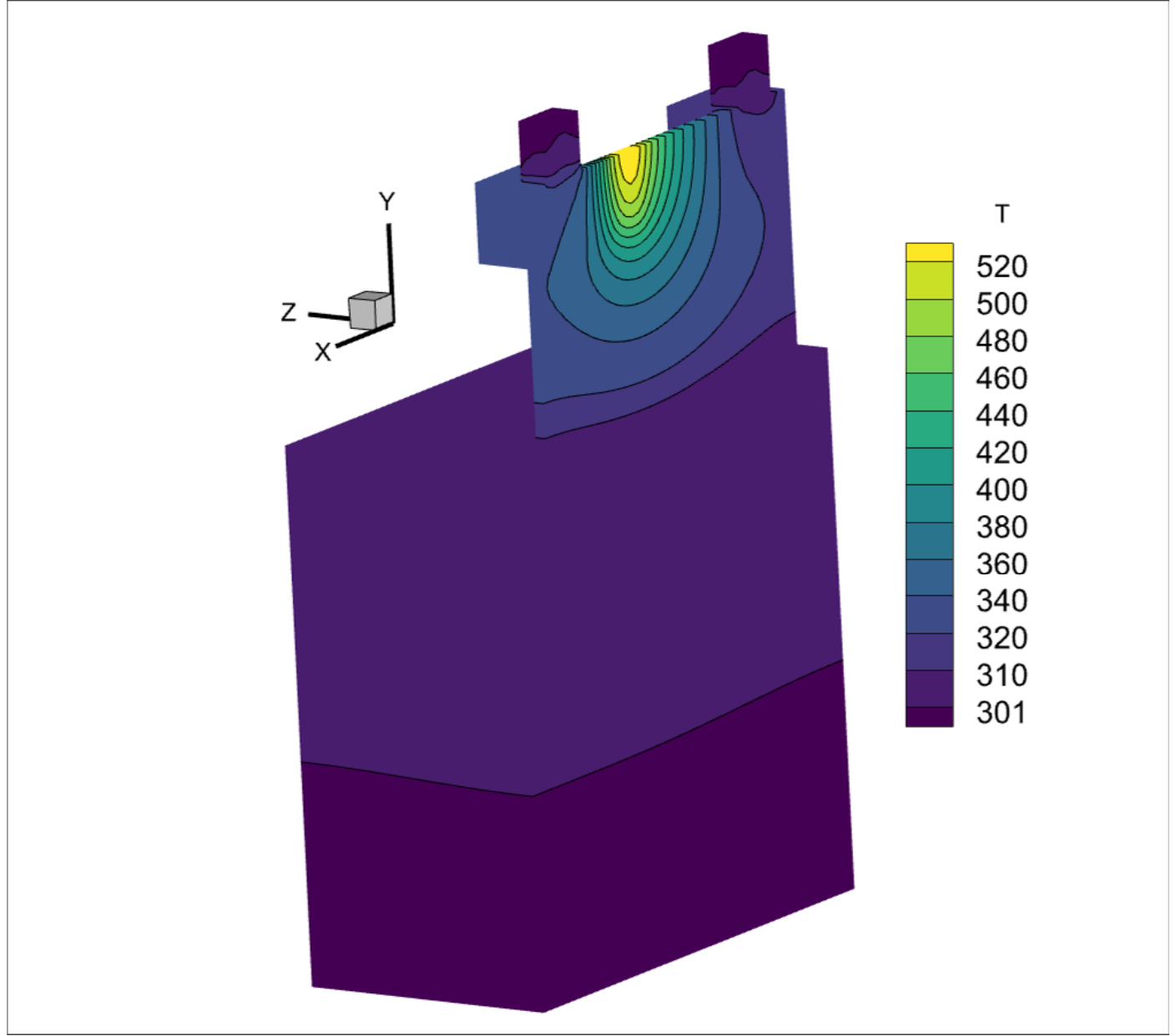}}  
\end{minipage}
~
\begin{minipage}[b]{0.3\linewidth}
\centering
\subfloat[XY plane, BTE solution]{\includegraphics[scale=0.28,viewport=130 10 640 580,clip=true]{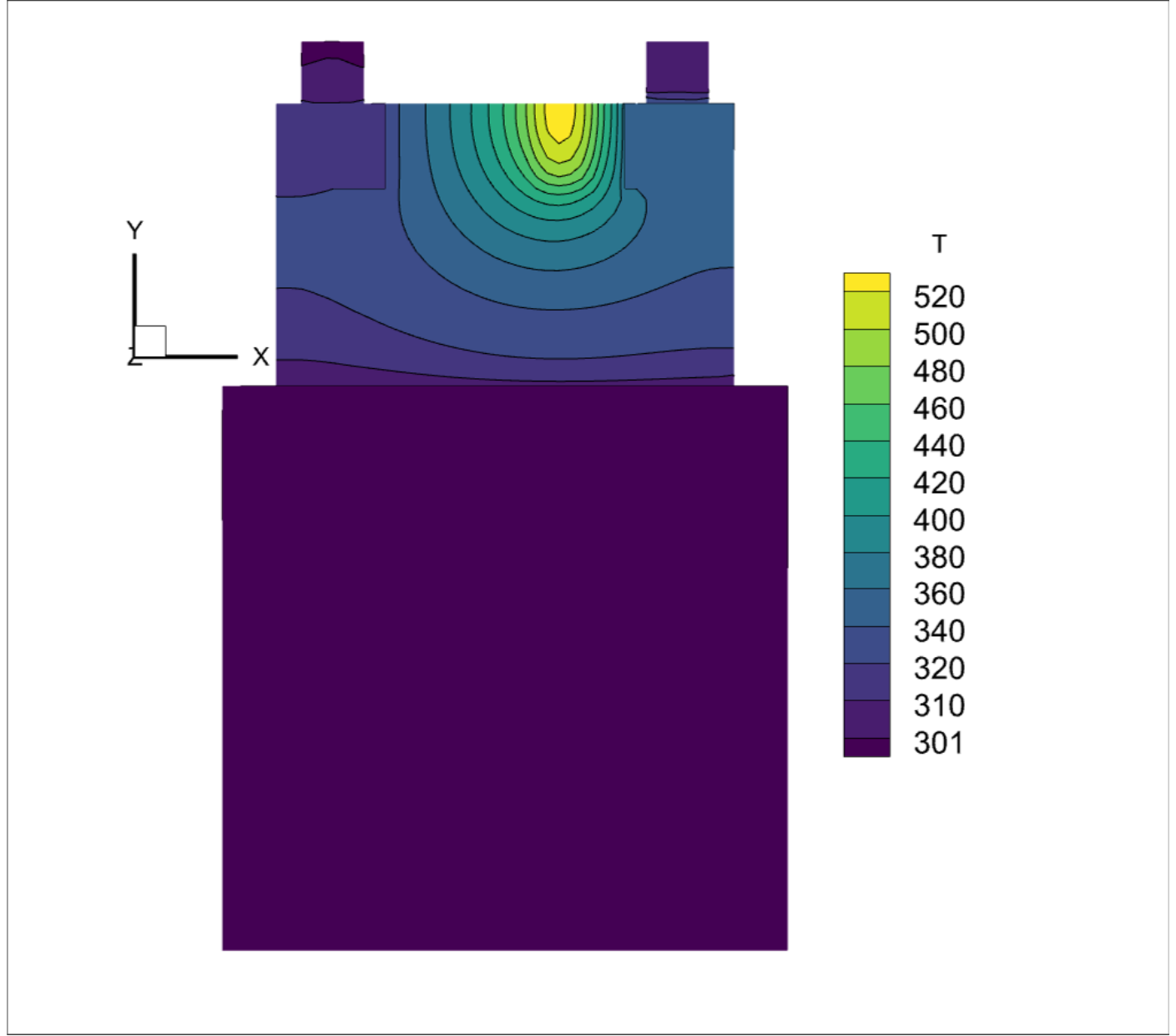}} \\
\subfloat[XY plane, Effective Fourier's law]{\includegraphics[scale=0.28,viewport=170 10 680 580,clip=true]{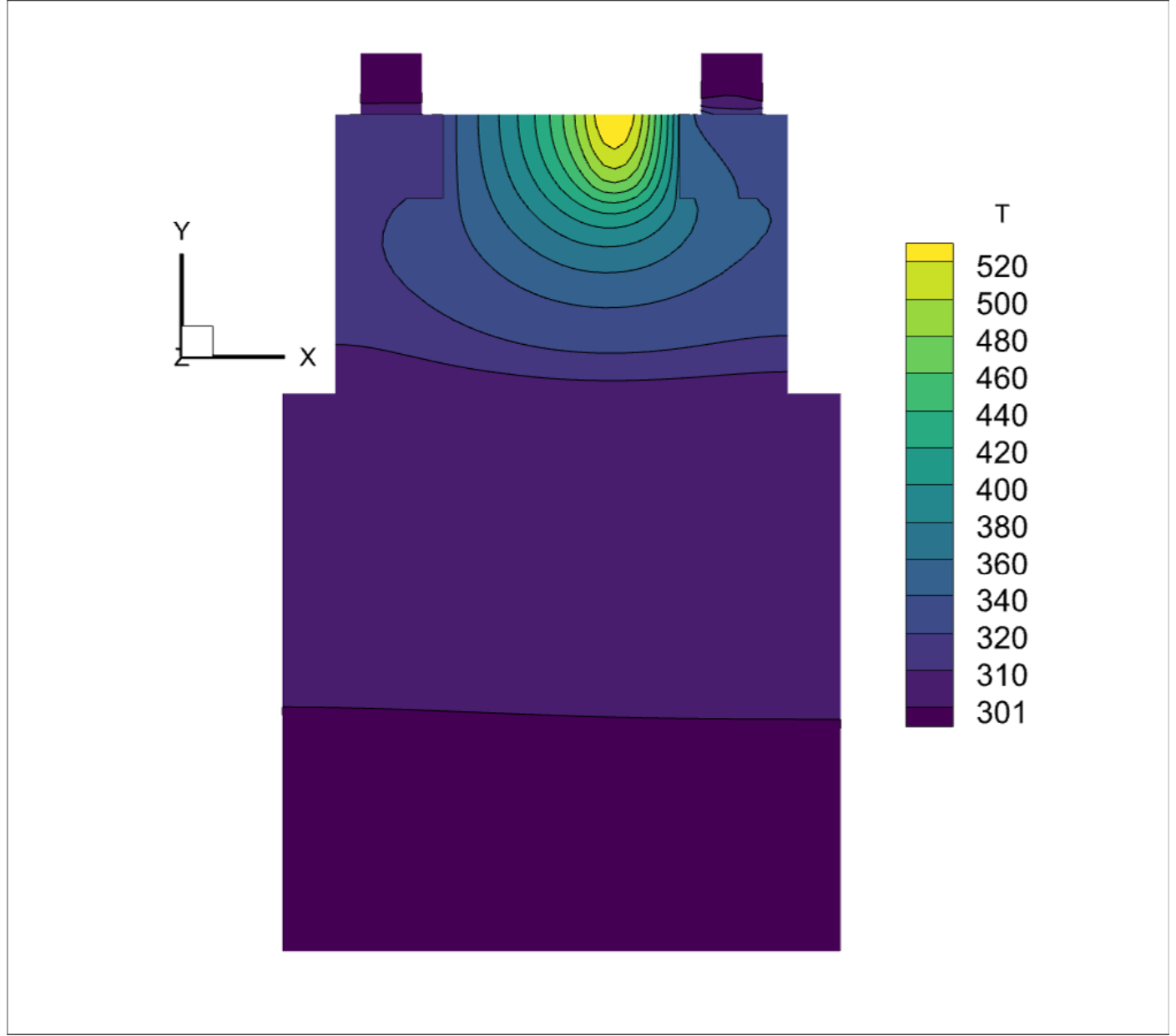}}
\end{minipage}
\caption{(a) Half of a single-fin bulk FET geometry~\cite{3DFINFET_2014_mc}, where the whole back surface of this structure is symmetric. (b) $48 \times 44 \times 39$ non-uniform discretized cells are used for the computational domain. Steady temperature distributions of half of a single-fin bulk FET predicted by (c,d) synthetic iterative scheme and (e,f) effective Fourier's law.}
\label{FINFET3D2022_temperature}
\end{figure}
\begin{figure}[htb]
\centering
\subfloat[Double-fin bulk FET]{\includegraphics[width=0.33\textwidth]{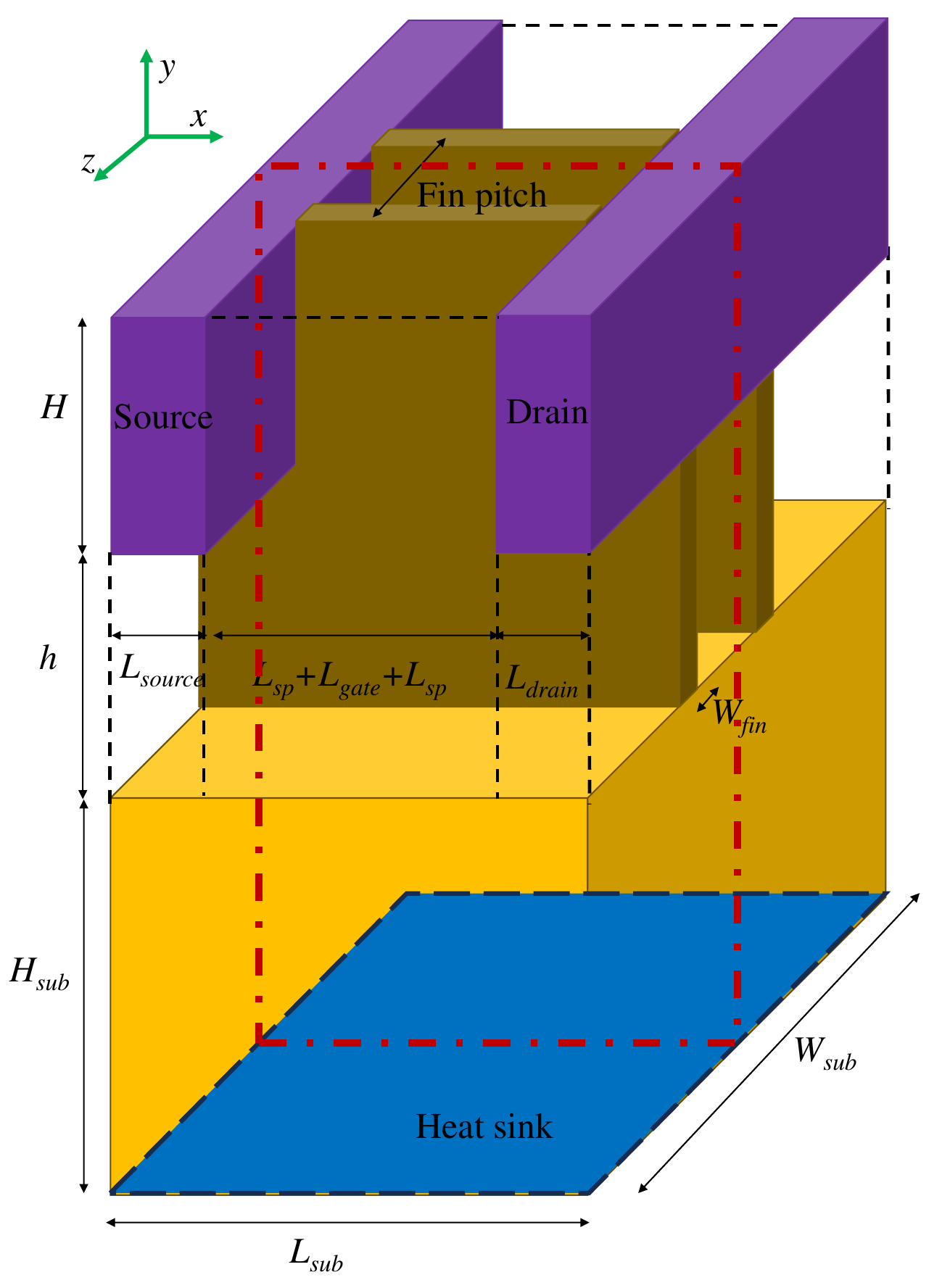 }}~~~~~~
\subfloat[Discretized cells]{\includegraphics[width=0.32\textwidth]{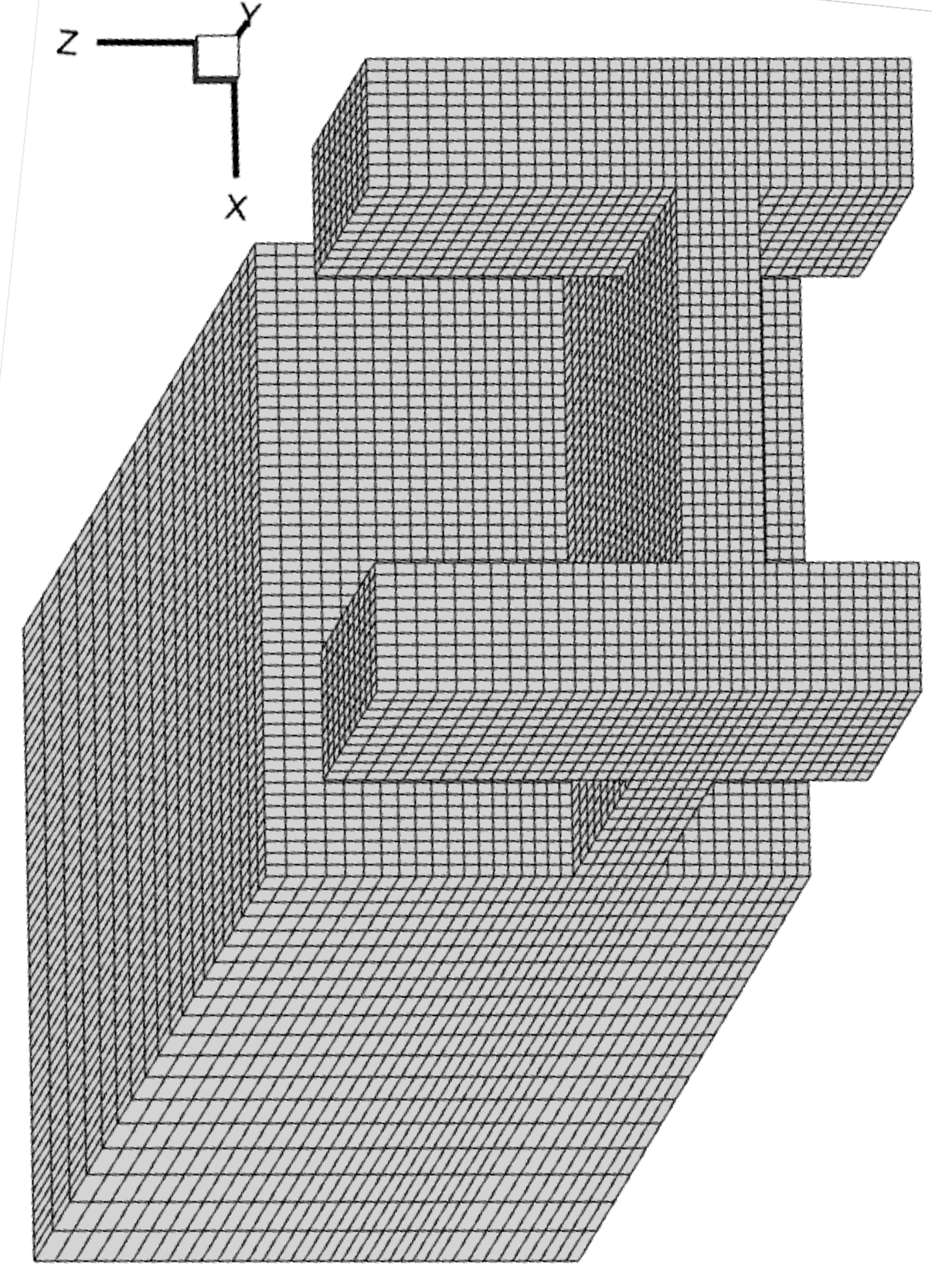 }}  \\
\subfloat[BTE, $10$-nm node]{\includegraphics[scale=0.28,viewport=100 0 600 580,clip=true]{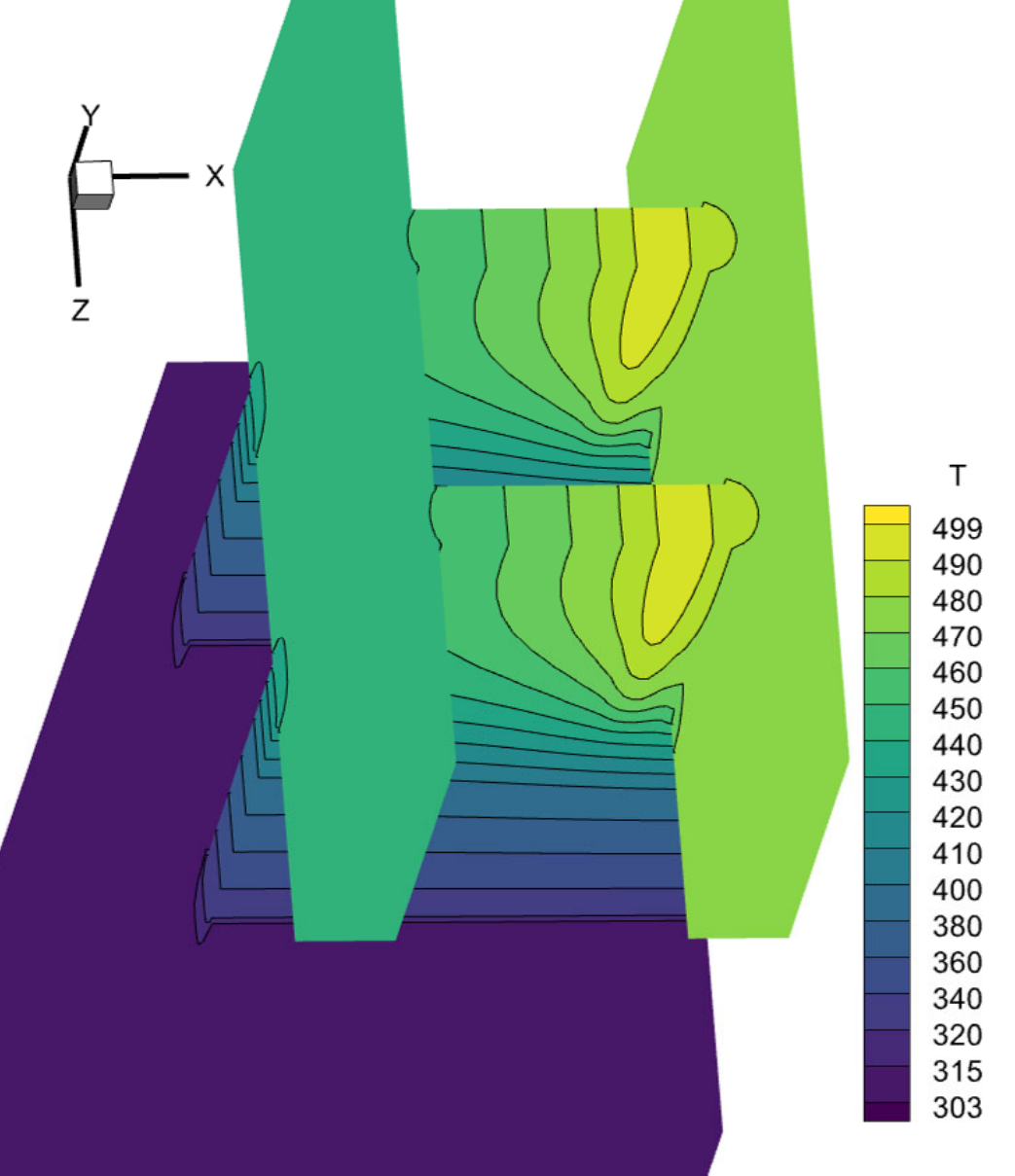}} ~
\subfloat[BTE, $7$-nm node]{\includegraphics[scale=0.28,viewport=100 0 600 580,clip=true]{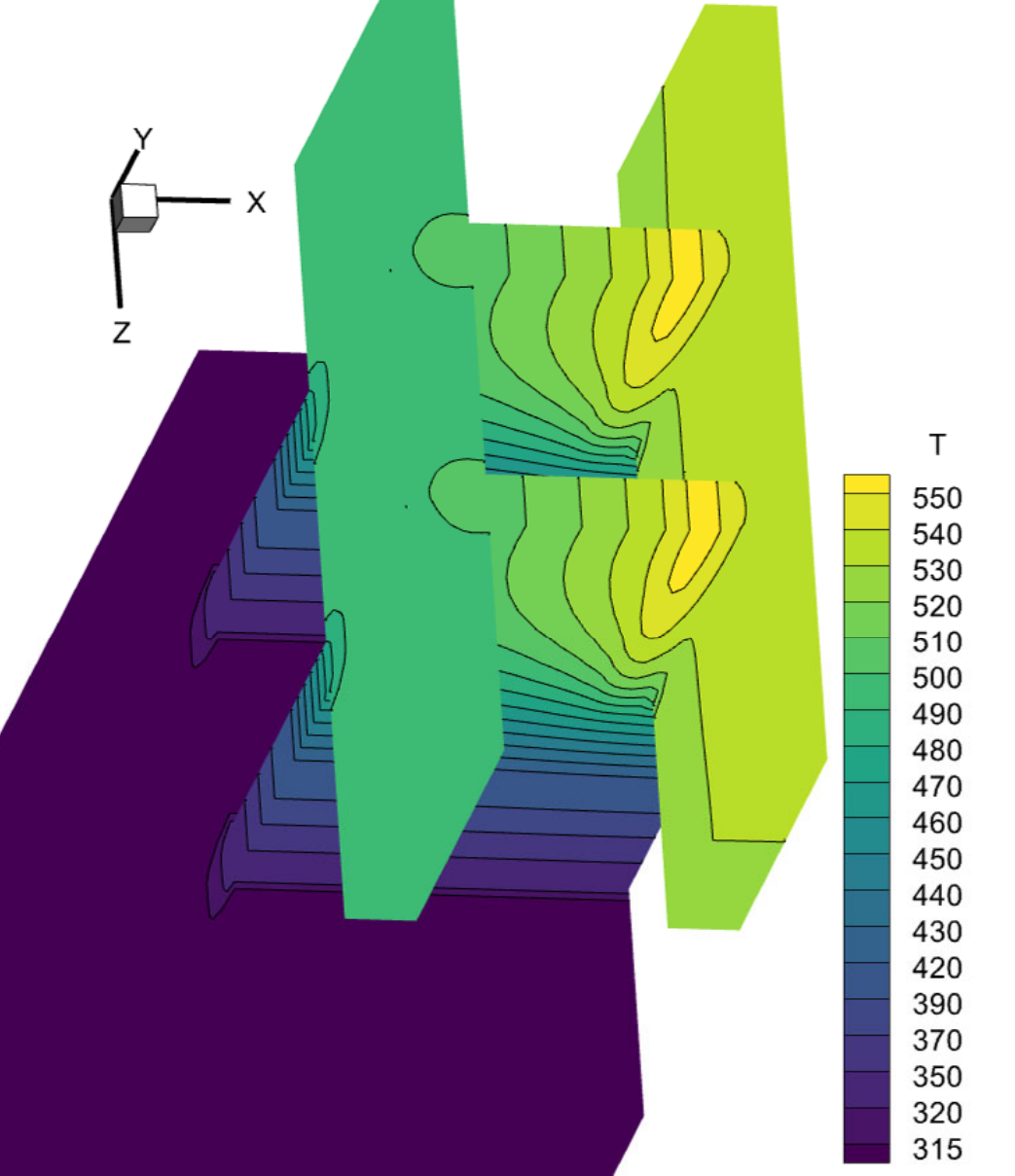}} ~
\subfloat[BTE, $5$-nm node]{\includegraphics[scale=0.28,viewport=100 0 600 580,clip=true]{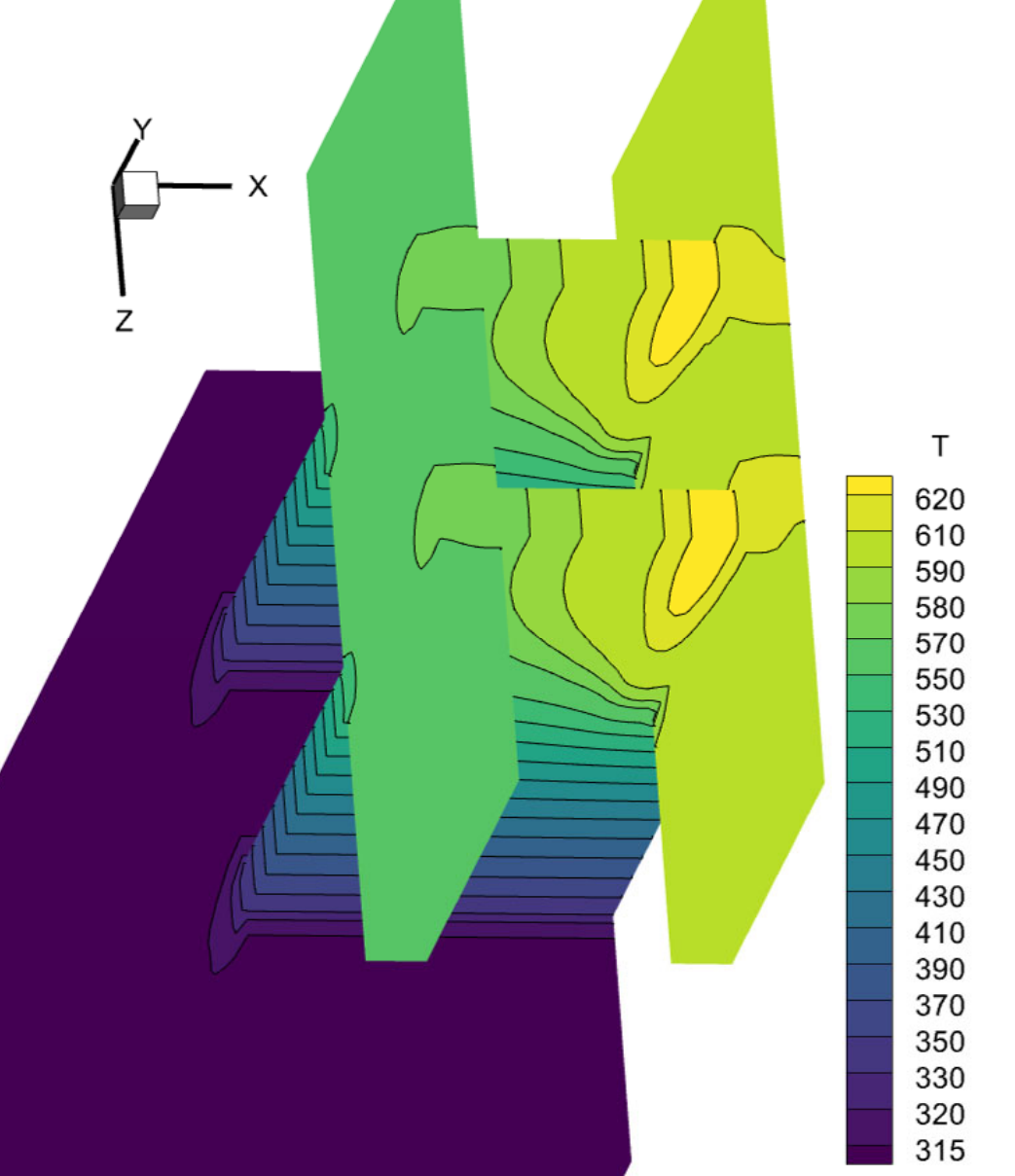}}  \\
\subfloat[Effective Fourier's law, $10$-nm node]{\includegraphics[scale=0.28,viewport=100 0 600 580,clip=true]{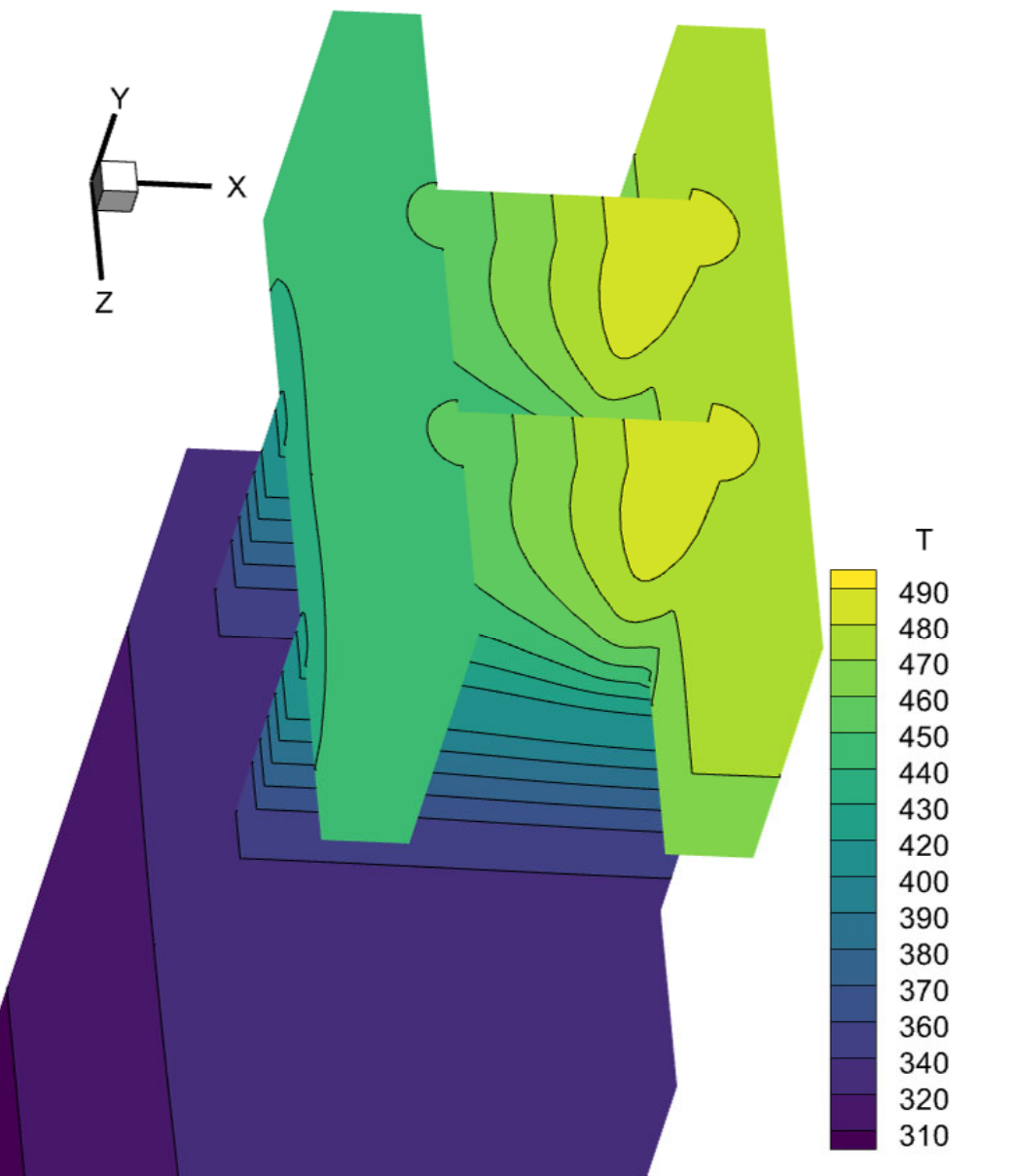}} ~
\subfloat[Effective Fourier's law, $7$-nm node]{\includegraphics[scale=0.28,viewport=100 0 600 580,clip=true]{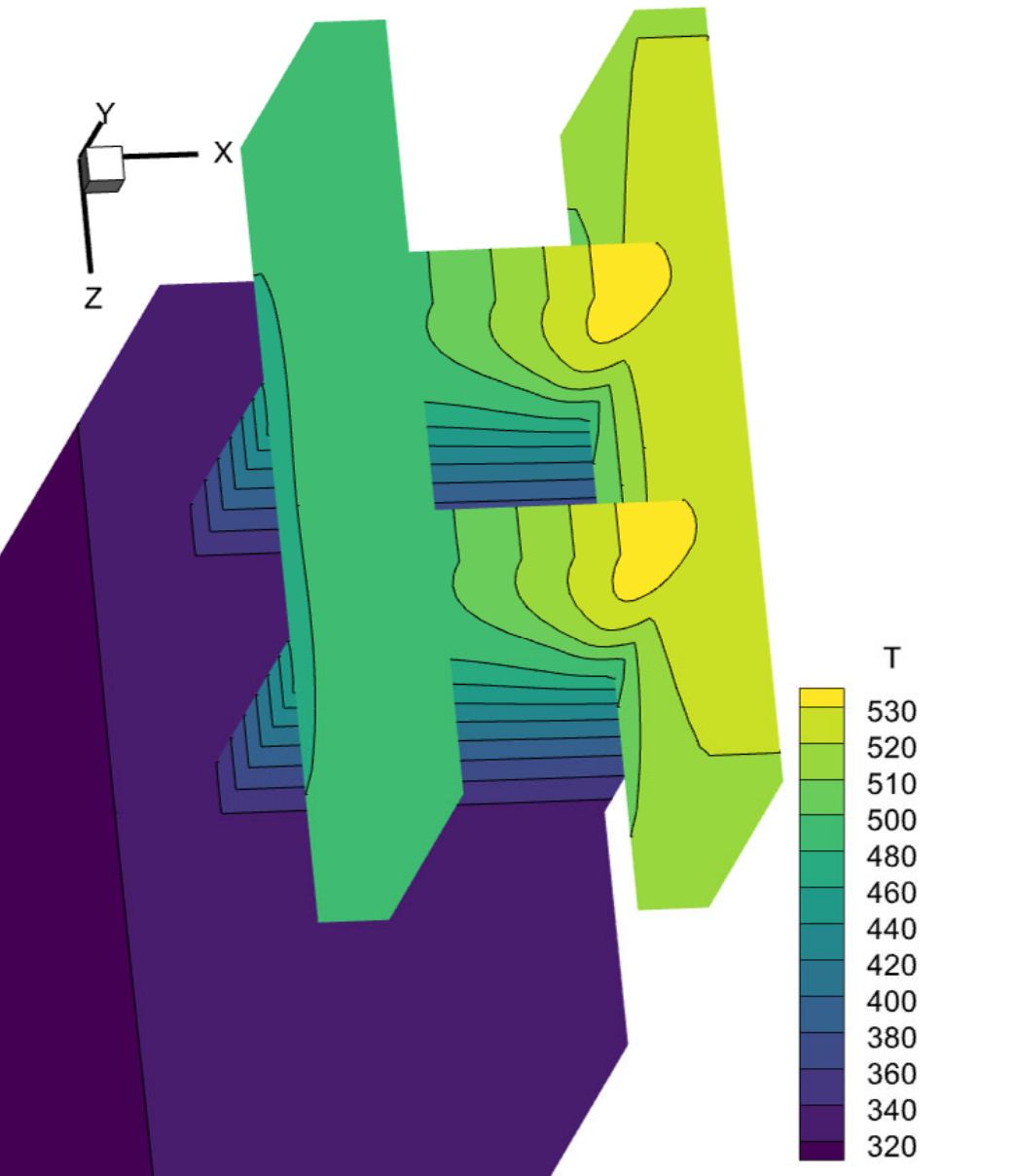}} ~
\subfloat[Effective Fourier's law, $5$-nm node]{\includegraphics[scale=0.28,viewport=100 0 600 580,clip=true]{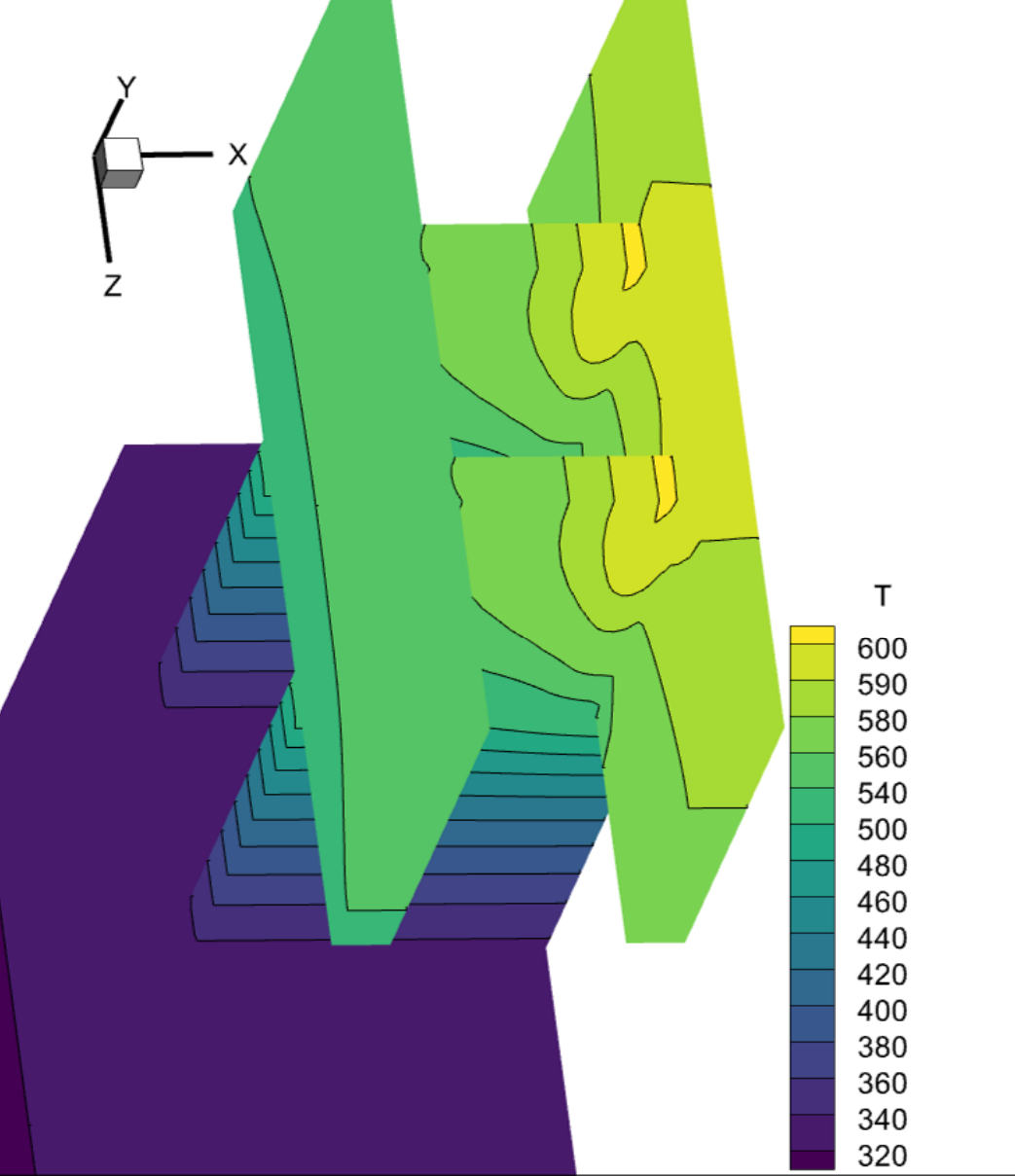}}  \\
\caption{(a) Schematic of a double-fin bulk FET structure unit in sub$-10$ nm advanced technology nodes. The front, back, left and right of the whole geometric structure unit are symmetric interfaces, and the whole structure unit is symmetrical about the red dashed line plane. (b) Non-uniform discretized cells for half of the structure unit in $7$-nm node. Steady temperature distributions of a double-fin bulk FET structure unit predicted by (c,d,e) synthetic iterative scheme and (f,g,h) effective Fourier's law in (c,f)$10$-nm, (d,g) $7$-nm, (e,h) $5$-nm advanced technology nodes.}
\label{twobulkFinFET_temperature}
\end{figure}
\begin{figure}[htb]
\centering
\subfloat[GAAFET]{\includegraphics[width=0.73\textwidth]{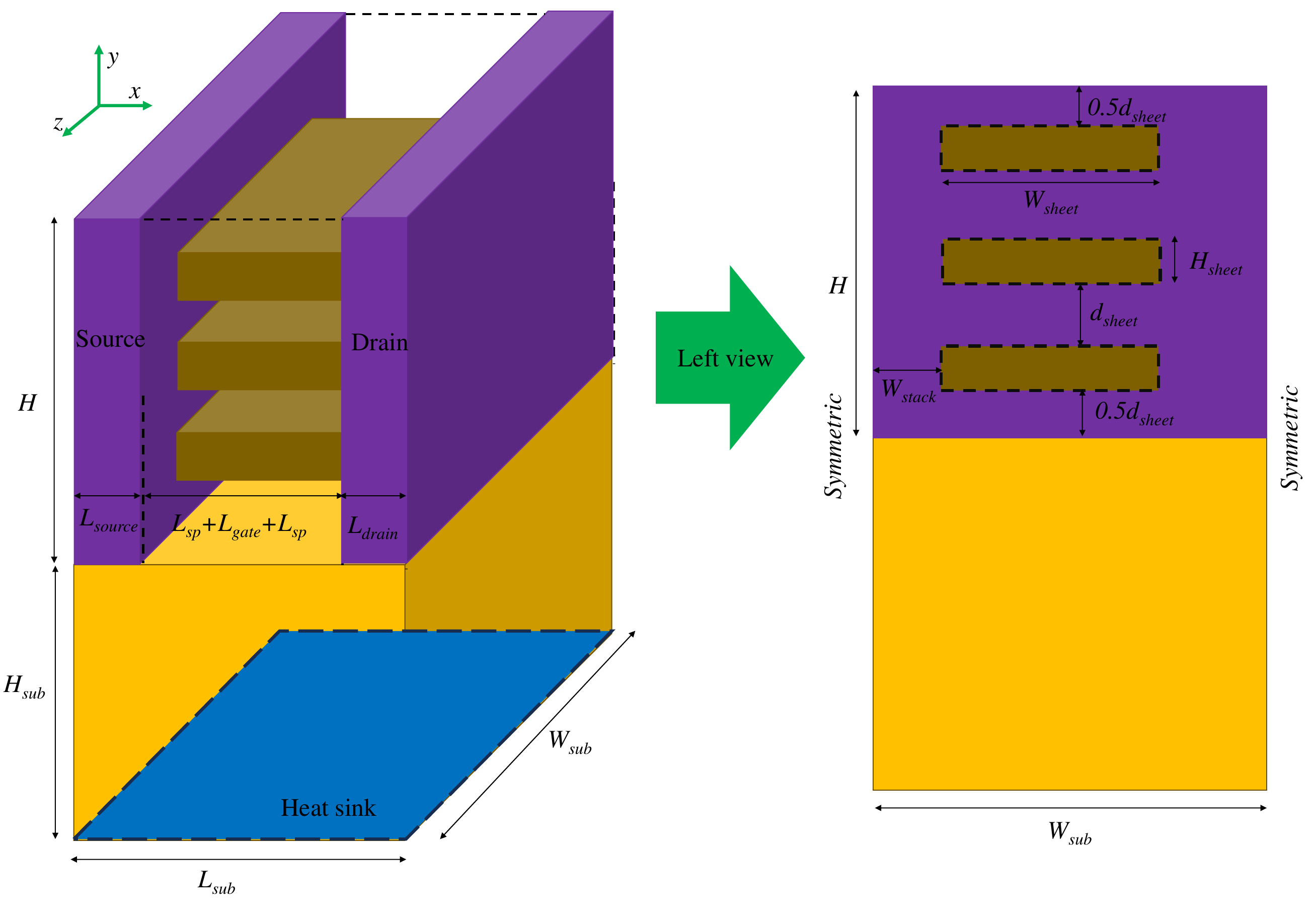 }}   \\   
\subfloat[$2W_{stack}=25$ nm, BTE]{\includegraphics[scale=0.25,viewport=90 0 750 580,clip=true]{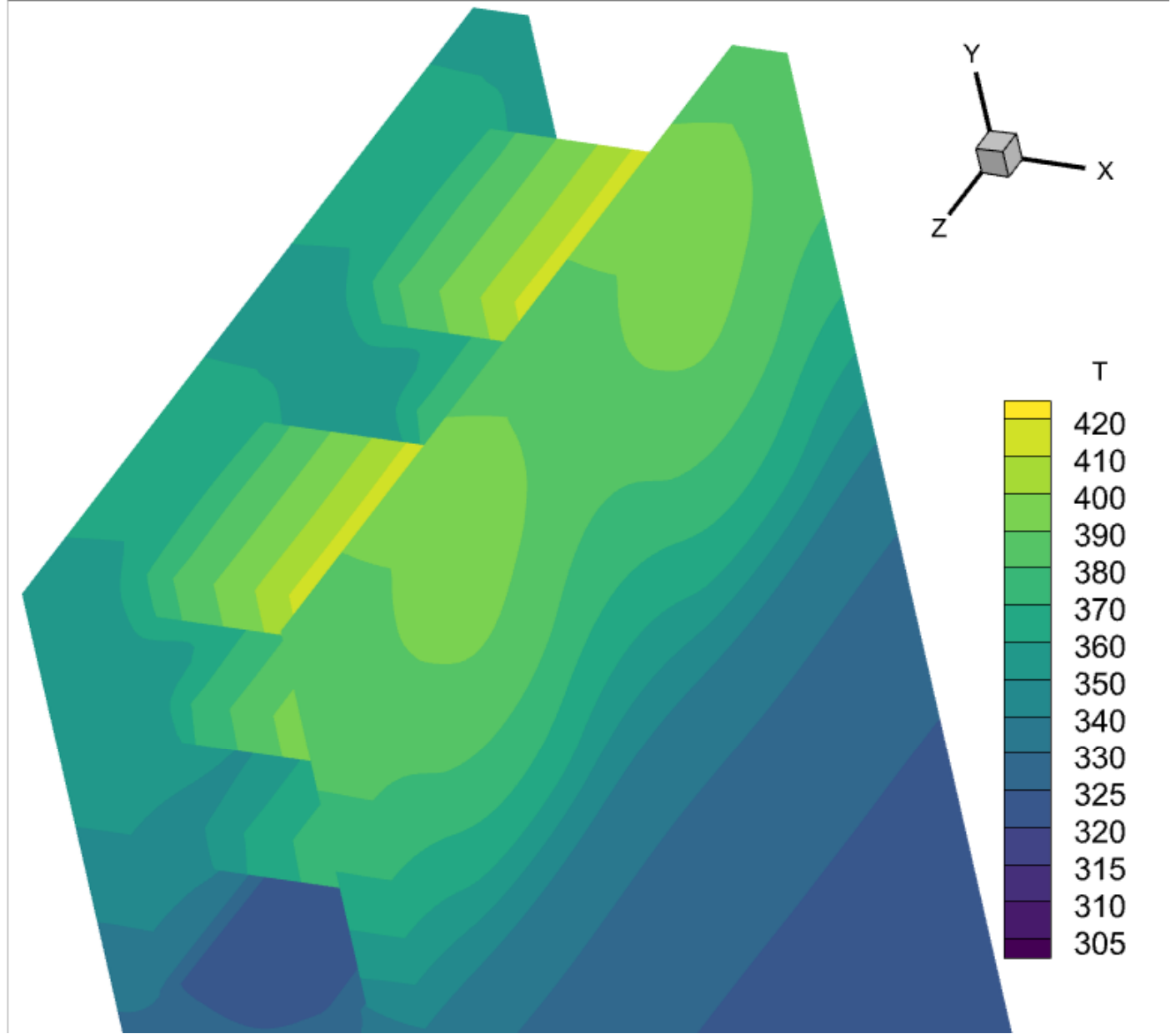}}
\subfloat[$2W_{stack}=0$, BTE]{\includegraphics[scale=0.25,viewport=90 0 750 580,clip=true]{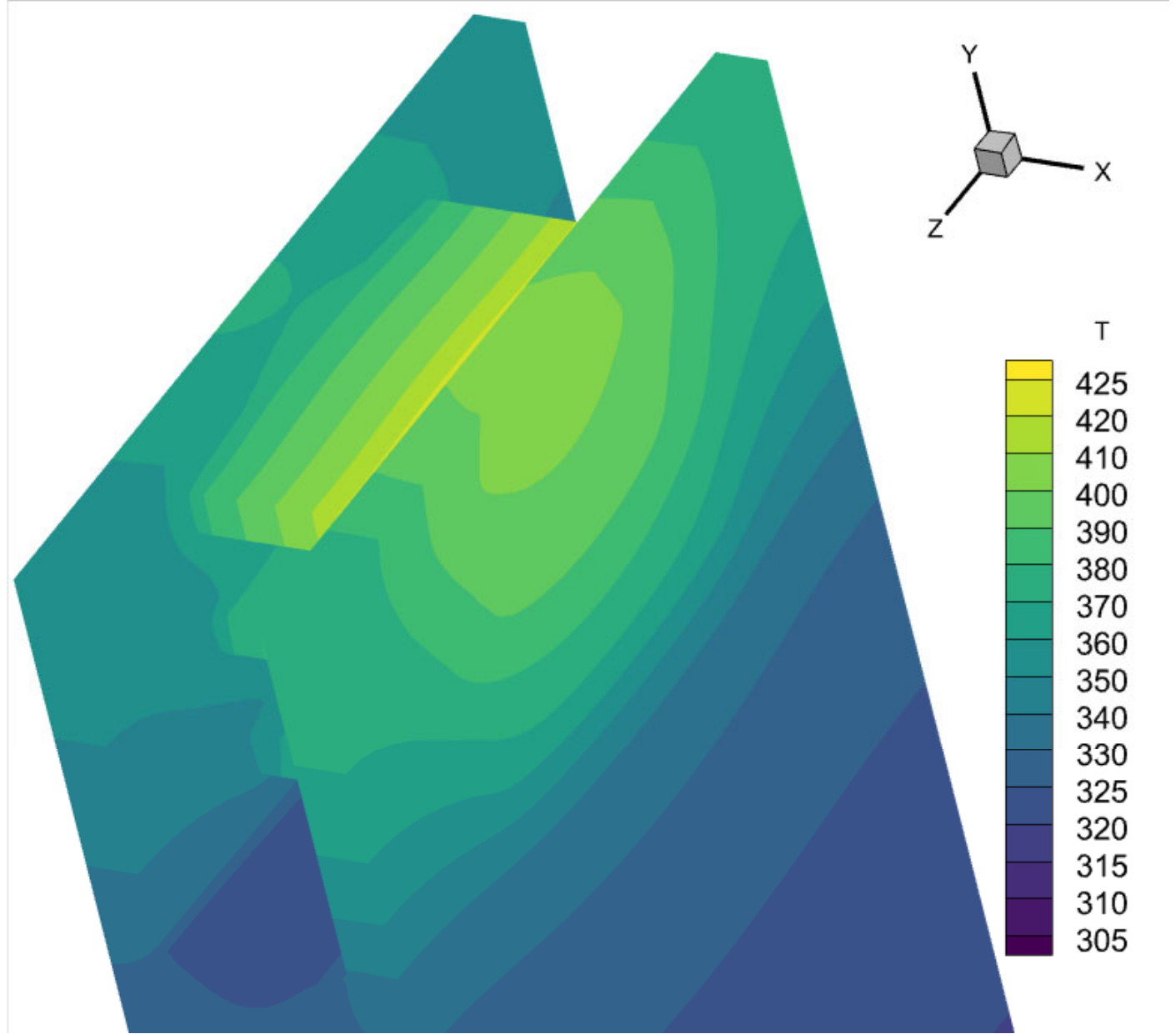}} \\
\subfloat[$2W_{stack}=25$ nm, Effective Fourier's law]{\includegraphics[scale=0.25,viewport=90 0 750 580,clip=true]{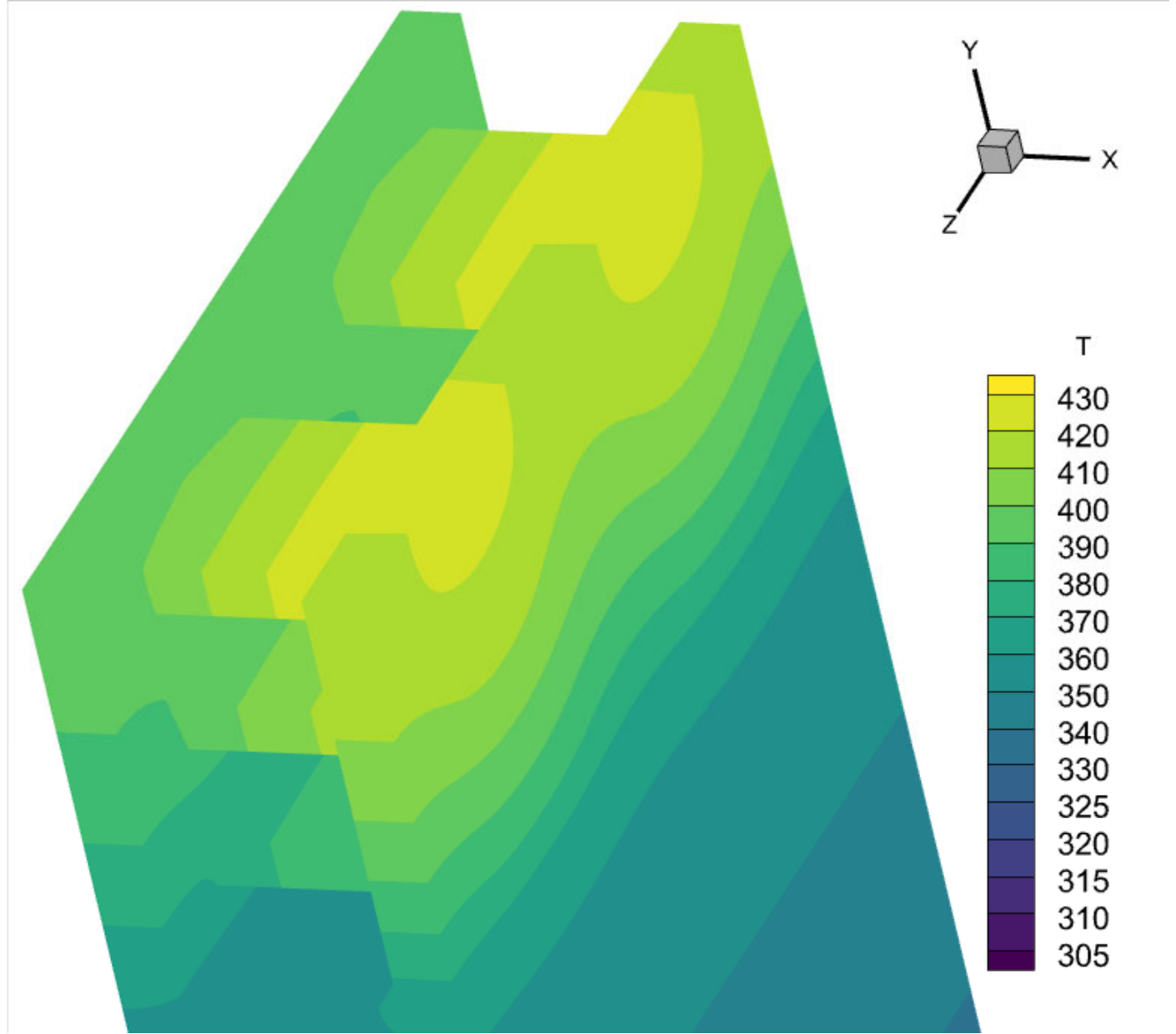}}  ~~
\subfloat[$2W_{stack}=0$, Effective Fourier's law]{\includegraphics[scale=0.25,viewport=90 0 750  580,clip=true]{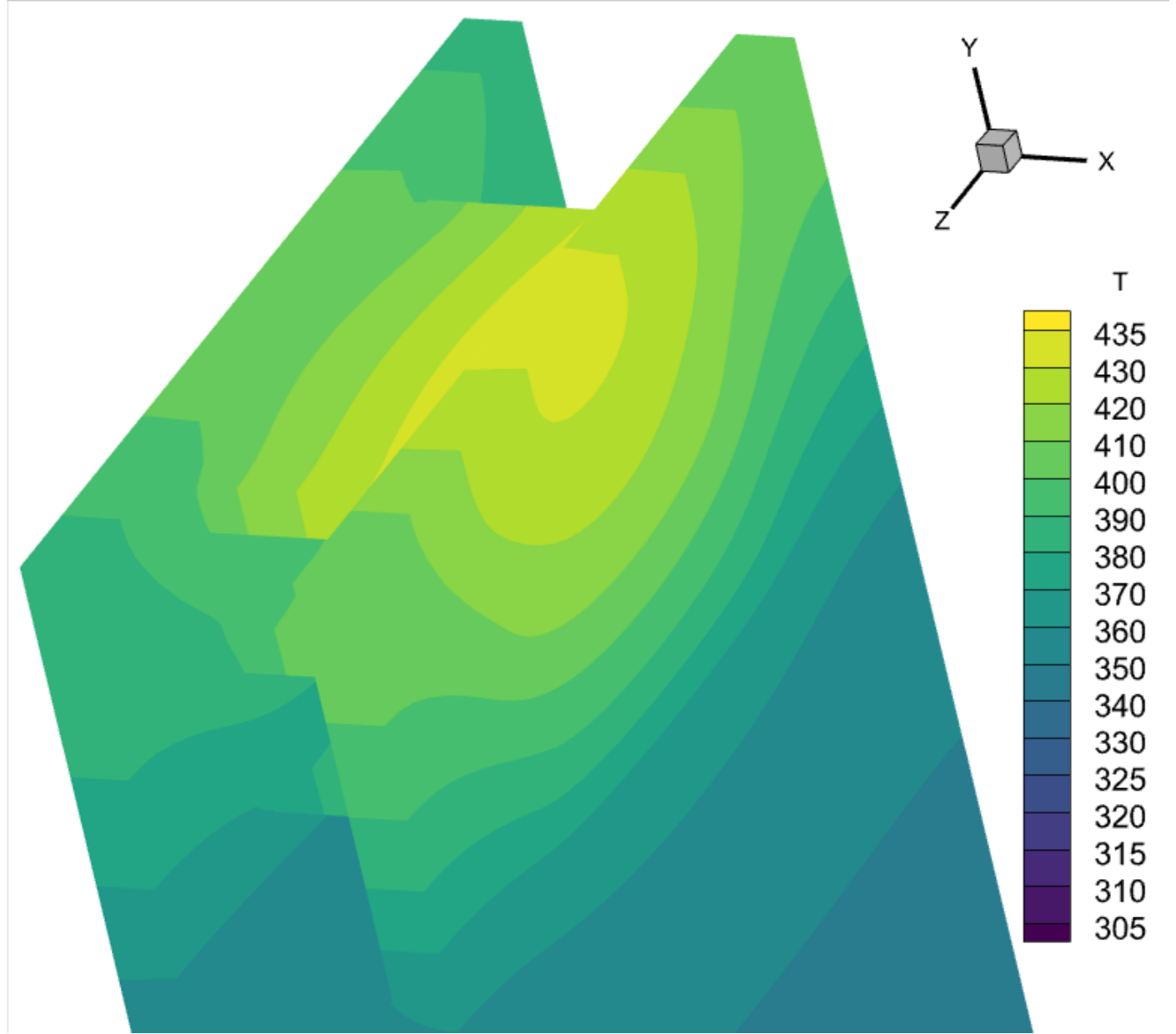}}   \\
\caption{(a) Schematic of 3D horizontally stacked GAA nanosheet FETs, where the front, back, left and right of the whole geometry are symmetric interfaces. Steady temperature distributions of 3D horizontally stacked GAA nanosheet FETs predicted by (b,c) synthetic iterative scheme and (d,e) effective Fourier's law.  }
\label{GAAsheetFET_temperature}
\end{figure}

\subsection{Heat conduction in 3D FinFET/GAAFET}

Firstly heat conduction in a single-fin bulk FET is simulated.  
Half of geometry is shown in~\cref{FINFET3D2022_temperature}(a), whose back surface is symmetric.
All geometry parameters are the same as those mentioned in previous paper~\cite{3DFINFET_2014_mc}.
Geometry sizes in the $xy$ plane are the same as those shown in~\cref{FINFET2022_thermal}(a) if looking along the $z$ direction.
The widths of fin, source/drain, top heat sink and substrate are $W_{fin} =4$ nm, $W_{source}=25$ nm, $W_{sink}=11$ nm and $W_{sub}=100$ nm, respectively.
The bottom surface of the substrate, front/left/right/top surfaces of the top heat sink are all isothermal boundary conditions with $300$ K.
The front/left/right surfaces of the substrate are symmetric.
Diffusely reflecting adiabatic boundary conditions are adopted for other surfaces of this structure.
The position and power density of the Joule heating zone in the $xy$ plane are the same as that shown in~\cref{FINFET2022_thermal}(a), and the width of the Joule heating zone is the same as that of the fin.

{\color{black}{In previous studies~\cite{3DFINFET_2014_mc}, the maximum temperature in the hotspot region predicted by Monte Carlo method and classical Fourier's law with bulk thermal conductivity is $493$ K and $398$ K, respectively, which indicates that the classical Fourier's law significantly underestimates the peak temperature.}}
Considering that the width of the fin is much smaller than its length and height, the heat conduction in the $xy$ plane will be suppressed by the width, so we set $\alpha = 15.0/146.0$~\eqref{eq:effectivekappa}.
In other spatial areas, its value is the same as those used in~\cref{FINFET2022_thermal}(a).
Non-uniform discretized cells are used and shown in~\cref{FINFET3D2022_temperature}(b).
Steady temperature distributions are shown in~\cref{FINFET3D2022_temperature}(c-f).
It can be found that the temperature rise of the 3D structure is about $50$ K higher than that in quasi-2D structure (\cref{FINFET2022_thermal}), which indicates that the smallest fin width affects significantly the hotspot temperature.
The temperature rise of substrate is below $2$ K and most heat is blocked in the fin areas.
In addition, the temperature rise predicted by the effective Fourier's law is coarsely consistent with the numerical results of BTE although there are some obvious differences in other areas.

Secondly the heat conduction in a double-fin bulk FET~\cite{bulkfinfet2019,3DFINFETtransient} structure unit in sub$-10$ nm advanced technology nodes is studied, as shown in~\cref{twobulkFinFET_temperature}.
Front, back, left and right of the whole structure unit are symmetric interfaces, and the whole structure unit is symmetrical about the red dashed line plane.
In order to reduce computational cost, only half of the structure unit is calculated because the structure unit is symmetric about the red dashed plane.
Detailed geometrical parameters are shown in~\cref{FinFETsub10nodes}~\cite{bulkfinfet2019}.
Note that the symmetric boundaries are used in the present simulations so that the lengths of source and drain are only a half of real transistors.
Gate is at the center of source and drain, and Joule heating area is generated at the right edge of the gate with power density $\dot{S}=7.1\times 10^{18}$ W/m$^3$, length $L_{hot}=4$ nm and same width of the fin.
In the following simulations, the heat generation power density is controlled equally, and the steady-state heat dissipation of double-fin bulk FET structure in $10$-nm, $7$-nm, and $5$-nm advanced technology nodes are shown in~\cref{twobulkFinFET_temperature}, where $\alpha=30/146.0$ in the substrate and $15.0/146.0$ in the above the substrate in the numerical solutions of the effective Fourier's law.
When the transistors sizes decreases, the hotspot temperature increases significantly.
It can be found that the temperature distributions predicted by BTE is obviously different from the effective Fourier's law near the hotspot areas and the deviations increase when the transistors sizes decreases.
That's because that the small fin width and close distance between two fins significantly increases the nonlinear relationship between the heat flux and temperature near the connected areas between double fin.

Thirdly heat conduction in a horizontally stacked GAA nanosheet FET structure~\cite{Nanosheet_FINFET2018,IBM_2017_nanosheet} is studied.
Half of a structure unit is shown in~\cref{GAAsheetFET_temperature}(a), where the front, back, left and right of the whole geometry are symmetric interfaces.
Geometry parameters are listed in~\cref{GAAsheetFETparameters}.
In a structure unit, there are $6$ nanosheets with width $25$ nm when $2W_{stack}=23$ nm and $3$ nanosheets with width $50$ nm when $2W_{stack}=0$.
The bottom of substrate is the heat sink with isothermal boundary conditions at $300$ K and the other boundaries are all diffusely reflecting adiabatic boundary conditions.
Gate is at the center of source and drain, and Joule heating area is generated at the right edge of the gate with power density $\dot{S}=7.1\times 10^{18}$ W/m$^3$, length $5$ nm and same height/width as nanosheets.
The three-dimensional temperature contour is shown in~\cref{GAAsheetFET_temperature}, where the effective thermal conductivity is the same as those used in double-fin bulk FET.
It can be found that the effective Fourier's law predict a about $10$ K higher hotspot temperature than phonon BTE.
Main reason of the numerical deviations is highly anisotropic thermal conduction due to the complex stacked typological structures from heat source to heat sink in GAAFET.

{\color{black}{These results in this section show that unlike the classical Fourier law which vastly underestimate the peak temperature, the effective Fourier's law is possible to capture the highest temperature rise by empirically adjusting the effective thermal conductivity. 
However, some local nonlinear thermal conduction phenomena still cannot be accurately described due to the linear assumption of heat flux and temperature gradient.
In addition, it is most noteworthy that in engineering thermal applications, how to choose a reasonable effective thermal conductivity coefficient is still a difficult and empirical problem.}}

\section{CONCLUSION}
\label{sec:conclusions}

{\color{black}{A synthetic iterative scheme is developed for thermal applications in hotspot systems with large temperature variance, where the Newton method is used to deal with the nonlinear relationship between the equilibrium state and temperature.
A macroscopic iteration is introduced for preprocessing based on the iterative solutions of the stationary phonon BTE.
Macroscopic and mesoscopic physical evolution processes are connected by the heat flux, which is no longer calculated by classical Fourier's law but obtained by taking the moment of phonon distribution function.
These two processes exchange information from different scales, such that the present scheme could efficiently deal with heat conduction problems from ballistic to diffusive regime.
Numerical results show that the present scheme could efficiently capture the multiscale heat conduction in hotspot systems with large temperature variances.
The heat dissipation efficiency of silicon-based hotspot systems is high if the external heat source mainly heats the longitude acoustic phonons.
In addition, a comparison is made between the solutions of phonon BTE and effective Fourier's law by several heat dissipations problems.
Numerical results show that the effective Fourier's law is possible to capture the highest temperature rise by empirically adjusting the effective thermal conductivity. 
However, some local nonlinear thermal conduction phenomena still cannot be accurately described due to the linear assumption of heat flux and temperature gradient.
The present work could provide theoretical guidance for the practical multiscale thermal applications.}}

\section*{Conflict of interest}

No conflict of interest declared.

\section*{Author Statements}

\textbf{Chuang Zhang}: Supervision, Conceptualization, Investigation, Methodology, Numerical analysis, Writing - original draft.
\textbf{Qin Lou}: Methodology, Numerical analysis, Funding acquisition, Writing-review \& editing.
\textbf{Hong Liang}: Conceptualization, Methodology, Numerical analysis, Writing-review \& editing.

\section*{Acknowledgments}

This work is supported by the National Natural Science Foundation of China (52376068, 12147122).
C.Z. acknowledges Dr. Yang Shen in Tsinghua University for the communications on heat dissipations in FinFETs and GAAFETs, acknowledges Dr. Xiao Wan in Huazhong University of Science and Technology and Prof. Samuel Huberman in McGill University for communications on selective phonon excitation.
The authors acknowledge Beijng PARATERA Tech CO.,Ltd. for providing HPC resources that have contributed to the research results reported within this paper.

\appendix

\section{Geometric parameters of FinFET and GAAFET}
\label{sec:GAAFETsub5nmnodes}

Geometric parameters of FinFET and GAAFET used in our present simulations are shown in Tables.~\ref{FinFETsub10nodes} and~\ref{GAAsheetFETparameters}.

\begin{table}[htb]
\caption{Geometrical parameters of a double-fin bulk FET structure unit (\cref{{twobulkFinFET_temperature}}) in sub$-10$ nm advanced technology nodes~\cite{bulkfinfet2019,3DFINFETtransient}.}
\centering
\begin{tabular}{*{4}{c}}
\hline
\hline
\multirow{2}{*}{{\shortstack{ Geometry parameters (nm)  }}}  & \multicolumn{3}{c}{Advanced technology nodes}   \\
\cline{2-4}
& $10$-nm node &   $7$-nm node &  $ 5$-nm node    \\
\hline
Fin length $L_{sub}$  & 54 & 44 &  36    \\
\hline
Fin pitch  & 34 & 30 &  26    \\
\hline
$W_{sub}$  & 102 & 90 &  84    \\
\hline
Gate length $L_{gate}$  & 18 & 16 &  14   \\
\hline
Spacer length $L_{sp}$ & 7 & 6 &  5    \\
\hline
$L_{source}=L_{drain}$  & 11 & 8 &  6   \\
\hline
Fin width $W_{fin}$ & 7 & 6 &  5    \\
\hline
$H$  & 46 & 46 &  46    \\
\hline
Box height $h$  & 50 & 50 &  50    \\
\hline
$H_{sub}$  & 200 & 200 &  200    \\
\hline
\hline
\end{tabular}
\label{FinFETsub10nodes}
\end{table}
\begin{table}[htb]
\caption{Geometrical parameters of 3D horizontally stacked GAA nanosheet FETs (\cref{GAAsheetFET_temperature}).}
\centering
\begin{tabular}{*{3}{c}}
\hline
\hline
Symbols  &  Physical meanings  & Size (nm)  \\
\hline
$L_{sub}$ &  Substrate length  &  34    \\
\hline
$H_{sub}$ &  Substrate height  &  200    \\
\hline
$W_{sub}$ &  Substrate width  &  48   \\
\hline
$2W_{stack}$ &  Horizontal distance of nanosheet  &   0 or 23  \\
\hline
$H_{sheet}$ &  Thickness of nanosheet  &  5   \\
\hline
$W_{sheet}$ &  Width of nanosheet  &  25   \\
\hline
$d_{sheet}$ &  Vertical distance of nanosheet  & 10  \\
\hline
$L_{gate}$ &  Gate length  & 12   \\
\hline
$L_{source}$ and $L_{drain}$ & Source/drain length  &  6   \\
\hline
$H$ & Source/drain height &  45   \\
\hline
$L_{sp}$ &  Space length between source/drain and gate &  5   \\
\hline
\hline
\end{tabular}
\label{GAAsheetFETparameters}
\end{table}

\bibliography{phonon}

\begin{thebibliography}{70}%
\makeatletter
\providecommand \@ifxundefined [1]{%
 \@ifx{#1\undefined}
}%
\providecommand \@ifnum [1]{%
 \ifnum #1\expandafter \@firstoftwo
 \else \expandafter \@secondoftwo
 \fi
}%
\providecommand \@ifx [1]{%
 \ifx #1\expandafter \@firstoftwo
 \else \expandafter \@secondoftwo
 \fi
}%
\providecommand \natexlab [1]{#1}%
\providecommand \enquote  [1]{``#1''}%
\providecommand \bibnamefont  [1]{#1}%
\providecommand \bibfnamefont [1]{#1}%
\providecommand \citenamefont [1]{#1}%
\providecommand \href@noop [0]{\@secondoftwo}%
\providecommand \href [0]{\begingroup \@sanitize@url \@href}%
\providecommand \@href[1]{\@@startlink{#1}\@@href}%
\providecommand \@@href[1]{\endgroup#1\@@endlink}%
\providecommand \@sanitize@url [0]{\catcode `\\12\catcode `\$12\catcode
  `\&12\catcode `\#12\catcode `\^12\catcode `\_12\catcode `\%12\relax}%
\providecommand \@@startlink[1]{}%
\providecommand \@@endlink[0]{}%
\providecommand \url  [0]{\begingroup\@sanitize@url \@url }%
\providecommand \@url [1]{\endgroup\@href {#1}{\urlprefix }}%
\providecommand \urlprefix  [0]{URL }%
\providecommand \Eprint [0]{\href }%
\providecommand \doibase [0]{http://dx.doi.org/}%
\providecommand \selectlanguage [0]{\@gobble}%
\providecommand \bibinfo  [0]{\@secondoftwo}%
\providecommand \bibfield  [0]{\@secondoftwo}%
\providecommand \translation [1]{[#1]}%
\providecommand \BibitemOpen [0]{}%
\providecommand \bibitemStop [0]{}%
\providecommand \bibitemNoStop [0]{.\EOS\space}%
\providecommand \EOS [0]{\spacefactor3000\relax}%
\providecommand \BibitemShut  [1]{\csname bibitem#1\endcsname}%
\let\auto@bib@innerbib\@empty
\bibitem [{\citenamefont {IEEE}(2023)}]{IRDS2023}%
  \BibitemOpen
  \bibfield  {author} {\bibinfo {author} {\bibnamefont {IEEE}},\ }\href
  {https://irds.ieee.org/editions/2023} {\emph {\bibinfo {title} {International
  Roadmap for Devices and Systems ({IRDS™})}}}\ (\bibinfo  {publisher}
  {IEEE},\ \bibinfo {year} {2023})\BibitemShut {NoStop}%
\bibitem [{\citenamefont {Warzoha}\ \emph {et~al.}(2021)\citenamefont
  {Warzoha}, \citenamefont {Wilson}, \citenamefont {Donovan}, \citenamefont
  {Donmezer}, \citenamefont {Giri}, \citenamefont {Hopkins}, \citenamefont
  {Choi}, \citenamefont {Pahinkar}, \citenamefont {Shi}, \citenamefont
  {Graham}, \citenamefont {Tian},\ and\ \citenamefont
  {Ruppalt}}]{warzoha_applications_2021}%
  \BibitemOpen
  \bibfield  {author} {\bibinfo {author} {\bibfnamefont {R.~J.}\ \bibnamefont
  {Warzoha}}, \bibinfo {author} {\bibfnamefont {A.~A.}\ \bibnamefont {Wilson}},
  \bibinfo {author} {\bibfnamefont {B.~F.}\ \bibnamefont {Donovan}}, \bibinfo
  {author} {\bibfnamefont {N.}~\bibnamefont {Donmezer}}, \bibinfo {author}
  {\bibfnamefont {A.}~\bibnamefont {Giri}}, \bibinfo {author} {\bibfnamefont
  {P.~E.}\ \bibnamefont {Hopkins}}, \bibinfo {author} {\bibfnamefont
  {S.}~\bibnamefont {Choi}}, \bibinfo {author} {\bibfnamefont {D.}~\bibnamefont
  {Pahinkar}}, \bibinfo {author} {\bibfnamefont {J.}~\bibnamefont {Shi}},
  \bibinfo {author} {\bibfnamefont {S.}~\bibnamefont {Graham}}, \bibinfo
  {author} {\bibfnamefont {Z.}~\bibnamefont {Tian}}, \ and\ \bibinfo {author}
  {\bibfnamefont {L.}~\bibnamefont {Ruppalt}},\ }\href {\doibase
  10.1115/1.4049293} {\bibfield  {journal} {\bibinfo  {journal} {J Electron.
  Packaging}\ }\textbf {\bibinfo {volume} {143}},\ \bibinfo {pages} {020804}
  (\bibinfo {year} {2021})}\BibitemShut {NoStop}%
\bibitem [{\citenamefont {Pop}(2010)}]{pop_energy_2010}%
  \BibitemOpen
  \bibfield  {author} {\bibinfo {author} {\bibfnamefont {E.}~\bibnamefont
  {Pop}},\ }\href {\doibase 10.1007/s12274-010-1019-z} {\bibfield  {journal}
  {\bibinfo  {journal} {Nano Res.}\ }\textbf {\bibinfo {volume} {3}},\ \bibinfo
  {pages} {147} (\bibinfo {year} {2010})}\BibitemShut {NoStop}%
\bibitem [{\citenamefont {Tang}\ and\ \citenamefont
  {Cao}(2023)}]{TANG2023123497}%
  \BibitemOpen
  \bibfield  {author} {\bibinfo {author} {\bibfnamefont {D.-S.}\ \bibnamefont
  {Tang}}\ and\ \bibinfo {author} {\bibfnamefont {B.-Y.}\ \bibnamefont {Cao}},\
  }\href {\doibase https://doi.org/10.1016/j.ijheatmasstransfer.2022.123497}
  {\bibfield  {journal} {\bibinfo  {journal} {Int. J. Heat Mass Transfer}\
  }\textbf {\bibinfo {volume} {200}},\ \bibinfo {pages} {123497} (\bibinfo
  {year} {2023})}\BibitemShut {NoStop}%
\bibitem [{\citenamefont {Kuo}\ \emph {et~al.}(2023)\citenamefont {Kuo},
  \citenamefont {Lee}, \citenamefont {Su},\ and\ \citenamefont
  {Lin}}]{TSMC_2023_self_heating}%
  \BibitemOpen
  \bibfield  {author} {\bibinfo {author} {\bibfnamefont {J.}~\bibnamefont
  {Kuo}}, \bibinfo {author} {\bibfnamefont {W.}~\bibnamefont {Lee}}, \bibinfo
  {author} {\bibfnamefont {K.}~\bibnamefont {Su}}, \ and\ \bibinfo {author}
  {\bibfnamefont {C.}~\bibnamefont {Lin}},\ }in\ \href {\doibase
  10.23919/SNW57900.2023.10183933} {\emph {\bibinfo {booktitle} {2023 Silicon
  Nanoelectronics Workshop (SNW)}}}\ (\bibinfo {year} {2023})\ pp.\ \bibinfo
  {pages} {89--90}\BibitemShut {NoStop}%
\bibitem [{\citenamefont {Kim}\ \emph {et~al.}(2023)\citenamefont {Kim},
  \citenamefont {Park}, \citenamefont {Choi}, \citenamefont {Kim},
  \citenamefont {Kim}, \citenamefont {Shim}, \citenamefont {Chung},\ and\
  \citenamefont {Jung}}]{MBCFET2023_Samsung}%
  \BibitemOpen
  \bibfield  {author} {\bibinfo {author} {\bibfnamefont {S.}~\bibnamefont
  {Kim}}, \bibinfo {author} {\bibfnamefont {H.}~\bibnamefont {Park}}, \bibinfo
  {author} {\bibfnamefont {E.}~\bibnamefont {Choi}}, \bibinfo {author}
  {\bibfnamefont {Y.~H.}\ \bibnamefont {Kim}}, \bibinfo {author} {\bibfnamefont
  {D.}~\bibnamefont {Kim}}, \bibinfo {author} {\bibfnamefont {H.}~\bibnamefont
  {Shim}}, \bibinfo {author} {\bibfnamefont {S.}~\bibnamefont {Chung}}, \ and\
  \bibinfo {author} {\bibfnamefont {P.}~\bibnamefont {Jung}},\ }in\ \href
  {\doibase 10.1109/IRPS48203.2023.10117953} {\emph {\bibinfo {booktitle} {2023
  IEEE International Reliability Physics Symposium (IRPS)}}}\ (\bibinfo {year}
  {2023})\ pp.\ \bibinfo {pages} {1--8}\BibitemShut {NoStop}%
\bibitem [{\citenamefont {Landon}\ \emph {et~al.}(2023)\citenamefont {Landon},
  \citenamefont {Jiang}, \citenamefont {Pantuso}, \citenamefont {Meric},
  \citenamefont {Komeyli}, \citenamefont {Hicks},\ and\ \citenamefont
  {Schroeder}}]{intel_2023_GAAFET}%
  \BibitemOpen
  \bibfield  {author} {\bibinfo {author} {\bibfnamefont {C.}~\bibnamefont
  {Landon}}, \bibinfo {author} {\bibfnamefont {L.}~\bibnamefont {Jiang}},
  \bibinfo {author} {\bibfnamefont {D.}~\bibnamefont {Pantuso}}, \bibinfo
  {author} {\bibfnamefont {I.}~\bibnamefont {Meric}}, \bibinfo {author}
  {\bibfnamefont {K.}~\bibnamefont {Komeyli}}, \bibinfo {author} {\bibfnamefont
  {J.}~\bibnamefont {Hicks}}, \ and\ \bibinfo {author} {\bibfnamefont
  {D.}~\bibnamefont {Schroeder}},\ }in\ \href {\doibase
  10.1109/IRPS48203.2023.10117903} {\emph {\bibinfo {booktitle} {2023 IEEE
  International Reliability Physics Symposium (IRPS)}}}\ (\bibinfo {year}
  {2023})\ pp.\ \bibinfo {pages} {1--5}\BibitemShut {NoStop}%
\bibitem [{\citenamefont {Chang}\ \emph {et~al.}(2023)\citenamefont {Chang},
  \citenamefont {Oprins}, \citenamefont {Lofrano}, \citenamefont {Cherman},
  \citenamefont {Vermeersch}, \citenamefont {Fortuny}, \citenamefont {Park},
  \citenamefont {Tokei},\ and\ \citenamefont {De~Wolf}}]{BEOL2023_IMEC}%
  \BibitemOpen
  \bibfield  {author} {\bibinfo {author} {\bibfnamefont {X.}~\bibnamefont
  {Chang}}, \bibinfo {author} {\bibfnamefont {H.}~\bibnamefont {Oprins}},
  \bibinfo {author} {\bibfnamefont {M.}~\bibnamefont {Lofrano}}, \bibinfo
  {author} {\bibfnamefont {V.}~\bibnamefont {Cherman}}, \bibinfo {author}
  {\bibfnamefont {B.}~\bibnamefont {Vermeersch}}, \bibinfo {author}
  {\bibfnamefont {J.~D.}\ \bibnamefont {Fortuny}}, \bibinfo {author}
  {\bibfnamefont {S.}~\bibnamefont {Park}}, \bibinfo {author} {\bibfnamefont
  {Z.}~\bibnamefont {Tokei}}, \ and\ \bibinfo {author} {\bibfnamefont
  {I.}~\bibnamefont {De~Wolf}},\ }in\ \href {\doibase
  10.1109/IITC/MAM57687.2023.10154768} {\emph {\bibinfo {booktitle} {2023 IEEE
  International Interconnect Technology Conference (IITC) and IEEE Materials
  for Advanced Metallization Conference (MAM)(IITC/MAM)}}}\ (\bibinfo {year}
  {2023})\ pp.\ \bibinfo {pages} {1--3}\BibitemShut {NoStop}%
\bibitem [{\citenamefont {Pop}\ \emph {et~al.}(2004)\citenamefont {Pop},
  \citenamefont {Dutton},\ and\ \citenamefont {Goodson}}]{pop2004analytic}%
  \BibitemOpen
  \bibfield  {author} {\bibinfo {author} {\bibfnamefont {E.}~\bibnamefont
  {Pop}}, \bibinfo {author} {\bibfnamefont {R.~W.}\ \bibnamefont {Dutton}}, \
  and\ \bibinfo {author} {\bibfnamefont {K.~E.}\ \bibnamefont {Goodson}},\
  }\href {\doibase 10.1063/1.1788838} {\bibfield  {journal} {\bibinfo
  {journal} {J. Appl. Phys.}\ }\textbf {\bibinfo {volume} {96}},\ \bibinfo
  {pages} {4998} (\bibinfo {year} {2004})}\BibitemShut {NoStop}%
\bibitem [{\citenamefont {Chen}(2005)}]{ChenG05Oxford}%
  \BibitemOpen
  \bibfield  {author} {\bibinfo {author} {\bibfnamefont {G.}~\bibnamefont
  {Chen}},\ }\href
  {https://global.oup.com/ushe/product/nanoscale-energy-transport-and-conversion-9780195159424?cc=cn&lang=en&}
  {\emph {\bibinfo {title} {Nanoscale energy transport and conversion: {A}
  parallel treatment of electrons, molecules, phonons, and photons}}}\
  (\bibinfo  {publisher} {Oxford University Press},\ \bibinfo {year}
  {2005})\BibitemShut {NoStop}%
\bibitem [{\citenamefont {Xu}\ \emph {et~al.}(2023)\citenamefont {Xu},
  \citenamefont {Hu},\ and\ \citenamefont {Bao}}]{PhysRevApplied.19.014007}%
  \BibitemOpen
  \bibfield  {author} {\bibinfo {author} {\bibfnamefont {J.}~\bibnamefont
  {Xu}}, \bibinfo {author} {\bibfnamefont {Y.}~\bibnamefont {Hu}}, \ and\
  \bibinfo {author} {\bibfnamefont {H.}~\bibnamefont {Bao}},\ }\href {\doibase
  10.1103/PhysRevApplied.19.014007} {\bibfield  {journal} {\bibinfo  {journal}
  {Phys. Rev. Appl.}\ }\textbf {\bibinfo {volume} {19}},\ \bibinfo {pages}
  {014007} (\bibinfo {year} {2023})}\BibitemShut {NoStop}%
\bibitem [{\citenamefont {Wan}\ \emph {et~al.}(2024)\citenamefont {Wan},
  \citenamefont {Pan}, \citenamefont {Zong}, \citenamefont {Qin}, \citenamefont
  {Lü}, \citenamefont {Volz}, \citenamefont {Zhang},\ and\ \citenamefont
  {Yang}}]{wan2024manipulating}%
  \BibitemOpen
  \bibfield  {author} {\bibinfo {author} {\bibfnamefont {X.}~\bibnamefont
  {Wan}}, \bibinfo {author} {\bibfnamefont {D.}~\bibnamefont {Pan}}, \bibinfo
  {author} {\bibfnamefont {Z.}~\bibnamefont {Zong}}, \bibinfo {author}
  {\bibfnamefont {Y.}~\bibnamefont {Qin}}, \bibinfo {author} {\bibfnamefont
  {J.-T.}\ \bibnamefont {Lü}}, \bibinfo {author} {\bibfnamefont
  {S.}~\bibnamefont {Volz}}, \bibinfo {author} {\bibfnamefont {L.}~\bibnamefont
  {Zhang}}, \ and\ \bibinfo {author} {\bibfnamefont {N.}~\bibnamefont {Yang}},\
  }\href {\doibase 10.1021/acs.nanolett.4c00478} {\bibfield  {journal}
  {\bibinfo  {journal} {Nano Letters}\ }\textbf {\bibinfo {volume} {24}},\
  \bibinfo {pages} {6889} (\bibinfo {year} {2024})}\BibitemShut {NoStop}%
\bibitem [{\citenamefont {Sekiguchi}\ \emph {et~al.}(2021)\citenamefont
  {Sekiguchi}, \citenamefont {Hirori}, \citenamefont {Yumoto}, \citenamefont
  {Shimazaki}, \citenamefont {Nakamura}, \citenamefont {Wakamiya},\ and\
  \citenamefont {Kanemitsu}}]{PhysRevLett.126.077401}%
  \BibitemOpen
  \bibfield  {author} {\bibinfo {author} {\bibfnamefont {F.}~\bibnamefont
  {Sekiguchi}}, \bibinfo {author} {\bibfnamefont {H.}~\bibnamefont {Hirori}},
  \bibinfo {author} {\bibfnamefont {G.}~\bibnamefont {Yumoto}}, \bibinfo
  {author} {\bibfnamefont {A.}~\bibnamefont {Shimazaki}}, \bibinfo {author}
  {\bibfnamefont {T.}~\bibnamefont {Nakamura}}, \bibinfo {author}
  {\bibfnamefont {A.}~\bibnamefont {Wakamiya}}, \ and\ \bibinfo {author}
  {\bibfnamefont {Y.}~\bibnamefont {Kanemitsu}},\ }\href {\doibase
  10.1103/PhysRevLett.126.077401} {\bibfield  {journal} {\bibinfo  {journal}
  {Phys. Rev. Lett.}\ }\textbf {\bibinfo {volume} {126}},\ \bibinfo {pages}
  {077401} (\bibinfo {year} {2021})}\BibitemShut {NoStop}%
\bibitem [{\citenamefont {Chiloyan}\ \emph {et~al.}(2020)\citenamefont
  {Chiloyan}, \citenamefont {Huberman}, \citenamefont {Maznev}, \citenamefont
  {Nelson},\ and\ \citenamefont {Chen}}]{APLnonthermal2020}%
  \BibitemOpen
  \bibfield  {author} {\bibinfo {author} {\bibfnamefont {V.}~\bibnamefont
  {Chiloyan}}, \bibinfo {author} {\bibfnamefont {S.}~\bibnamefont {Huberman}},
  \bibinfo {author} {\bibfnamefont {A.~A.}\ \bibnamefont {Maznev}}, \bibinfo
  {author} {\bibfnamefont {K.~A.}\ \bibnamefont {Nelson}}, \ and\ \bibinfo
  {author} {\bibfnamefont {G.}~\bibnamefont {Chen}},\ }\href {\doibase
  10.1063/1.5139069} {\bibfield  {journal} {\bibinfo  {journal} {Appl. Phys.
  Lett.}\ }\textbf {\bibinfo {volume} {116}},\ \bibinfo {pages} {163102}
  (\bibinfo {year} {2020})}\BibitemShut {NoStop}%
\bibitem [{\citenamefont {Chen}\ \emph {et~al.}(2018)\citenamefont {Chen},
  \citenamefont {Hua}, \citenamefont {Zhang}, \citenamefont {Ravichandran},\
  and\ \citenamefont {Minnich}}]{PhysRevApplied.10.054068}%
  \BibitemOpen
  \bibfield  {author} {\bibinfo {author} {\bibfnamefont {X.}~\bibnamefont
  {Chen}}, \bibinfo {author} {\bibfnamefont {C.}~\bibnamefont {Hua}}, \bibinfo
  {author} {\bibfnamefont {H.}~\bibnamefont {Zhang}}, \bibinfo {author}
  {\bibfnamefont {N.~K.}\ \bibnamefont {Ravichandran}}, \ and\ \bibinfo
  {author} {\bibfnamefont {A.~J.}\ \bibnamefont {Minnich}},\ }\href {\doibase
  10.1103/PhysRevApplied.10.054068} {\bibfield  {journal} {\bibinfo  {journal}
  {Phys. Rev. Applied}\ }\textbf {\bibinfo {volume} {10}},\ \bibinfo {pages}
  {054068} (\bibinfo {year} {2018})}\BibitemShut {NoStop}%
\bibitem [{\citenamefont {Frazer}\ \emph {et~al.}(2019)\citenamefont {Frazer},
  \citenamefont {Knobloch}, \citenamefont {Hoogeboom-Pot}, \citenamefont
  {Nardi}, \citenamefont {Chao}, \citenamefont {Falcone}, \citenamefont
  {Murnane}, \citenamefont {Kapteyn},\ and\ \citenamefont
  {Hernandez-Charpak}}]{PhysRevApplied.11.024042}%
  \BibitemOpen
  \bibfield  {author} {\bibinfo {author} {\bibfnamefont {T.~D.}\ \bibnamefont
  {Frazer}}, \bibinfo {author} {\bibfnamefont {J.~L.}\ \bibnamefont
  {Knobloch}}, \bibinfo {author} {\bibfnamefont {K.~M.}\ \bibnamefont
  {Hoogeboom-Pot}}, \bibinfo {author} {\bibfnamefont {D.}~\bibnamefont
  {Nardi}}, \bibinfo {author} {\bibfnamefont {W.}~\bibnamefont {Chao}},
  \bibinfo {author} {\bibfnamefont {R.~W.}\ \bibnamefont {Falcone}}, \bibinfo
  {author} {\bibfnamefont {M.~M.}\ \bibnamefont {Murnane}}, \bibinfo {author}
  {\bibfnamefont {H.~C.}\ \bibnamefont {Kapteyn}}, \ and\ \bibinfo {author}
  {\bibfnamefont {J.~N.}\ \bibnamefont {Hernandez-Charpak}},\ }\href {\doibase
  10.1103/PhysRevApplied.11.024042} {\bibfield  {journal} {\bibinfo  {journal}
  {Phys. Rev. Applied}\ }\textbf {\bibinfo {volume} {11}},\ \bibinfo {pages}
  {024042} (\bibinfo {year} {2019})}\BibitemShut {NoStop}%
\bibitem [{\citenamefont {Zhang}\ \emph {et~al.}(2022)\citenamefont {Zhang},
  \citenamefont {Ma}, \citenamefont {Shang}, \citenamefont {Wan}, \citenamefont
  {Lü}, \citenamefont {Guo}, \citenamefont {Li},\ and\ \citenamefont
  {Yang}}]{chuang2021graded}%
  \BibitemOpen
  \bibfield  {author} {\bibinfo {author} {\bibfnamefont {C.}~\bibnamefont
  {Zhang}}, \bibinfo {author} {\bibfnamefont {D.}~\bibnamefont {Ma}}, \bibinfo
  {author} {\bibfnamefont {M.}~\bibnamefont {Shang}}, \bibinfo {author}
  {\bibfnamefont {X.}~\bibnamefont {Wan}}, \bibinfo {author} {\bibfnamefont
  {J.-T.}\ \bibnamefont {Lü}}, \bibinfo {author} {\bibfnamefont
  {Z.}~\bibnamefont {Guo}}, \bibinfo {author} {\bibfnamefont {B.}~\bibnamefont
  {Li}}, \ and\ \bibinfo {author} {\bibfnamefont {N.}~\bibnamefont {Yang}},\
  }\href {\doibase https://doi.org/10.1016/j.mtphys.2022.100605} {\bibfield
  {journal} {\bibinfo  {journal} {Mater. Today Phys.}\ }\textbf {\bibinfo
  {volume} {22}},\ \bibinfo {pages} {100605} (\bibinfo {year}
  {2022})}\BibitemShut {NoStop}%
\bibitem [{\citenamefont {Chen}(2021)}]{chen_non-fourier_2021}%
  \BibitemOpen
  \bibfield  {author} {\bibinfo {author} {\bibfnamefont {G.}~\bibnamefont
  {Chen}},\ }\href {\doibase 10.1038/s42254-021-00334-1} {\bibfield  {journal}
  {\bibinfo  {journal} {Nat. Rev. Phys.}\ }\textbf {\bibinfo {volume} {3}},\
  \bibinfo {pages} {555} (\bibinfo {year} {2021})}\BibitemShut {NoStop}%
\bibitem [{\citenamefont {Ziabari}\ \emph {et~al.}(2018)\citenamefont
  {Ziabari}, \citenamefont {Torres}, \citenamefont {Vermeersch}, \citenamefont
  {Xuan}, \citenamefont {Cartoix{\`a}}, \citenamefont {Torell{\'o}},
  \citenamefont {Bahk}, \citenamefont {Koh}, \citenamefont {Parsa},
  \citenamefont {Ye}, \citenamefont {Alvarez},\ and\ \citenamefont
  {Shakouri}}]{ziabari2018a}%
  \BibitemOpen
  \bibfield  {author} {\bibinfo {author} {\bibfnamefont {A.}~\bibnamefont
  {Ziabari}}, \bibinfo {author} {\bibfnamefont {P.}~\bibnamefont {Torres}},
  \bibinfo {author} {\bibfnamefont {B.}~\bibnamefont {Vermeersch}}, \bibinfo
  {author} {\bibfnamefont {Y.}~\bibnamefont {Xuan}}, \bibinfo {author}
  {\bibfnamefont {X.}~\bibnamefont {Cartoix{\`a}}}, \bibinfo {author}
  {\bibfnamefont {A.}~\bibnamefont {Torell{\'o}}}, \bibinfo {author}
  {\bibfnamefont {J.-H.}\ \bibnamefont {Bahk}}, \bibinfo {author}
  {\bibfnamefont {Y.~R.}\ \bibnamefont {Koh}}, \bibinfo {author} {\bibfnamefont
  {M.}~\bibnamefont {Parsa}}, \bibinfo {author} {\bibfnamefont {P.~D.}\
  \bibnamefont {Ye}}, \bibinfo {author} {\bibfnamefont {F.~X.}\ \bibnamefont
  {Alvarez}}, \ and\ \bibinfo {author} {\bibfnamefont {A.}~\bibnamefont
  {Shakouri}},\ }\href {\doibase 10.1038/s41467-017-02652-4} {\bibfield
  {journal} {\bibinfo  {journal} {Nat. Commun.}\ }\textbf {\bibinfo {volume}
  {9}},\ \bibinfo {pages} {255} (\bibinfo {year} {2018})}\BibitemShut {NoStop}%
\bibitem [{\citenamefont {Beardo}\ \emph {et~al.}(2021)\citenamefont {Beardo},
  \citenamefont {Knobloch}, \citenamefont {Sendra}, \citenamefont {Bafaluy},
  \citenamefont {Frazer}, \citenamefont {Chao}, \citenamefont
  {Hernandez-Charpak}, \citenamefont {Kapteyn}, \citenamefont {Abad},
  \citenamefont {Murnane}, \citenamefont {Alvarez},\ and\ \citenamefont
  {Camacho}}]{beardo_general_2021}%
  \BibitemOpen
  \bibfield  {author} {\bibinfo {author} {\bibfnamefont {A.}~\bibnamefont
  {Beardo}}, \bibinfo {author} {\bibfnamefont {J.~L.}\ \bibnamefont
  {Knobloch}}, \bibinfo {author} {\bibfnamefont {L.}~\bibnamefont {Sendra}},
  \bibinfo {author} {\bibfnamefont {J.}~\bibnamefont {Bafaluy}}, \bibinfo
  {author} {\bibfnamefont {T.~D.}\ \bibnamefont {Frazer}}, \bibinfo {author}
  {\bibfnamefont {W.}~\bibnamefont {Chao}}, \bibinfo {author} {\bibfnamefont
  {J.~N.}\ \bibnamefont {Hernandez-Charpak}}, \bibinfo {author} {\bibfnamefont
  {H.~C.}\ \bibnamefont {Kapteyn}}, \bibinfo {author} {\bibfnamefont
  {B.}~\bibnamefont {Abad}}, \bibinfo {author} {\bibfnamefont {M.~M.}\
  \bibnamefont {Murnane}}, \bibinfo {author} {\bibfnamefont {F.~X.}\
  \bibnamefont {Alvarez}}, \ and\ \bibinfo {author} {\bibfnamefont
  {J.}~\bibnamefont {Camacho}},\ }\href {\doibase 10.1021/acsnano.1c01946}
  {\bibfield  {journal} {\bibinfo  {journal} {ACS Nano}\ }\textbf {\bibinfo
  {volume} {15}},\ \bibinfo {pages} {13019} (\bibinfo {year}
  {2021})}\BibitemShut {NoStop}%
\bibitem [{\citenamefont {Chang}\ \emph {et~al.}(2008)\citenamefont {Chang},
  \citenamefont {Okawa}, \citenamefont {Garcia}, \citenamefont {Majumdar},\
  and\ \citenamefont {Zettl}}]{chang_breakdown_2008}%
  \BibitemOpen
  \bibfield  {author} {\bibinfo {author} {\bibfnamefont {C.~W.}\ \bibnamefont
  {Chang}}, \bibinfo {author} {\bibfnamefont {D.}~\bibnamefont {Okawa}},
  \bibinfo {author} {\bibfnamefont {H.}~\bibnamefont {Garcia}}, \bibinfo
  {author} {\bibfnamefont {A.}~\bibnamefont {Majumdar}}, \ and\ \bibinfo
  {author} {\bibfnamefont {A.}~\bibnamefont {Zettl}},\ }\href {\doibase
  10.1103/PhysRevLett.101.075903} {\bibfield  {journal} {\bibinfo  {journal}
  {Phys. Rev. Lett.}\ }\textbf {\bibinfo {volume} {101}},\ \bibinfo {pages}
  {075903} (\bibinfo {year} {2008})}\BibitemShut {NoStop}%
\bibitem [{\citenamefont {L{\"u}}\ \emph {et~al.}(2002)\citenamefont {L{\"u}},
  \citenamefont {Shen},\ and\ \citenamefont {Chu}}]{lu2002size}%
  \BibitemOpen
  \bibfield  {author} {\bibinfo {author} {\bibfnamefont {X.}~\bibnamefont
  {L{\"u}}}, \bibinfo {author} {\bibfnamefont {W.}~\bibnamefont {Shen}}, \ and\
  \bibinfo {author} {\bibfnamefont {J.}~\bibnamefont {Chu}},\ }\href {\doibase
  10.1063/1.1427134} {\bibfield  {journal} {\bibinfo  {journal} {J. Appl.
  Phys.}\ }\textbf {\bibinfo {volume} {91}},\ \bibinfo {pages} {1542} (\bibinfo
  {year} {2002})}\BibitemShut {NoStop}%
\bibitem [{\citenamefont {Xu}\ \emph {et~al.}(2020)\citenamefont {Xu},
  \citenamefont {Fan}, \citenamefont {Wang}, \citenamefont {Zhang},\ and\
  \citenamefont {Wang}}]{xu_raman-based_2020}%
  \BibitemOpen
  \bibfield  {author} {\bibinfo {author} {\bibfnamefont {S.}~\bibnamefont
  {Xu}}, \bibinfo {author} {\bibfnamefont {A.}~\bibnamefont {Fan}}, \bibinfo
  {author} {\bibfnamefont {H.}~\bibnamefont {Wang}}, \bibinfo {author}
  {\bibfnamefont {X.}~\bibnamefont {Zhang}}, \ and\ \bibinfo {author}
  {\bibfnamefont {X.}~\bibnamefont {Wang}},\ }\href {\doibase
  10.1016/j.ijheatmasstransfer.2020.119751} {\bibfield  {journal} {\bibinfo
  {journal} {Int. J. Heat Mass Transfer}\ }\textbf {\bibinfo {volume} {154}},\
  \bibinfo {pages} {119751} (\bibinfo {year} {2020})}\BibitemShut {NoStop}%
\bibitem [{\citenamefont {Jiang}\ \emph {et~al.}(2018)\citenamefont {Jiang},
  \citenamefont {Qian},\ and\ \citenamefont {Yang}}]{jiang_tutorial_2018}%
  \BibitemOpen
  \bibfield  {author} {\bibinfo {author} {\bibfnamefont {P.}~\bibnamefont
  {Jiang}}, \bibinfo {author} {\bibfnamefont {X.}~\bibnamefont {Qian}}, \ and\
  \bibinfo {author} {\bibfnamefont {R.}~\bibnamefont {Yang}},\ }\href {\doibase
  10.1063/1.5046944} {\bibfield  {journal} {\bibinfo  {journal} {J. Appl.
  Phys.}\ }\textbf {\bibinfo {volume} {124}},\ \bibinfo {pages} {161103}
  (\bibinfo {year} {2018})}\BibitemShut {NoStop}%
\bibitem [{\citenamefont {Liu}\ \emph {et~al.}(2022)\citenamefont {Liu},
  \citenamefont {Fan}, \citenamefont {Hu}, \citenamefont {Li}, \citenamefont
  {Liu},\ and\ \citenamefont {Kang}}]{xiaoyanliu_2022_review_thermal}%
  \BibitemOpen
  \bibfield  {author} {\bibinfo {author} {\bibfnamefont {X.}~\bibnamefont
  {Liu}}, \bibinfo {author} {\bibfnamefont {M.}~\bibnamefont {Fan}}, \bibinfo
  {author} {\bibfnamefont {Y.}~\bibnamefont {Hu}}, \bibinfo {author}
  {\bibfnamefont {H.}~\bibnamefont {Li}}, \bibinfo {author} {\bibfnamefont
  {F.}~\bibnamefont {Liu}}, \ and\ \bibinfo {author} {\bibfnamefont
  {J.}~\bibnamefont {Kang}},\ }in\ \href {\doibase
  10.1109/IEDM45625.2022.10019403} {\emph {\bibinfo {booktitle} {2022
  International Electron Devices Meeting (IEDM)}}}\ (\bibinfo {year} {2022})\
  pp.\ \bibinfo {pages} {15.4.1--15.4.4}\BibitemShut {NoStop}%
\bibitem [{\citenamefont {Sendra}\ \emph {et~al.}(2021)\citenamefont {Sendra},
  \citenamefont {Beardo}, \citenamefont {Torres}, \citenamefont {Bafaluy},
  \citenamefont {Alvarez},\ and\ \citenamefont
  {Camacho}}]{PhysRevB.103.L140301}%
  \BibitemOpen
  \bibfield  {author} {\bibinfo {author} {\bibfnamefont {L.}~\bibnamefont
  {Sendra}}, \bibinfo {author} {\bibfnamefont {A.}~\bibnamefont {Beardo}},
  \bibinfo {author} {\bibfnamefont {P.}~\bibnamefont {Torres}}, \bibinfo
  {author} {\bibfnamefont {J.}~\bibnamefont {Bafaluy}}, \bibinfo {author}
  {\bibfnamefont {F.~X.}\ \bibnamefont {Alvarez}}, \ and\ \bibinfo {author}
  {\bibfnamefont {J.}~\bibnamefont {Camacho}},\ }\href {\doibase
  10.1103/PhysRevB.103.L140301} {\bibfield  {journal} {\bibinfo  {journal}
  {Phys. Rev. B}\ }\textbf {\bibinfo {volume} {103}},\ \bibinfo {pages}
  {L140301} (\bibinfo {year} {2021})}\BibitemShut {NoStop}%
\bibitem [{\citenamefont {Bao}\ \emph {et~al.}(2018)\citenamefont {Bao},
  \citenamefont {Chen}, \citenamefont {Gu},\ and\ \citenamefont
  {Cao}}]{esee8c149}%
  \BibitemOpen
  \bibfield  {author} {\bibinfo {author} {\bibfnamefont {H.}~\bibnamefont
  {Bao}}, \bibinfo {author} {\bibfnamefont {J.}~\bibnamefont {Chen}}, \bibinfo
  {author} {\bibfnamefont {X.}~\bibnamefont {Gu}}, \ and\ \bibinfo {author}
  {\bibfnamefont {B.}~\bibnamefont {Cao}},\ }\href {\doibase
  10.30919/esee8c149} {\bibfield  {journal} {\bibinfo  {journal} {ES Energy.
  Environ.}\ }\textbf {\bibinfo {volume} {1}},\ \bibinfo {pages} {16} (\bibinfo
  {year} {2018})}\BibitemShut {NoStop}%
\bibitem [{\citenamefont {Guo}\ and\ \citenamefont
  {Wang}(2015)}]{WANGMR15APPLICATION}%
  \BibitemOpen
  \bibfield  {author} {\bibinfo {author} {\bibfnamefont {Y.}~\bibnamefont
  {Guo}}\ and\ \bibinfo {author} {\bibfnamefont {M.}~\bibnamefont {Wang}},\
  }\href {\doibase 10.1016/j.physrep.2015.07.003} {\bibfield  {journal}
  {\bibinfo  {journal} {Phys. Rep.}\ }\textbf {\bibinfo {volume} {595}},\
  \bibinfo {pages} {1 } (\bibinfo {year} {2015})}\BibitemShut {NoStop}%
\bibitem [{\citenamefont {de~Tomas}\ \emph {et~al.}(2014)\citenamefont
  {de~Tomas}, \citenamefont {Cantarero}, \citenamefont {Lopeandia},\ and\
  \citenamefont {Alvarez}}]{de_tomas_kinetic_2014}%
  \BibitemOpen
  \bibfield  {author} {\bibinfo {author} {\bibfnamefont {C.}~\bibnamefont
  {de~Tomas}}, \bibinfo {author} {\bibfnamefont {A.}~\bibnamefont {Cantarero}},
  \bibinfo {author} {\bibfnamefont {A.~F.}\ \bibnamefont {Lopeandia}}, \ and\
  \bibinfo {author} {\bibfnamefont {F.~X.}\ \bibnamefont {Alvarez}},\ }\href
  {\doibase 10.1063/1.4871672} {\bibfield  {journal} {\bibinfo  {journal} {J.
  Appl. Phys.}\ }\textbf {\bibinfo {volume} {115}},\ \bibinfo {pages} {164314}
  (\bibinfo {year} {2014})}\BibitemShut {NoStop}%
\bibitem [{\citenamefont {Kim}\ \emph {et~al.}(2018)\citenamefont {Kim},
  \citenamefont {Son}, \citenamefont {Myeong}, \citenamefont {Kang},
  \citenamefont {Jeon},\ and\ \citenamefont
  {Shin}}]{Three_Stacked_NanoplateFET2018}%
  \BibitemOpen
  \bibfield  {author} {\bibinfo {author} {\bibfnamefont {H.}~\bibnamefont
  {Kim}}, \bibinfo {author} {\bibfnamefont {D.}~\bibnamefont {Son}}, \bibinfo
  {author} {\bibfnamefont {I.}~\bibnamefont {Myeong}}, \bibinfo {author}
  {\bibfnamefont {M.}~\bibnamefont {Kang}}, \bibinfo {author} {\bibfnamefont
  {J.}~\bibnamefont {Jeon}}, \ and\ \bibinfo {author} {\bibfnamefont
  {H.}~\bibnamefont {Shin}},\ }\href {\doibase 10.1109/TED.2018.2862918}
  {\bibfield  {journal} {\bibinfo  {journal} {IEEE T. Electron Dev.}\ }\textbf
  {\bibinfo {volume} {65}},\ \bibinfo {pages} {4520} (\bibinfo {year}
  {2018})}\BibitemShut {NoStop}%
\bibitem [{\citenamefont {Zhao}\ \emph {et~al.}(2022)\citenamefont {Zhao},
  \citenamefont {Zhao}, \citenamefont {He},\ and\ \citenamefont
  {Du}}]{comparison_FIN_GAAFET}%
  \BibitemOpen
  \bibfield  {author} {\bibinfo {author} {\bibfnamefont {P.}~\bibnamefont
  {Zhao}}, \bibinfo {author} {\bibfnamefont {S.-H.}\ \bibnamefont {Zhao}},
  \bibinfo {author} {\bibfnamefont {Y.-D.}\ \bibnamefont {He}}, \ and\ \bibinfo
  {author} {\bibfnamefont {G.}~\bibnamefont {Du}},\ }in\ \href {\doibase
  10.1109/ICSICT55466.2022.9963426} {\emph {\bibinfo {booktitle} {2022 IEEE
  16th International Conference on Solid-State \& Integrated Circuit Technology
  (ICSICT)}}}\ (\bibinfo {year} {2022})\ pp.\ \bibinfo {pages}
  {1--3}\BibitemShut {NoStop}%
\bibitem [{\citenamefont {Kumar}\ \emph
  {et~al.}(2023{\natexlab{a}})\citenamefont {Kumar}, \citenamefont {Patel},
  \citenamefont {Datta},\ and\ \citenamefont {Dasgupta}}]{KUMAR2023100056}%
  \BibitemOpen
  \bibfield  {author} {\bibinfo {author} {\bibfnamefont {V.}~\bibnamefont
  {Kumar}}, \bibinfo {author} {\bibfnamefont {J.}~\bibnamefont {Patel}},
  \bibinfo {author} {\bibfnamefont {A.}~\bibnamefont {Datta}}, \ and\ \bibinfo
  {author} {\bibfnamefont {S.}~\bibnamefont {Dasgupta}},\ }\href {\doibase
  https://doi.org/10.1016/j.memori.2023.100056} {\bibfield  {journal} {\bibinfo
   {journal} {Memories - Materials, Devices, Circuits and Systems}\ }\textbf
  {\bibinfo {volume} {4}},\ \bibinfo {pages} {100056} (\bibinfo {year}
  {2023}{\natexlab{a}})}\BibitemShut {NoStop}%
\bibitem [{\citenamefont {Stettler}\ \emph {et~al.}(2021)\citenamefont
  {Stettler}, \citenamefont {Cea}, \citenamefont {Hasan}, \citenamefont
  {Jiang}, \citenamefont {Keys}, \citenamefont {Landon}, \citenamefont
  {Marepalli}, \citenamefont {Pantuso},\ and\ \citenamefont
  {Weber}}]{TCAD_application_intel_2021_review}%
  \BibitemOpen
  \bibfield  {author} {\bibinfo {author} {\bibfnamefont {M.~A.}\ \bibnamefont
  {Stettler}}, \bibinfo {author} {\bibfnamefont {S.~M.}\ \bibnamefont {Cea}},
  \bibinfo {author} {\bibfnamefont {S.}~\bibnamefont {Hasan}}, \bibinfo
  {author} {\bibfnamefont {L.}~\bibnamefont {Jiang}}, \bibinfo {author}
  {\bibfnamefont {P.~H.}\ \bibnamefont {Keys}}, \bibinfo {author}
  {\bibfnamefont {C.~D.}\ \bibnamefont {Landon}}, \bibinfo {author}
  {\bibfnamefont {P.}~\bibnamefont {Marepalli}}, \bibinfo {author}
  {\bibfnamefont {D.}~\bibnamefont {Pantuso}}, \ and\ \bibinfo {author}
  {\bibfnamefont {C.~E.}\ \bibnamefont {Weber}},\ }\href {\doibase
  10.1109/TED.2021.3076976} {\bibfield  {journal} {\bibinfo  {journal} {IEEE
  Transactions on Electron Devices}\ }\textbf {\bibinfo {volume} {68}},\
  \bibinfo {pages} {5350} (\bibinfo {year} {2021})}\BibitemShut {NoStop}%
\bibitem [{\citenamefont {Ni}\ \emph {et~al.}(2012)\citenamefont {Ni},
  \citenamefont {Aksamija}, \citenamefont {Murthy},\ and\ \citenamefont
  {Ravaioli}}]{ni2012coupled}%
  \BibitemOpen
  \bibfield  {author} {\bibinfo {author} {\bibfnamefont {C.}~\bibnamefont
  {Ni}}, \bibinfo {author} {\bibfnamefont {Z.}~\bibnamefont {Aksamija}},
  \bibinfo {author} {\bibfnamefont {J.~Y.}\ \bibnamefont {Murthy}}, \ and\
  \bibinfo {author} {\bibfnamefont {U.}~\bibnamefont {Ravaioli}},\ }\href
  {\doibase 10.1007/s10825-012-0387-x} {\bibfield  {journal} {\bibinfo
  {journal} {Journal of Computational Electronics}\ }\textbf {\bibinfo {volume}
  {11}},\ \bibinfo {pages} {93–105} (\bibinfo {year} {2012})}\BibitemShut
  {NoStop}%
\bibitem [{\citenamefont {Kumar}\ \emph
  {et~al.}(2023{\natexlab{b}})\citenamefont {Kumar}, \citenamefont {Datta},\
  and\ \citenamefont {Dasgupta}}]{IEEE2023_5nm}%
  \BibitemOpen
  \bibfield  {author} {\bibinfo {author} {\bibfnamefont {V.}~\bibnamefont
  {Kumar}}, \bibinfo {author} {\bibfnamefont {A.}~\bibnamefont {Datta}}, \ and\
  \bibinfo {author} {\bibfnamefont {S.}~\bibnamefont {Dasgupta}},\ }in\ \href
  {\doibase 10.1109/ITherm55368.2023.10177526} {\emph {\bibinfo {booktitle}
  {2023 22nd IEEE Intersociety Conference on Thermal and Thermomechanical
  Phenomena in Electronic Systems (ITherm)}}}\ (\bibinfo {year} {2023})\ pp.\
  \bibinfo {pages} {1--8}\BibitemShut {NoStop}%
\bibitem [{\citenamefont {Murthy}\ \emph {et~al.}(2005)\citenamefont {Murthy},
  \citenamefont {Narumanchi}, \citenamefont {Pascual-Gutierrez}, \citenamefont
  {Wang}, \citenamefont {Ni},\ and\ \citenamefont {Mathur}}]{MurthyJY05Review}%
  \BibitemOpen
  \bibfield  {author} {\bibinfo {author} {\bibfnamefont {J.~Y.}\ \bibnamefont
  {Murthy}}, \bibinfo {author} {\bibfnamefont {S.~V.~J.}\ \bibnamefont
  {Narumanchi}}, \bibinfo {author} {\bibfnamefont {J.~A.}\ \bibnamefont
  {Pascual-Gutierrez}}, \bibinfo {author} {\bibfnamefont {T.}~\bibnamefont
  {Wang}}, \bibinfo {author} {\bibfnamefont {C.}~\bibnamefont {Ni}}, \ and\
  \bibinfo {author} {\bibfnamefont {S.~R.}\ \bibnamefont {Mathur}},\ }\href
  {\doibase 10.1615/IntJMultCompEng.v3.i1.20} {\bibfield  {journal} {\bibinfo
  {journal} {Int. J. Multiscale Computat. Eng.}\ }\textbf {\bibinfo {volume}
  {3}},\ \bibinfo {pages} {5} (\bibinfo {year} {2005})}\BibitemShut {NoStop}%
\bibitem [{\citenamefont {Pop}\ \emph {et~al.}(2006)\citenamefont {Pop},
  \citenamefont {Sinha},\ and\ \citenamefont
  {Goodson}}]{generation_thermalFET}%
  \BibitemOpen
  \bibfield  {author} {\bibinfo {author} {\bibfnamefont {E.}~\bibnamefont
  {Pop}}, \bibinfo {author} {\bibfnamefont {S.}~\bibnamefont {Sinha}}, \ and\
  \bibinfo {author} {\bibfnamefont {K.}~\bibnamefont {Goodson}},\ }\href
  {\doibase 10.1109/JPROC.2006.879794} {\bibfield  {journal} {\bibinfo
  {journal} {Proceedings of the IEEE}\ }\textbf {\bibinfo {volume} {94}},\
  \bibinfo {pages} {1587} (\bibinfo {year} {2006})}\BibitemShut {NoStop}%
\bibitem [{\citenamefont {Thu Trang~Nghiêm}\ \emph {et~al.}(2014)\citenamefont
  {Thu Trang~Nghiêm}, \citenamefont {Saint-Martin},\ and\ \citenamefont
  {Dollfus}}]{JAP_2014_double_gate_BTE}%
  \BibitemOpen
  \bibfield  {author} {\bibinfo {author} {\bibfnamefont {T.}~\bibnamefont {Thu
  Trang~Nghiêm}}, \bibinfo {author} {\bibfnamefont {J.}~\bibnamefont
  {Saint-Martin}}, \ and\ \bibinfo {author} {\bibfnamefont {P.}~\bibnamefont
  {Dollfus}},\ }\href {\doibase 10.1063/1.4893646} {\bibfield  {journal}
  {\bibinfo  {journal} {Journal of Applied Physics}\ }\textbf {\bibinfo
  {volume} {116}},\ \bibinfo {pages} {074514} (\bibinfo {year}
  {2014})}\BibitemShut {NoStop}%
\bibitem [{\citenamefont {Sverdrup}\ \emph {et~al.}(2001)\citenamefont
  {Sverdrup}, \citenamefont {Ju},\ and\ \citenamefont
  {Goodson}}]{SverdrupPG01subcontinuum}%
  \BibitemOpen
  \bibfield  {author} {\bibinfo {author} {\bibfnamefont {P.~G.}\ \bibnamefont
  {Sverdrup}}, \bibinfo {author} {\bibfnamefont {Y.~S.}\ \bibnamefont {Ju}}, \
  and\ \bibinfo {author} {\bibfnamefont {K.~E.}\ \bibnamefont {Goodson}},\
  }\href {\doibase 10.1115/1.1337651} {\bibfield  {journal} {\bibinfo
  {journal} {J. Heat Transfer}\ }\textbf {\bibinfo {volume} {123}},\ \bibinfo
  {pages} {130} (\bibinfo {year} {2001})}\BibitemShut {NoStop}%
\bibitem [{\citenamefont {Vallabhaneni}\ \emph {et~al.}(2016)\citenamefont
  {Vallabhaneni}, \citenamefont {Singh}, \citenamefont {Bao}, \citenamefont
  {Murthy},\ and\ \citenamefont {Ruan}}]{PhysRevB.93.125432}%
  \BibitemOpen
  \bibfield  {author} {\bibinfo {author} {\bibfnamefont {A.~K.}\ \bibnamefont
  {Vallabhaneni}}, \bibinfo {author} {\bibfnamefont {D.}~\bibnamefont {Singh}},
  \bibinfo {author} {\bibfnamefont {H.}~\bibnamefont {Bao}}, \bibinfo {author}
  {\bibfnamefont {J.}~\bibnamefont {Murthy}}, \ and\ \bibinfo {author}
  {\bibfnamefont {X.}~\bibnamefont {Ruan}},\ }\href {\doibase
  10.1103/PhysRevB.93.125432} {\bibfield  {journal} {\bibinfo  {journal} {Phys.
  Rev. B}\ }\textbf {\bibinfo {volume} {93}},\ \bibinfo {pages} {125432}
  (\bibinfo {year} {2016})}\BibitemShut {NoStop}%
\bibitem [{\citenamefont {Huberman}\ \emph {et~al.}(2023)\citenamefont
  {Huberman}, \citenamefont {Haibeh}, \citenamefont {Zhang},\ and\
  \citenamefont {Song}}]{APS_second_sound}%
  \BibitemOpen
  \bibfield  {author} {\bibinfo {author} {\bibfnamefont {S.}~\bibnamefont
  {Huberman}}, \bibinfo {author} {\bibfnamefont {J.~A.}\ \bibnamefont
  {Haibeh}}, \bibinfo {author} {\bibfnamefont {C.}~\bibnamefont {Zhang}}, \
  and\ \bibinfo {author} {\bibfnamefont {Q.}~\bibnamefont {Song}},\ }\href
  {https://meetings.aps.org/Meeting/MAR23/Session/LL06.5} {\bibfield  {journal}
  {\bibinfo  {journal} {Bulletin of the American Physical Society}\ } (\bibinfo
  {year} {2023})}\BibitemShut {NoStop}%
\bibitem [{\citenamefont {Hao}\ \emph {et~al.}(2009)\citenamefont {Hao},
  \citenamefont {Chen},\ and\ \citenamefont {Jeng}}]{MC_porous_gang2009JAP}%
  \BibitemOpen
  \bibfield  {author} {\bibinfo {author} {\bibfnamefont {Q.}~\bibnamefont
  {Hao}}, \bibinfo {author} {\bibfnamefont {G.}~\bibnamefont {Chen}}, \ and\
  \bibinfo {author} {\bibfnamefont {M.-S.}\ \bibnamefont {Jeng}},\ }\href
  {\doibase 10.1063/1.3266169} {\bibfield  {journal} {\bibinfo  {journal} {J.
  Appl. Phys.}\ }\textbf {\bibinfo {volume} {106}},\ \bibinfo {pages} {114321}
  (\bibinfo {year} {2009})}\BibitemShut {NoStop}%
\bibitem [{\citenamefont {Chen}\ \emph {et~al.}(2023)\citenamefont {Chen},
  \citenamefont {Hu}, \citenamefont {Wang},\ and\ \citenamefont
  {Tang}}]{CHEN2023108592}%
  \BibitemOpen
  \bibfield  {author} {\bibinfo {author} {\bibfnamefont {G.}~\bibnamefont
  {Chen}}, \bibinfo {author} {\bibfnamefont {B.}~\bibnamefont {Hu}}, \bibinfo
  {author} {\bibfnamefont {Z.}~\bibnamefont {Wang}}, \ and\ \bibinfo {author}
  {\bibfnamefont {D.}~\bibnamefont {Tang}},\ }\href {\doibase
  https://doi.org/10.1016/j.ijthermalsci.2023.108592} {\bibfield  {journal}
  {\bibinfo  {journal} {Int. J. Therm. Sci.}\ }\textbf {\bibinfo {volume}
  {194}},\ \bibinfo {pages} {108592} (\bibinfo {year} {2023})}\BibitemShut
  {NoStop}%
\bibitem [{\citenamefont {Shen}\ \emph {et~al.}(2023)\citenamefont {Shen},
  \citenamefont {Yang},\ and\ \citenamefont {Cao}}]{SHEN2023_IJHMT}%
  \BibitemOpen
  \bibfield  {author} {\bibinfo {author} {\bibfnamefont {Y.}~\bibnamefont
  {Shen}}, \bibinfo {author} {\bibfnamefont {H.-A.}\ \bibnamefont {Yang}}, \
  and\ \bibinfo {author} {\bibfnamefont {B.-Y.}\ \bibnamefont {Cao}},\ }\href
  {\doibase https://doi.org/10.1016/j.ijheatmasstransfer.2023.124284}
  {\bibfield  {journal} {\bibinfo  {journal} {Int. J. Heat Mass Transfer}\
  }\textbf {\bibinfo {volume} {211}},\ \bibinfo {pages} {124284} (\bibinfo
  {year} {2023})}\BibitemShut {NoStop}%
\bibitem [{\citenamefont {Pathak}\ \emph {et~al.}(2021)\citenamefont {Pathak},
  \citenamefont {Pawnday}, \citenamefont {Roy}, \citenamefont {Aref},
  \citenamefont {Dargush},\ and\ \citenamefont {Bansal}}]{PATHAK2021108003}%
  \BibitemOpen
  \bibfield  {author} {\bibinfo {author} {\bibfnamefont {A.}~\bibnamefont
  {Pathak}}, \bibinfo {author} {\bibfnamefont {A.}~\bibnamefont {Pawnday}},
  \bibinfo {author} {\bibfnamefont {A.~P.}\ \bibnamefont {Roy}}, \bibinfo
  {author} {\bibfnamefont {A.~J.}\ \bibnamefont {Aref}}, \bibinfo {author}
  {\bibfnamefont {G.~F.}\ \bibnamefont {Dargush}}, \ and\ \bibinfo {author}
  {\bibfnamefont {D.}~\bibnamefont {Bansal}},\ }\href {\doibase
  https://doi.org/10.1016/j.cpc.2021.108003} {\bibfield  {journal} {\bibinfo
  {journal} {Comput. Phys. Commun.}\ }\textbf {\bibinfo {volume} {265}},\
  \bibinfo {pages} {108003} (\bibinfo {year} {2021})}\BibitemShut {NoStop}%
\bibitem [{\citenamefont {Adisusilo}\ \emph {et~al.}(2014)\citenamefont
  {Adisusilo}, \citenamefont {Kukita},\ and\ \citenamefont
  {Kamakura}}]{3DFINFET_2014_mc}%
  \BibitemOpen
  \bibfield  {author} {\bibinfo {author} {\bibfnamefont {I.~N.}\ \bibnamefont
  {Adisusilo}}, \bibinfo {author} {\bibfnamefont {K.}~\bibnamefont {Kukita}}, \
  and\ \bibinfo {author} {\bibfnamefont {Y.}~\bibnamefont {Kamakura}},\ }in\
  \href {\doibase 10.1109/SISPAD.2014.6931552} {\emph {\bibinfo {booktitle}
  {2014 International Conference on Simulation of Semiconductor Processes and
  Devices (SISPAD)}}}\ (\bibinfo {year} {2014})\ pp.\ \bibinfo {pages}
  {17--20}\BibitemShut {NoStop}%
\bibitem [{\citenamefont {Medlar}\ and\ \citenamefont
  {Hensel}(2022{\natexlab{a}})}]{3DFINFETtransient}%
  \BibitemOpen
  \bibfield  {author} {\bibinfo {author} {\bibfnamefont {M.~P.}\ \bibnamefont
  {Medlar}}\ and\ \bibinfo {author} {\bibfnamefont {E.~C.}\ \bibnamefont
  {Hensel}},\ }\href {\doibase 10.1115/1.4056002} {\bibfield  {journal}
  {\bibinfo  {journal} {ASME Journal of Heat and Mass Transfer}\ }\textbf
  {\bibinfo {volume} {145}},\ \bibinfo {pages} {022501} (\bibinfo {year}
  {2022}{\natexlab{a}})}\BibitemShut {NoStop}%
\bibitem [{\citenamefont {Medlar}\ and\ \citenamefont
  {Hensel}(2022{\natexlab{b}})}]{JAP_2022_PHONONTHERMAL}%
  \BibitemOpen
  \bibfield  {author} {\bibinfo {author} {\bibfnamefont {M.~P.}\ \bibnamefont
  {Medlar}}\ and\ \bibinfo {author} {\bibfnamefont {E.~C.}\ \bibnamefont
  {Hensel}},\ }\href {\doibase 10.1115/1.4054600} {\bibfield  {journal}
  {\bibinfo  {journal} {Journal of Heat Transfer}\ }\textbf {\bibinfo {volume}
  {144}},\ \bibinfo {pages} {082503} (\bibinfo {year}
  {2022}{\natexlab{b}})}\BibitemShut {NoStop}%
\bibitem [{\citenamefont {Hao}\ \emph {et~al.}(2017)\citenamefont {Hao},
  \citenamefont {Zhao},\ and\ \citenamefont {Xiao}}]{JAP_qinghao_2017}%
  \BibitemOpen
  \bibfield  {author} {\bibinfo {author} {\bibfnamefont {Q.}~\bibnamefont
  {Hao}}, \bibinfo {author} {\bibfnamefont {H.}~\bibnamefont {Zhao}}, \ and\
  \bibinfo {author} {\bibfnamefont {Y.}~\bibnamefont {Xiao}},\ }\href {\doibase
  10.1063/1.4983761} {\bibfield  {journal} {\bibinfo  {journal} {J. Appl.
  Phys.}\ }\textbf {\bibinfo {volume} {121}},\ \bibinfo {pages} {204501}
  (\bibinfo {year} {2017})}\BibitemShut {NoStop}%
\bibitem [{\citenamefont {Hao}\ \emph {et~al.}(2018)\citenamefont {Hao},
  \citenamefont {Zhao}, \citenamefont {Xiao},\ and\ \citenamefont
  {Kronenfeld}}]{HAO2018496}%
  \BibitemOpen
  \bibfield  {author} {\bibinfo {author} {\bibfnamefont {Q.}~\bibnamefont
  {Hao}}, \bibinfo {author} {\bibfnamefont {H.}~\bibnamefont {Zhao}}, \bibinfo
  {author} {\bibfnamefont {Y.}~\bibnamefont {Xiao}}, \ and\ \bibinfo {author}
  {\bibfnamefont {M.~B.}\ \bibnamefont {Kronenfeld}},\ }\href {\doibase
  https://doi.org/10.1016/j.ijheatmasstransfer.2017.09.048} {\bibfield
  {journal} {\bibinfo  {journal} {Int. J. Heat Mass Transfer}\ }\textbf
  {\bibinfo {volume} {116}},\ \bibinfo {pages} {496} (\bibinfo {year}
  {2018})}\BibitemShut {NoStop}%
\bibitem [{\citenamefont {Terris}\ \emph {et~al.}(2009)\citenamefont {Terris},
  \citenamefont {Joulain}, \citenamefont {Lemonnier},\ and\ \citenamefont
  {Lacroix}}]{terris2009modeling}%
  \BibitemOpen
  \bibfield  {author} {\bibinfo {author} {\bibfnamefont {D.}~\bibnamefont
  {Terris}}, \bibinfo {author} {\bibfnamefont {K.}~\bibnamefont {Joulain}},
  \bibinfo {author} {\bibfnamefont {D.}~\bibnamefont {Lemonnier}}, \ and\
  \bibinfo {author} {\bibfnamefont {D.}~\bibnamefont {Lacroix}},\ }\href
  {\doibase 10.1063/1.3086409} {\bibfield  {journal} {\bibinfo  {journal} {J.
  Appl. Phys.}\ }\textbf {\bibinfo {volume} {105}},\ \bibinfo {pages} {073516}
  (\bibinfo {year} {2009})}\BibitemShut {NoStop}%
\bibitem [{\citenamefont {Ali}\ \emph {et~al.}(2014)\citenamefont {Ali},
  \citenamefont {Kollu}, \citenamefont {Mazumder}, \citenamefont {Sadayappan},\
  and\ \citenamefont {Mittal}}]{SyedAA14LargeScale}%
  \BibitemOpen
  \bibfield  {author} {\bibinfo {author} {\bibfnamefont {S.~A.}\ \bibnamefont
  {Ali}}, \bibinfo {author} {\bibfnamefont {G.}~\bibnamefont {Kollu}}, \bibinfo
  {author} {\bibfnamefont {S.}~\bibnamefont {Mazumder}}, \bibinfo {author}
  {\bibfnamefont {P.}~\bibnamefont {Sadayappan}}, \ and\ \bibinfo {author}
  {\bibfnamefont {A.}~\bibnamefont {Mittal}},\ }\href {\doibase
  10.1016/j.ijthermalsci.2014.07.019} {\bibfield  {journal} {\bibinfo
  {journal} {Int. J. Therm. Sci}\ }\textbf {\bibinfo {volume} {86}},\ \bibinfo
  {pages} {341 } (\bibinfo {year} {2014})}\BibitemShut {NoStop}%
\bibitem [{\citenamefont {Adams}\ and\ \citenamefont
  {Larsen}(2002)}]{ADAMS02fastiterative}%
  \BibitemOpen
  \bibfield  {author} {\bibinfo {author} {\bibfnamefont {M.~L.}\ \bibnamefont
  {Adams}}\ and\ \bibinfo {author} {\bibfnamefont {E.~W.}\ \bibnamefont
  {Larsen}},\ }\href {\doibase 10.1016/S0149-1970(01)00023-3} {\bibfield
  {journal} {\bibinfo  {journal} {Prog. Nucl. Energ.}\ }\textbf {\bibinfo
  {volume} {40}},\ \bibinfo {pages} {3 } (\bibinfo {year} {2002})}\BibitemShut
  {NoStop}%
\bibitem [{\citenamefont {Zhang}\ \emph {et~al.}(2017)\citenamefont {Zhang},
  \citenamefont {Guo},\ and\ \citenamefont {Chen}}]{Chuang17gray}%
  \BibitemOpen
  \bibfield  {author} {\bibinfo {author} {\bibfnamefont {C.}~\bibnamefont
  {Zhang}}, \bibinfo {author} {\bibfnamefont {Z.}~\bibnamefont {Guo}}, \ and\
  \bibinfo {author} {\bibfnamefont {S.}~\bibnamefont {Chen}},\ }\href {\doibase
  10.1103/PhysRevE.96.063311} {\bibfield  {journal} {\bibinfo  {journal} {Phys.
  Rev. E}\ }\textbf {\bibinfo {volume} {96}},\ \bibinfo {pages} {063311}
  (\bibinfo {year} {2017})}\BibitemShut {NoStop}%
\bibitem [{\citenamefont {Zhang}\ \emph {et~al.}(2019)\citenamefont {Zhang},
  \citenamefont {Guo},\ and\ \citenamefont {Chen}}]{ZHANG20191366}%
  \BibitemOpen
  \bibfield  {author} {\bibinfo {author} {\bibfnamefont {C.}~\bibnamefont
  {Zhang}}, \bibinfo {author} {\bibfnamefont {Z.}~\bibnamefont {Guo}}, \ and\
  \bibinfo {author} {\bibfnamefont {S.}~\bibnamefont {Chen}},\ }\href {\doibase
  10.1016/j.ijheatmasstransfer.2018.10.141} {\bibfield  {journal} {\bibinfo
  {journal} {Int. J. Heat Mass Transfer}\ }\textbf {\bibinfo {volume} {130}},\
  \bibinfo {pages} {1366} (\bibinfo {year} {2019})}\BibitemShut {NoStop}%
\bibitem [{\citenamefont {Zhang}\ \emph {et~al.}(2021)\citenamefont {Zhang},
  \citenamefont {Chen}, \citenamefont {Guo},\ and\ \citenamefont
  {Wu}}]{zhang2021e}%
  \BibitemOpen
  \bibfield  {author} {\bibinfo {author} {\bibfnamefont {C.}~\bibnamefont
  {Zhang}}, \bibinfo {author} {\bibfnamefont {S.}~\bibnamefont {Chen}},
  \bibinfo {author} {\bibfnamefont {Z.}~\bibnamefont {Guo}}, \ and\ \bibinfo
  {author} {\bibfnamefont {L.}~\bibnamefont {Wu}},\ }\href {\doibase
  10.1016/j.ijheatmasstransfer.2021.121308} {\bibfield  {journal} {\bibinfo
  {journal} {Int. J. Heat Mass Transfer}\ }\textbf {\bibinfo {volume} {174}},\
  \bibinfo {pages} {121308} (\bibinfo {year} {2021})}\BibitemShut {NoStop}%
\bibitem [{\citenamefont {Zhang}\ \emph {et~al.}(2023)\citenamefont {Zhang},
  \citenamefont {Huberman}, \citenamefont {Song}, \citenamefont {Zhao},
  \citenamefont {Chen},\ and\ \citenamefont {Wu}}]{ZHANG2023124715}%
  \BibitemOpen
  \bibfield  {author} {\bibinfo {author} {\bibfnamefont {C.}~\bibnamefont
  {Zhang}}, \bibinfo {author} {\bibfnamefont {S.}~\bibnamefont {Huberman}},
  \bibinfo {author} {\bibfnamefont {X.}~\bibnamefont {Song}}, \bibinfo {author}
  {\bibfnamefont {J.}~\bibnamefont {Zhao}}, \bibinfo {author} {\bibfnamefont
  {S.}~\bibnamefont {Chen}}, \ and\ \bibinfo {author} {\bibfnamefont
  {L.}~\bibnamefont {Wu}},\ }\href {\doibase
  https://doi.org/10.1016/j.ijheatmasstransfer.2023.124715} {\bibfield
  {journal} {\bibinfo  {journal} {Int. J. Heat Mass Transfer}\ }\textbf
  {\bibinfo {volume} {217}},\ \bibinfo {pages} {124715} (\bibinfo {year}
  {2023})}\BibitemShut {NoStop}%
\bibitem [{\citenamefont {Hu}\ \emph {et~al.}(2024)\citenamefont {Hu},
  \citenamefont {Jia}, \citenamefont {Xu}, \citenamefont {Sheng}, \citenamefont
  {Wen}, \citenamefont {Lin}, \citenamefont {Shen},\ and\ \citenamefont
  {Bao}}]{Hu_2024}%
  \BibitemOpen
  \bibfield  {author} {\bibinfo {author} {\bibfnamefont {Y.}~\bibnamefont
  {Hu}}, \bibinfo {author} {\bibfnamefont {R.}~\bibnamefont {Jia}}, \bibinfo
  {author} {\bibfnamefont {J.}~\bibnamefont {Xu}}, \bibinfo {author}
  {\bibfnamefont {Y.}~\bibnamefont {Sheng}}, \bibinfo {author} {\bibfnamefont
  {M.}~\bibnamefont {Wen}}, \bibinfo {author} {\bibfnamefont {J.}~\bibnamefont
  {Lin}}, \bibinfo {author} {\bibfnamefont {Y.}~\bibnamefont {Shen}}, \ and\
  \bibinfo {author} {\bibfnamefont {H.}~\bibnamefont {Bao}},\ }\href {\doibase
  10.1088/1361-648X/acfdea} {\bibfield  {journal} {\bibinfo  {journal} {Journal
  of Physics: Condensed Matter}\ }\textbf {\bibinfo {volume} {36}},\ \bibinfo
  {pages} {025901} (\bibinfo {year} {2024})}\BibitemShut {NoStop}%
\bibitem [{\citenamefont {Sheng}\ \emph {et~al.}(2024)\citenamefont {Sheng},
  \citenamefont {Wang}, \citenamefont {Hu}, \citenamefont {Xu}, \citenamefont
  {Ji},\ and\ \citenamefont {Bao}}]{baohua_IEEE_2024}%
  \BibitemOpen
  \bibfield  {author} {\bibinfo {author} {\bibfnamefont {Y.}~\bibnamefont
  {Sheng}}, \bibinfo {author} {\bibfnamefont {S.}~\bibnamefont {Wang}},
  \bibinfo {author} {\bibfnamefont {Y.}~\bibnamefont {Hu}}, \bibinfo {author}
  {\bibfnamefont {J.}~\bibnamefont {Xu}}, \bibinfo {author} {\bibfnamefont
  {Z.}~\bibnamefont {Ji}}, \ and\ \bibinfo {author} {\bibfnamefont
  {H.}~\bibnamefont {Bao}},\ }\href {\doibase 10.1109/TED.2024.3357440}
  {\bibfield  {journal} {\bibinfo  {journal} {IEEE Transactions on Electron
  Devices}\ }\textbf {\bibinfo {volume} {71}},\ \bibinfo {pages} {1769}
  (\bibinfo {year} {2024})}\BibitemShut {NoStop}%
\bibitem [{\citenamefont {Zhang}\ and\ \citenamefont
  {Guo}(2019)}]{zhang_discrete_2019}%
  \BibitemOpen
  \bibfield  {author} {\bibinfo {author} {\bibfnamefont {C.}~\bibnamefont
  {Zhang}}\ and\ \bibinfo {author} {\bibfnamefont {Z.}~\bibnamefont {Guo}},\
  }\href {\doibase 10.1016/j.ijheatmasstransfer.2019.02.056} {\bibfield
  {journal} {\bibinfo  {journal} {Int. J. Heat Mass Transfer}\ }\textbf
  {\bibinfo {volume} {134}},\ \bibinfo {pages} {1127} (\bibinfo {year}
  {2019})}\BibitemShut {NoStop}%
\bibitem [{\citenamefont {Kubo~R.}\ and\ \citenamefont
  {N.}(1991)}]{Kubo1991statistical}%
  \BibitemOpen
  \bibfield  {author} {\bibinfo {author} {\bibfnamefont {T.~M.}\ \bibnamefont
  {Kubo~R.}}\ and\ \bibinfo {author} {\bibfnamefont {H.}~\bibnamefont {N.}},\
  }\href {https://doi.org/10.1007/978-3-642-58244-8} {\emph {\bibinfo {title}
  {Statistical Physics II Nonequilibrium Statistical Mechanics}}},\ Springer
  Series in Solid State Sciences\ (\bibinfo  {publisher} {Springer, Berlin,
  Heidelberg},\ \bibinfo {year} {1991})\BibitemShut {NoStop}%
\bibitem [{\citenamefont {Moore}(1968)}]{Moore68Newton}%
  \BibitemOpen
  \bibfield  {author} {\bibinfo {author} {\bibfnamefont {R.~H.}\ \bibnamefont
  {Moore}},\ }\href {\doibase 10.1007/BF02170993} {\bibfield  {journal}
  {\bibinfo  {journal} {Numer. Math.}\ }\textbf {\bibinfo {volume} {12}},\
  \bibinfo {pages} {23} (\bibinfo {year} {1968})}\BibitemShut {NoStop}%
\bibitem [{\citenamefont {Lacroix}\ \emph {et~al.}(2005)\citenamefont
  {Lacroix}, \citenamefont {Joulain},\ and\ \citenamefont
  {Lemonnier}}]{Lacroix05}%
  \BibitemOpen
  \bibfield  {author} {\bibinfo {author} {\bibfnamefont {D.}~\bibnamefont
  {Lacroix}}, \bibinfo {author} {\bibfnamefont {K.}~\bibnamefont {Joulain}}, \
  and\ \bibinfo {author} {\bibfnamefont {D.}~\bibnamefont {Lemonnier}},\ }\href
  {\doibase 10.1103/PhysRevB.72.064305} {\bibfield  {journal} {\bibinfo
  {journal} {Phys. Rev. B}\ }\textbf {\bibinfo {volume} {72}},\ \bibinfo
  {pages} {064305} (\bibinfo {year} {2005})}\BibitemShut {NoStop}%
\bibitem [{\citenamefont {Majumdar}(1993)}]{MajumdarA93Film}%
  \BibitemOpen
  \bibfield  {author} {\bibinfo {author} {\bibfnamefont {A.}~\bibnamefont
  {Majumdar}},\ }\href {\doibase 10.1115/1.2910673} {\bibfield  {journal}
  {\bibinfo  {journal} {J. Heat Transfer}\ }\textbf {\bibinfo {volume} {115}},\
  \bibinfo {pages} {7} (\bibinfo {year} {1993})}\BibitemShut {NoStop}%
\bibitem [{\citenamefont {Yang}\ \emph {et~al.}(2005)\citenamefont {Yang},
  \citenamefont {Chen}, \citenamefont {Laroche},\ and\ \citenamefont
  {Taur}}]{YangRg05BDE}%
  \BibitemOpen
  \bibfield  {author} {\bibinfo {author} {\bibfnamefont {R.}~\bibnamefont
  {Yang}}, \bibinfo {author} {\bibfnamefont {G.}~\bibnamefont {Chen}}, \bibinfo
  {author} {\bibfnamefont {M.}~\bibnamefont {Laroche}}, \ and\ \bibinfo
  {author} {\bibfnamefont {Y.}~\bibnamefont {Taur}},\ }\href {\doibase
  10.1115/1.1857941} {\bibfield  {journal} {\bibinfo  {journal} {J. Heat
  Transfer}\ }\textbf {\bibinfo {volume} {127}},\ \bibinfo {pages} {298}
  (\bibinfo {year} {2005})}\BibitemShut {NoStop}%
\bibitem [{\citenamefont {Yoon}\ \emph {et~al.}(2019)\citenamefont {Yoon},
  \citenamefont {Jeong}, \citenamefont {Lee},\ and\ \citenamefont
  {Baek}}]{bulkfinfet2019}%
  \BibitemOpen
  \bibfield  {author} {\bibinfo {author} {\bibfnamefont {J.-S.}\ \bibnamefont
  {Yoon}}, \bibinfo {author} {\bibfnamefont {J.}~\bibnamefont {Jeong}},
  \bibinfo {author} {\bibfnamefont {S.}~\bibnamefont {Lee}}, \ and\ \bibinfo
  {author} {\bibfnamefont {R.-H.}\ \bibnamefont {Baek}},\ }\href {\doibase
  10.1109/ACCESS.2019.2920902} {\bibfield  {journal} {\bibinfo  {journal} {IEEE
  Access}\ }\textbf {\bibinfo {volume} {7}},\ \bibinfo {pages} {75762}
  (\bibinfo {year} {2019})}\BibitemShut {NoStop}%
\bibitem [{\citenamefont {Cai}\ \emph {et~al.}(2018)\citenamefont {Cai},
  \citenamefont {Chen}, \citenamefont {Du}, \citenamefont {Zhang},\ and\
  \citenamefont {Liu}}]{Nanosheet_FINFET2018}%
  \BibitemOpen
  \bibfield  {author} {\bibinfo {author} {\bibfnamefont {L.}~\bibnamefont
  {Cai}}, \bibinfo {author} {\bibfnamefont {W.}~\bibnamefont {Chen}}, \bibinfo
  {author} {\bibfnamefont {G.}~\bibnamefont {Du}}, \bibinfo {author}
  {\bibfnamefont {X.}~\bibnamefont {Zhang}}, \ and\ \bibinfo {author}
  {\bibfnamefont {X.}~\bibnamefont {Liu}},\ }\href {\doibase
  10.1109/TED.2018.2825498} {\bibfield  {journal} {\bibinfo  {journal} {IEEE
  Transactions on Electron Devices}\ }\textbf {\bibinfo {volume} {65}},\
  \bibinfo {pages} {2647} (\bibinfo {year} {2018})}\BibitemShut {NoStop}%
\bibitem [{\citenamefont {Loubet}\ \emph {et~al.}(2017)\citenamefont {Loubet},
  \citenamefont {Hook}, \citenamefont {Montanini}, \citenamefont {Yeung},
  \citenamefont {Kanakasabapathy}, \citenamefont {Guillom}, \citenamefont
  {Yamashita}, \citenamefont {Zhang}, \citenamefont {Miao}, \citenamefont
  {Wang}, \citenamefont {Young}, \citenamefont {Chao}, \citenamefont {Kang},
  \citenamefont {Liu}, \citenamefont {Fan}, \citenamefont {Hamieh},
  \citenamefont {Sieg}, \citenamefont {Mignot}, \citenamefont {Xu},
  \citenamefont {Seo}, \citenamefont {Yoo}, \citenamefont {Mochizuki},
  \citenamefont {Sankarapandian}, \citenamefont {Kwon}, \citenamefont {Carr},
  \citenamefont {Greene}, \citenamefont {Park}, \citenamefont {Frougier},
  \citenamefont {Galatage}, \citenamefont {Bao}, \citenamefont {Shearer},
  \citenamefont {Conti}, \citenamefont {Song}, \citenamefont {Lee},
  \citenamefont {Kong}, \citenamefont {Xu}, \citenamefont {Arceo},
  \citenamefont {Bi}, \citenamefont {Xu}, \citenamefont {Muthinti},
  \citenamefont {Li}, \citenamefont {Wong}, \citenamefont {Brown},
  \citenamefont {Oldiges}, \citenamefont {Robison}, \citenamefont {Arnold},
  \citenamefont {Felix}, \citenamefont {Skordas}, \citenamefont {Gaudiello},
  \citenamefont {Standaert}, \citenamefont {Jagannathan}, \citenamefont
  {Corliss}, \citenamefont {Na}, \citenamefont {Knorr}, \citenamefont {Wu},
  \citenamefont {Gupta}, \citenamefont {Lian}, \citenamefont {Divakaruni},
  \citenamefont {Gow}, \citenamefont {Labelle}, \citenamefont {Lee},
  \citenamefont {Paruchuri}, \citenamefont {Bu},\ and\ \citenamefont
  {Khare}}]{IBM_2017_nanosheet}%
  \BibitemOpen
  \bibfield  {author} {\bibinfo {author} {\bibfnamefont {N.}~\bibnamefont
  {Loubet}}, \bibinfo {author} {\bibfnamefont {T.}~\bibnamefont {Hook}},
  \bibinfo {author} {\bibfnamefont {P.}~\bibnamefont {Montanini}}, \bibinfo
  {author} {\bibfnamefont {C.-W.}\ \bibnamefont {Yeung}}, \bibinfo {author}
  {\bibfnamefont {S.}~\bibnamefont {Kanakasabapathy}}, \bibinfo {author}
  {\bibfnamefont {M.}~\bibnamefont {Guillom}}, \bibinfo {author} {\bibfnamefont
  {T.}~\bibnamefont {Yamashita}}, \bibinfo {author} {\bibfnamefont
  {J.}~\bibnamefont {Zhang}}, \bibinfo {author} {\bibfnamefont
  {X.}~\bibnamefont {Miao}}, \bibinfo {author} {\bibfnamefont {J.}~\bibnamefont
  {Wang}}, \bibinfo {author} {\bibfnamefont {A.}~\bibnamefont {Young}},
  \bibinfo {author} {\bibfnamefont {R.}~\bibnamefont {Chao}}, \bibinfo {author}
  {\bibfnamefont {M.}~\bibnamefont {Kang}}, \bibinfo {author} {\bibfnamefont
  {Z.}~\bibnamefont {Liu}}, \bibinfo {author} {\bibfnamefont {S.}~\bibnamefont
  {Fan}}, \bibinfo {author} {\bibfnamefont {B.}~\bibnamefont {Hamieh}},
  \bibinfo {author} {\bibfnamefont {S.}~\bibnamefont {Sieg}}, \bibinfo {author}
  {\bibfnamefont {Y.}~\bibnamefont {Mignot}}, \bibinfo {author} {\bibfnamefont
  {W.}~\bibnamefont {Xu}}, \bibinfo {author} {\bibfnamefont {S.-C.}\
  \bibnamefont {Seo}}, \bibinfo {author} {\bibfnamefont {J.}~\bibnamefont
  {Yoo}}, \bibinfo {author} {\bibfnamefont {S.}~\bibnamefont {Mochizuki}},
  \bibinfo {author} {\bibfnamefont {M.}~\bibnamefont {Sankarapandian}},
  \bibinfo {author} {\bibfnamefont {O.}~\bibnamefont {Kwon}}, \bibinfo {author}
  {\bibfnamefont {A.}~\bibnamefont {Carr}}, \bibinfo {author} {\bibfnamefont
  {A.}~\bibnamefont {Greene}}, \bibinfo {author} {\bibfnamefont
  {Y.}~\bibnamefont {Park}}, \bibinfo {author} {\bibfnamefont {J.}~\bibnamefont
  {Frougier}}, \bibinfo {author} {\bibfnamefont {R.}~\bibnamefont {Galatage}},
  \bibinfo {author} {\bibfnamefont {R.}~\bibnamefont {Bao}}, \bibinfo {author}
  {\bibfnamefont {J.}~\bibnamefont {Shearer}}, \bibinfo {author} {\bibfnamefont
  {R.}~\bibnamefont {Conti}}, \bibinfo {author} {\bibfnamefont
  {H.}~\bibnamefont {Song}}, \bibinfo {author} {\bibfnamefont {D.}~\bibnamefont
  {Lee}}, \bibinfo {author} {\bibfnamefont {D.}~\bibnamefont {Kong}}, \bibinfo
  {author} {\bibfnamefont {Y.}~\bibnamefont {Xu}}, \bibinfo {author}
  {\bibfnamefont {A.}~\bibnamefont {Arceo}}, \bibinfo {author} {\bibfnamefont
  {Z.}~\bibnamefont {Bi}}, \bibinfo {author} {\bibfnamefont {P.}~\bibnamefont
  {Xu}}, \bibinfo {author} {\bibfnamefont {R.}~\bibnamefont {Muthinti}},
  \bibinfo {author} {\bibfnamefont {J.}~\bibnamefont {Li}}, \bibinfo {author}
  {\bibfnamefont {R.}~\bibnamefont {Wong}}, \bibinfo {author} {\bibfnamefont
  {D.}~\bibnamefont {Brown}}, \bibinfo {author} {\bibfnamefont
  {P.}~\bibnamefont {Oldiges}}, \bibinfo {author} {\bibfnamefont
  {R.}~\bibnamefont {Robison}}, \bibinfo {author} {\bibfnamefont
  {J.}~\bibnamefont {Arnold}}, \bibinfo {author} {\bibfnamefont
  {N.}~\bibnamefont {Felix}}, \bibinfo {author} {\bibfnamefont
  {S.}~\bibnamefont {Skordas}}, \bibinfo {author} {\bibfnamefont
  {J.}~\bibnamefont {Gaudiello}}, \bibinfo {author} {\bibfnamefont
  {T.}~\bibnamefont {Standaert}}, \bibinfo {author} {\bibfnamefont
  {H.}~\bibnamefont {Jagannathan}}, \bibinfo {author} {\bibfnamefont
  {D.}~\bibnamefont {Corliss}}, \bibinfo {author} {\bibfnamefont {M.-H.}\
  \bibnamefont {Na}}, \bibinfo {author} {\bibfnamefont {A.}~\bibnamefont
  {Knorr}}, \bibinfo {author} {\bibfnamefont {T.}~\bibnamefont {Wu}}, \bibinfo
  {author} {\bibfnamefont {D.}~\bibnamefont {Gupta}}, \bibinfo {author}
  {\bibfnamefont {S.}~\bibnamefont {Lian}}, \bibinfo {author} {\bibfnamefont
  {R.}~\bibnamefont {Divakaruni}}, \bibinfo {author} {\bibfnamefont
  {T.}~\bibnamefont {Gow}}, \bibinfo {author} {\bibfnamefont {C.}~\bibnamefont
  {Labelle}}, \bibinfo {author} {\bibfnamefont {S.}~\bibnamefont {Lee}},
  \bibinfo {author} {\bibfnamefont {V.}~\bibnamefont {Paruchuri}}, \bibinfo
  {author} {\bibfnamefont {H.}~\bibnamefont {Bu}}, \ and\ \bibinfo {author}
  {\bibfnamefont {M.}~\bibnamefont {Khare}},\ }in\ \href {\doibase
  10.23919/VLSIT.2017.7998183} {\emph {\bibinfo {booktitle} {2017 Symposium on
  VLSI Technology}}}\ (\bibinfo {year} {2017})\ pp.\ \bibinfo {pages}
  {T230--T231}\BibitemShut {NoStop}%
\bibitem [{\citenamefont {Venkateswarlu}\ and\ \citenamefont
  {Nayak}(2020)}]{gaafet2020}%
  \BibitemOpen
  \bibfield  {author} {\bibinfo {author} {\bibfnamefont {S.}~\bibnamefont
  {Venkateswarlu}}\ and\ \bibinfo {author} {\bibfnamefont {K.}~\bibnamefont
  {Nayak}},\ }\href {\doibase 10.1109/TED.2020.3017567} {\bibfield  {journal}
  {\bibinfo  {journal} {IEEE Transactions on Electron Devices}\ }\textbf
  {\bibinfo {volume} {67}},\ \bibinfo {pages} {4493} (\bibinfo {year}
  {2020})}\BibitemShut {NoStop}%
\bibitem [{\citenamefont {Jain}\ \emph {et~al.}(2018)\citenamefont {Jain},
  \citenamefont {Gupta}, \citenamefont {Hook},\ and\ \citenamefont
  {Dixit}}]{finfet2018sub14}%
  \BibitemOpen
  \bibfield  {author} {\bibinfo {author} {\bibfnamefont {I.}~\bibnamefont
  {Jain}}, \bibinfo {author} {\bibfnamefont {A.}~\bibnamefont {Gupta}},
  \bibinfo {author} {\bibfnamefont {T.~B.}\ \bibnamefont {Hook}}, \ and\
  \bibinfo {author} {\bibfnamefont {A.}~\bibnamefont {Dixit}},\ }\href
  {\doibase 10.1109/TED.2018.2863730} {\bibfield  {journal} {\bibinfo
  {journal} {IEEE Transactions on Electron Devices}\ }\textbf {\bibinfo
  {volume} {65}},\ \bibinfo {pages} {4238} (\bibinfo {year}
  {2018})}\BibitemShut {NoStop}%
\end{thebibliography}%
\end{document}